\documentclass[fleqn,usenatbib]{mnras}

\usepackage{newtxtext,newtxmath}

\usepackage[T1]{fontenc}

\DeclareRobustCommand{\VAN}[3]{#2}
\let\VANthebibliography\thebibliography
\def\thebibliography{\DeclareRobustCommand{\VAN}[3]{##3}\VANthebibliography}

\usepackage{graphicx}	
\usepackage{amsmath} 
 
\usepackage{amssymb}	

\usepackage{siunitx}
\usepackage{xcolor}
\usepackage{multirow}
\usepackage{multicol}
\usepackage{hyperref}

\DeclareSIUnit \parsec {pc}

\newcommand{\add}[1]{{#1}}

\graphicspath{{Figures}}

\title[DEVILS: Cosmic evolution of SED metallicities]{DEVILS: Cosmic evolution of SED-derived metallicities and their connection to star-formation histories}

\author[J. E. Thorne et al.]{
Jessica E. Thorne,$^{1}$\thanks{E-mail: jessica.thorne@icrar.org}
Aaron S. G. Robotham,$^{1}$
Sabine Bellstedt,$^{1}$
Luke J. M. Davies,$^{1}$\newauthor
Robin H. W. Cook,$^{1}$
Luca Cortese,$^{1}$
Benne Holwerda,$^{2}$
Steven Phillipps,$^{3}$
Malgorzata Siudek$^{4,5}$
\\
$^{1}$ ICRAR, The University of Western Australia, 35 Stirling Highway, Crawley, WA 6009, Australia\\
$^{2}$ Department of Physics and Astronomy, University of Louisville, Natural Science Building 102, 40292 KY, Louisville, USA\\
$^{3}$ Astrophysics Group, School of Physics, University of Bristol, Bristol BS8 1TL, UK\\
$^{4}$Institut de F\'{\i}sica d'Altes Energies (IFAE), The Barcelona Institute of Science and Technology, 08193 Bellaterra (Barcelona), Spain\\
$^{5}$Institute of Space Sciences (ICE, CSIC), Campus UAB, Carrer de Magrans, 08193 Barcelona, Spain\\
}

\date{Accepted XXX. Received YYY; in original form ZZZ}

\pubyear{2022}

\begin{document}
\label{firstpage}
\pagerange{\pageref{firstpage}--\pageref{lastpage}}
\maketitle

\begin{abstract}
Gas-phase metallicities of galaxies are typically measured through \add{auroral or nebular} emission lines, but metallicity also leaves an imprint on the overall spectral energy distribution (SED) of a galaxy and can be estimated through SED fitting. 
We use the \textsc{ProSpect} SED fitting code with a flexible parametric star formation history and an evolving metallicity history to self-consistently measure metallicities, stellar mass, and other galaxy properties for $\sim90\,000$ galaxies from the Deep Extragalactic VIsible Legacy Survey (DEVILS) and Galaxy and Mass Assembly (GAMA) survey. 
We use these to trace the evolution of the mass-metallicity relation (MZR) and show that the MZR only evolves in normalisation by $\sim0.1\,$dex at stellar mass $M_\star = 10^{10.5}\,M_\odot$.
We find no difference in the MZR between galaxies with and without SED evidence of active galactic nuclei emission at low redshifts ($z<0.3$).
Our results suggest an anti-correlation between metallicity and star formation activity at fixed stellar mass for galaxies with $M_\star > 10^{10.5}\,M_\odot$ for $z<0.3$.
Using the star formation histories extracted using \textsc{ProSpect} we explore higher-order correlations of the MZR with properties of the star formation history including age, width, and shape. 
We find that at a given stellar mass, galaxies with higher metallicities formed most of their mass over shorter timescales, and before their peak star formation rate. 
\add{This work highlights the value of exploring the connection of a galaxy's current gas-phase metallicity to its star formation history in order to understand the physical processes shaping the MZR. }
\end{abstract}





\section{Introduction}\label{sec:Intro}
\add{The relationship between the stellar mass and gas phase metallicity (MZR) of galaxies has been well studied and is known to be positively correlated \citep[e.g.][]{LequeuxChemicalCompositionEvolution1979,TremontiOriginMassMetallicity2004}.
This correlation has been shown to hold to $z=3.5$ \citep{MaiolinoAMAZEevolutionmassmetallicity2008,MannucciLSDLymanbreakgalaxies2009,SandersMOSDEFSurveyEvolution2021,CullenNIRVANDELSSurveyrobust2021}}, and shows evidence of flattening at higher stellar masses ($M_\star > 10^{10}\,M_\odot$,  \citealt{Lara-LopezStudystarforminggalaxies2010,Lara-LopezGalaxyMassAssembly2013,Curtimassmetallicityfundamentalmetallicity2020a}).

The MZR provides essential insight into galaxy evolution as metallicity is a tracer of the fraction of baryonic mass that has been converted into stars but is also sensitive to metal loss as a result of stellar winds, supernovae and active galactic nuclei (AGN) feedback \citep{LarsonGaslossgroups1975,BrooksOriginEvolutionMassMetallicity2007,deRossiCluesoriginfundamental2007,KobayashiSimulationsCosmicChemical2007,DaveGalaxyevolutioncosmological2011,TumlinsonCircumgalacticMedium2017}.

\add{The MZR is thought to be shaped by two main physical processes: metal production and metal loss.
To first order, the correlation between stellar mass and gas-phase metallicity can be explained using a closed box scenario where a galaxy initially consists of metal-poor gas which collapses to form stars and no gas enters or leaves the galaxy (see the review by \citealt{TinsleyEvolutionStarsGas1980}). 
In this scenario, high-mass stars produce metals through nucleosynthesis and return these metals to the interstellar medium through stellar winds. 
With these assumptions, the gas-phase metallicity ($Z_\text{gas}$) increases as gas is processed through stars. 
This is often described as `chemical downsizing' and suggests that low mass systems are still at an early evolutionary stage and have yet to convert most of their gas into stars \citep{MaiolinoAMAZEevolutionmassmetallicity2008,Caluraevolutionmassmetallicityrelation2009,ValeAsarievolutionmassmetallicityrelation2009,ZahidMassMetallicityLuminosityMetallicityRelations2011,SomervillePhysicalModelsGalaxy2015,SpitoniConnectiongalacticdownsizing2020,ZhouSemianalyticspectralfitting2022}. 
}

\add{
Metal loss refers to outflows generated by starburst winds (or potentially AGN feedback, see \citealt{Camps-FarinaSignaturesAGNinducedMetal2021}) which eject metal-enriched gas into the inter-galactic medium \citep{GarnettLuminosityMetallicityRelationEffective2002, TremontiOriginMassMetallicity2004,DalcantonMetallicityGalaxyDisks2007, PeeplesConstraintsstarformation2011,ChisholmMetalenrichedgalacticoutflows2018}.
Outflows are often used to explain the lower metallicity in low-mass systems where gas can be efficiently expelled by winds and outflows due to a shallower gravitational potential\citep{DEugeniogasphasemetallicitiesstarforming2018}.
However, the importance of outflows might be lower at high redshift where gas fractions are significantly higher and metal dilution through gas accretion might play a larger role (see e.g. \citealt{ScovilleEvolutionInterstellarMedium2017,SandersCOEmissionMolecular2022}). 
}

\add{The fact that the physical processes driving the MZR are poorly understood has prompted a number of authors to consider potential higher dimension relations including additional physical properties which could provide better constraints on the processes driving metallicity evolution.
Of these, the three dimensional relationship between mass, metallicity and star formation rate (SFR) or specific SFR (sSFR) has been studied in the most detail  \citep[e.g.][]{EllisonCluesOriginMassMetallicity2008,Lara-LopezStudystarforminggalaxies2010,Mannuccifundamentalrelationmass2010,Yatesrelationmetallicitystellar2012,Lara-LopezGalaxyMassAssembly2013,HuntCoevolutionmetallicitystar2016,HirschauerMetalAbundancesKISS2018,Curtimassmetallicityfundamentalmetallicity2020a,BellstedtGalaxyMassAssembly2021}.
\cite{EllisonCluesOriginMassMetallicity2008} first showed an anti-correlation between metallicity and sSFR for galaxies at fixed stellar mass. 
\cite{Mannuccifundamentalrelationmass2010} also observed a secondary dependence of the MZR on the SFR. 
Both \cite{Yatesrelationmetallicitystellar2012} and \cite{Lara-LopezGalaxyMassAssembly2013} find a similar anti-correlation for low-mass galaxies, however, they demonstrate that for massive galaxies, the median metallicity is higher for galaxies with higher SFRs.  
\cite{Lara-LopezGalaxyMassAssembly2013} attribute this to a combination of downsizing and differing amounts of neutral gas, while \cite{Yatesrelationmetallicitystellar2012} suggest that accretion may play a significant role in regulating the gas-phase metallicities of massive galaxies. 
\citet{Curtimassmetallicityfundamentalmetallicity2020a} also find a strong anti-correlation between metallicity and SFR at low masses, however, they find that the anti-correlation weakens with increasing mass until disappearing at high masses ($M_\star > 10^{10.5}\,M_\odot$).
}

Another key element to understanding the MZR is understanding how it has evolved with time. 
Studies of metallicities in high redshift galaxies have shown that MZR has evolved in normalisation since earlier times, but has maintained a similar shape \citep[e.g.][]{Mannuccifundamentalrelationmass2010,HenryMetallicityEvolutionLowmass2013,SandersMOSDEFSurveyMass2015,LyMetalAbundancesCosmic2016,KashinoFMOSCOSMOSSurveyStarforming2017,SandersMOSDEFSurveyStellar2018,KashinoFMOSCOSMOSSurveyStarforming2019,Weldonstellarpopulationmetalpoor2020,SandersMOSDEFsurveydirectmethod2020,SandersMOSDEFSurveyEvolution2021,ChartabMOSDEFSurveyEnvironmental2021b,ToppingMOSDEFsurveymassmetallicity2021,Gillmanresolvedchemicalabundance2022}.
Understanding how mass, metallicity, and star formation evolve with time and in relation to one another is critical to understand the physical processes that govern the efficiency and timing of star formation in galaxies. 

\add{Gas-phase metallicities of galaxies are typically measured through auroral or nebular emission lines.}
However, metallicity also leaves an imprint on the overall spectral energy distribution of a galaxy. 
\cite{BellstedtGalaxyMassAssembly2021} used metallicities derived using SED fitting for 4,500 galaxies with $z < 0.06$ from the Galaxy and Mass Assembly (GAMA) survey to extract the MZR at $z\approx0$. 
Using 37\,809 galaxies from the Deep Extgragalactic VIsible Legacy Survey (DEVILS) and 50\,147 galaxies from the GAMA survey, we extend this analysis to $z\approx 1$ to investigate the evolution of the MZR and the impact of star formation history on the scatter of the MZR. 

The structure of this paper is as follows.
We describe the DEVILS and GAMA projects, related data sets, and the SED fitting method employed for this work in Section~\ref{sec:Data}.
We present the recovered MZR for $z = 0-1$ in Section~\ref{sec:MZR} including the MZR for AGN host galaxies and a comparison to simulations. 
In Section~\ref{sec:SFHs} we investigate the impact of star formation history on the scatter of the MZR. 
We summarise our results in Section~\ref{sec:conclusion}.
Throughout this work we use a \cite{ChabrierGalacticStellarSubstellar2003} IMF and all magnitudes are quoted in the AB system. 
We adopt the \cite{PlanckCollaborationPlanck2015results2016} cosmology with $H_0 = 67.8 \, \si{\kilo \meter \per \second \per \mega \parsec}$, $\Omega_{M} = 0.308$ and $\Omega_\Lambda = 0.692$. 

\section{Data} \label{sec:Data}

\subsection{Deep Extragalactic VIsible Legacy Survey}
For this work we use the Deep Extragalactic VIsible Legacy Survey (DEVILS; \citealt{DaviesDeepExtragalacticVIsible2018}). DEVILS is an optical spectroscopic redshift survey using the Anglo-Australian Telescope specifically designed to have high spectroscopic completeness over a large redshift range ($z < 1$) in three well-studied extragalactic fields: XMM-LSS/D02, ECDFS/D03, and COSMOS/D10 covering a total of 4.5 deg$^2$. 
DEVILS has measured spectroscopic redshifts to extend the redshift completeness limit by 1-2 mags in each of these regions to an 85\% completeness limit of $Y_{\text{mag}} = 20/19.6/20.7$ in D02/D03/D10 respectively.
This amounts to over 56\,000 spectroscopic redshifts with high completeness, allowing for robust characterization of group and pair environments in the distant universe. For a full description of the survey science goals, survey design, target selection and spectroscopic observations see \cite{DaviesDeepExtragalacticVIsible2018}. 

In \cite{ThorneDeepExtragalacticVIsible2021, ThorneDeepExtragalacticVIsible2022} we use the spectroscopic and photometric data from the D10-COSMOS field as it is the deepest field. 
We use the DEVILS photometry catalogue derived using the \textsc{ProFound} source extraction code \citep{RobothamProFoundSourceExtraction2018} and described in depth by \cite{DaviesDeepExtragalacticVIsible2021}.
\textsc{ProFound} is used for source finding and photometry extraction consistently across 22 bands spanning the FUV-FIR (1500\,\AA- 500\,$\mu$m) and includes GALEX \textit{FUV NUV} \citep{ZamojskiDeepGALEXImaging2007}, CFHT \textit{u} \citep{CapakFirstReleaseCOSMOS2007},  Subaru HSC \textit{griz} \citep{AiharaSeconddatarelease2019},  VISTA \textit{YJHK$_{s}$} \citep{McCrackenUltraVISTAnewultradeep2012}, Spitzer \textit{IRAC1 IRAC2 IRAC3 IRAC4 MIPS24 MIPS70} \citep{LaigleCOSMOS2015CATALOGEXPLORING2016,SandersSCOSMOSSpitzerLegacy2007}, and Herschel \textit{P100 P160 S250 S350 S500} \citep{LutzPACSEvolutionaryProbe2011,OliverHerschelMultitieredExtragalactic2012} bands.
\add{Briefly, photometry was extracted in two phases to account for large differences in resolution and depth between the FUV-NIR and MIR-FIR regimes.
Photometry for the \textit{FUV-IRAC4} bands was extracted using the segment mode in \textsc{ProFound} while the photometry for the \textit{MIPS24-S500} bands was extracted using the \texttt{FitMagPSF} mode in \textsc{ProFound}. 
Due to poor resolution and shallow imaging, FIR photometry was only extracted for optically selected objects with $Y<21.2$mag or that were detected in the MIPS24 imaging. 
If an object met the criteria for FIR photometry extraction it was passed through the \texttt{FitMagPSF} mode.
Objects that met the FIR photometry criteria but were not detected in a FIR band have a flux measurement of zero and a flux error measured from the sky noise. 
If no attempt was made to measure FIR photometry for an object then it will have no value for the flux and flux error.
5 per cent of all optically-detected objects met the FIR photometry criteria and therefore have FIR photometry measurements.
}

As described in \cite{ThorneDeepExtragalacticVIsible2021}, redshift catalogues for DEVILS \add{spanning $0<z<8$} have been compiled using photometric, grism, and spectroscopic redshifts. 
The redshift sources used for D10 are presented in table C1 of \cite{ThorneDeepExtragalacticVIsible2021} and include 3\,394 redshifts measured as part of the DEVILS program. 
We use these photometry and redshift measurements in this work.
As in \cite{ThorneDeepExtragalacticVIsible2021}, we remove all objects classed as stars (\textit{starflag} column), artefacts (\textit{artefactflag}) or that are masked (\textit{mask}). 
\add{Each of these flags are described in detail by \cite{DaviesDeepExtragalacticVIsible2021}, but briefly, stars and ambiguous objects are identified through size and colour, with cuts defined using the source type derived from the photometric redshift fitting code \textit{LePhare} from the COSMOS2015 catalogue \citep{LaigleCOSMOS2015CATALOGEXPLORING2016}.
Potential artefacts are flagged where flux is not associated with an astronomical source if the source is only detected in one optical/NIR band, if the source's Y-band R50 is less than half a pixel, or if the source has an r-Z colour $< -0.75$\,mag (an unphysical colour).
Masking is performed to remove ghosting around bright stars as this is a significant problem in the DEVILS imaging (see figure~8 of \citealt{DaviesDeepExtragalacticVIsible2021}).}
This results in 494\,084 objects, of which 24\,099 have spectroscopic redshifts and 7\,307 have grism redshifts.

\subsection{Galaxy and Mass Assembly Survey}
To supplement DEVILS at low-redshift, we also use the spectroscopic and photometric data from the 4th data release of the GAMA survey \citep{DriverGalaxyMassAssembly2022}.
GAMA was a large spectroscopic campaign on the Anglo-Australian Telescope targeting five fields (G02, G09, G12, G15, and G23), which gathered redshifts for $\sim300\,000$ galaxies across a total sky area of 230 square degrees.
GAMA targets were selected by size and colour above a magnitude limit of $r_\text{mag} \le 19.8$ (or $i_\text{mag} \le 19.0$ in G23). 

We use the far-UV to far-IR photometry derived using the \textsc{ProFound} source-finding software described in detail by \cite{BellstedtGalaxyMassAssembly2020a}.
The photometric bands from this data release include GALEX \textit{FUV} and \textit{NUV} \citep{MartinGalaxyEvolutionExplorer2005}; VST \textit{u, g, r, i} \citep{deJongKiloDegreeSurvey2013,deJongKiloDegreeSurvey2013a}; VISTA \textit{Z, Y, J, H, K$_S$} \citep{EdgeVISTAKilodegreeInfrared2013}; WISE \textit{W1, W2, W3, W4} \citep{WrightWidefieldInfraredSurvey2010} and Herschel P100, P160, S250, S350, and S500 \citep{EalesHerschelATLAS2010}. 
The photometry for GAMA is derived in much the same way as for DEVILS, with minor differences due to variations in depth between the surveys. 

For this work, we use the three equatorial fields (G09, G12, and G15) as well as G23.
We select all objects with $z>0$, \add{a redshift quality flag $nQ \ge 3$ (where spectroscopic redshifts are reasonably certain, $P>90\,$per cent), and are classified as a galaxy based on size and colour (\texttt{UBERCLASS=galaxy}).}
This results in a sample of 233\,762 galaxies all of which have spectroscopic redshifts.

\begin{figure*}
    \centering
    \includegraphics[width = \linewidth]{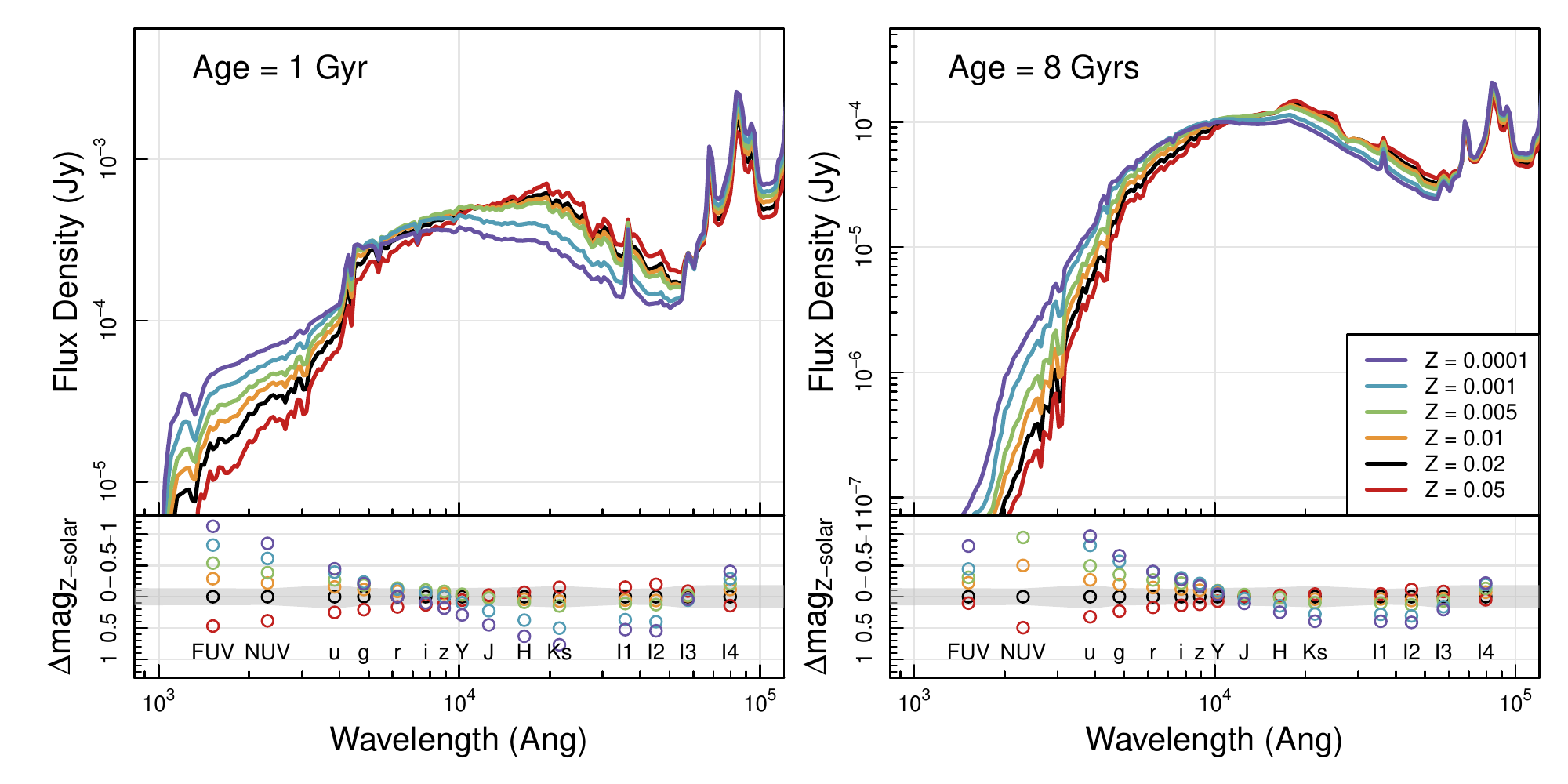}
    \caption{The impact of changing the final metallicity on the FUV-NIR SED of a young (left) and old (right) galaxies. 
    We show various metallicity values spanning the range of the \citet{BruzualStellarpopulationsynthesis2003} stellar templates as the different colour lines ranging from $Z=0.05$ (red) to $Z=0.0001$ (purple).
    We show $Z=0.02$ as the black line as this is the typically adopted solar metallicity when SED fitting.
    The bottom panels show the magnitude difference that would be measured between an SED of a galaxy with each given metallicity and with solar metallicity using the DEVILS filter set.
    The grey shaded region shows the typical errors and error floors in each band for galaxies in the DEVILS sample with $z<1.1$ \add{and highlights that differences in final metallicity are recoverable through broad-band SED fitting}.}
    \label{fig:MetallicityInfluence}
\end{figure*}

\subsection{SED Modelling}\label{Sec:SED}

\subsubsection{Impact of metallicity on a galaxy's SED}
The metallicity of stars in a galaxy impacts its SED in two ways.
Firstly, an increase in metallicity in a stellar population results in lower effective temperatures, including a cooler main sequence and giant branch.
Secondly, at fixed effective temperature, an increase in metallicity results in strong spectroscopic absorption features. 
Both of these effects result in an overall reddening of an SED with increasing metallicity (see \citealt{ConroyModelingPanchromaticSpectral2013} for more details).


To highlight the impact of metallicity on a galaxy's SED, Figure~\ref{fig:MetallicityInfluence} shows SEDs generated using \textsc{ProSpect} \citep{RobothamProSpectgeneratingspectral2020} for two mock galaxies with ages 1 and 8\,Gyrs and stellar masses of $M_\star = 10^{10}\,M_\odot$.
We assume that chemical enrichment is linearly mapped to the mass growth and show the resulting SED generated with a range of final metallicities spanning the limits of the \cite{BruzualStellarpopulationsynthesis2003} stellar population models. 
It is clear that for both stellar ages, the change in metallicity has a noticeable impact on the slope of the UV-NIR SED where the higher metallicity corresponds to a steeper (redder) slope. 
We also show the expected difference in magnitudes expected for a galaxy with each given metallicity in relation to the expected photometry for galaxy with solar metallicity ($Z=0.02$).
We use the DEVILS filter set (see Section~\ref{sec:Data} for more details) and show the typical photometry uncertainty and error floor for the DEVILS sample in each band as the grey shaded region. 
For galaxies with the same age but different metallicities, the difference in predicted photometry is larger than the average uncertainty. 
This is especially true for galaxies with lower metallicities, clearly showing that differences in metallicity result in signatures that are detectable with current imaging. 

The imprint of metallicity on a galaxy's SED can be detected through broad-band SED fitting as the differences in expected SED are larger than the uncertainties on current, high quality photometry. 
However, measuring gas or stellar metallicity from photometric data alone is difficult due to the age-metallicity-dust degeneracy (see \citealt{WortheyComprehensivestellarpopulation1994,PapovichStellarPopulationsEvolution2001}): a galaxy can appear red either because it does not form stars anymore, because it has a high metallicity, or because it is strongly attenuated. 
Because of this difficulty, most SED fitting codes take simple approaches when modelling the metallicity of galaxies.
Most commonly, SED fitting codes allow the metallicity value to be a modelled as a free parameter but will assume a constant value over the lifetime of the galaxy \citep[e.g.][]{LejaHotDustPanchromatic2018,CarnallInferringstarformation2018,JohnsonStellarPopulationInference2021}.  
In other cases they will take the even simpler approach of fixing the metallicity to the solar value \citep{BoquienCIGALEpythonCode2019}.
The consequence of making these assumptions is that the assumed metallicity evolution will generally affect other parameters of interest such as the stellar mass, dust opacity, and star formation rate (see, e.g., \citealt{WuytsRecoveringStellarPopulation2009,MarchesiniEvolutionStellarMass2009,PforrRecoveringgalaxystellar2012,BellstedtGalaxyMassAssembly2020b}). 
It is also important to note that the use of different stellar population synthesis (SPS) models will heavily impact the recovered metallicities as there is a large contribution in the NIR from asymptotic giant branch (AGB) stars which are treated very differently across different SPS models \citep{LeeAgeMetallicityEstimation2007,EminianPhysicalinterpretationnearinfrared2008}. 
We explore the impact of this in detail in Appendix~\ref{App:BPASSBC03} and find that when comparing the \citet[BC03]{BruzualStellarpopulationsynthesis2003} and \citet[BPASS]{EldridgeBinaryPopulationSpectral2017} SPS models, metallicities recovered using the BPASS models can be up to 1\,dex lower than metallicities recovered using BC03.

\subsubsection{SED fitting with ProSpect}\label{sec:ProSpect}

\begin{figure*}
    \centering
    \includegraphics[width = \linewidth]{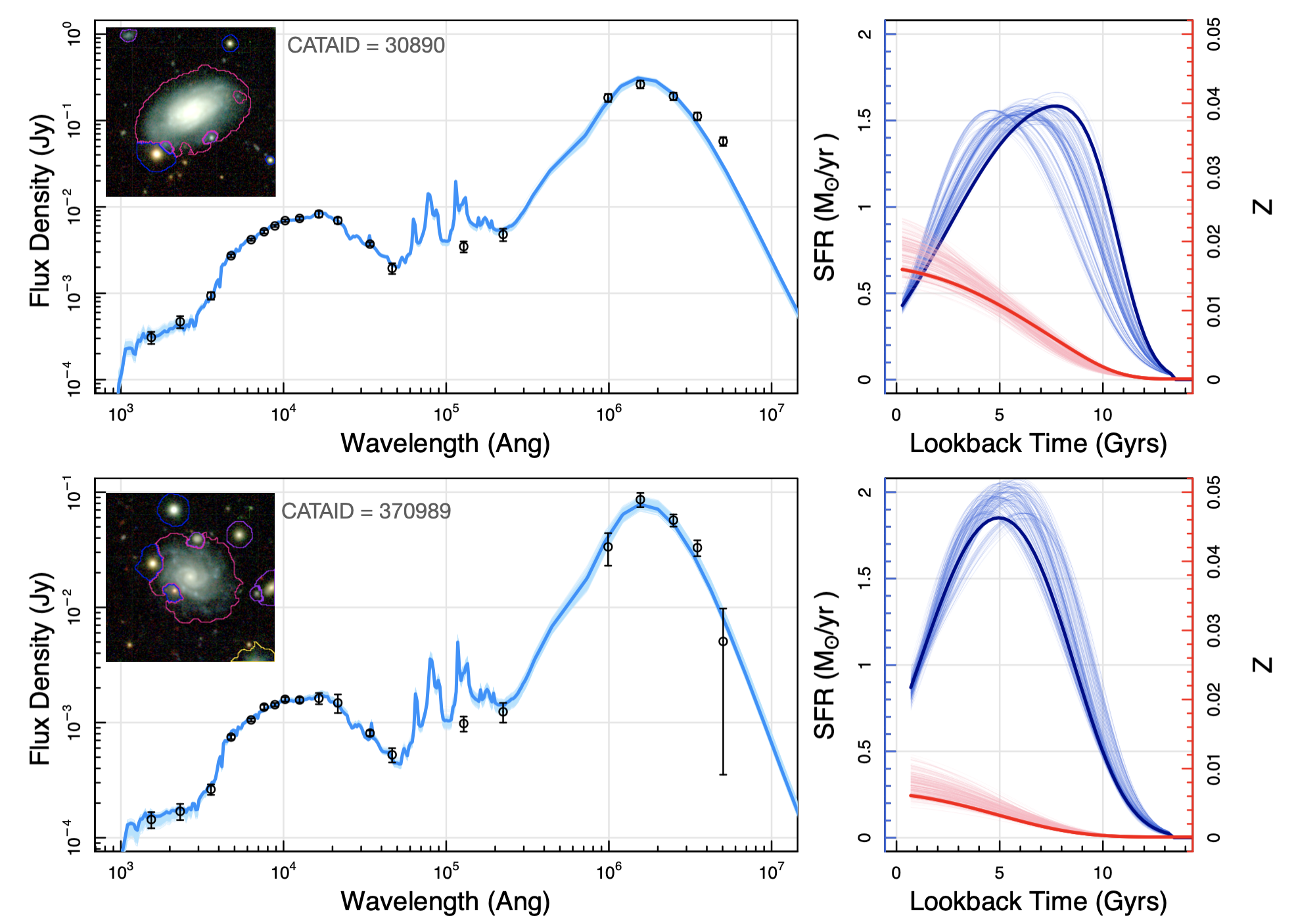}
    \caption[Example SED fits from \textsc{ProSpect}]{
    \add{Two example outputs from \textsc{ProSpect} for GAMA galaxies 30890 (upper) and 370989 (lower).
    Both galaxies have a stellar mass of $M_\star \approx 10^{9.8}\,M_\odot$, but galaxy 30890 has a final metallicity of \texttt{Zfinal} = 0.006, and galaxy 370989 has a metallicity of Z = 0.016.
    The left panel for both galaxies shows the input photometry (black circles and error bars), the best fit SED in blue and the SED generated from each step of the final MCMC chain in pale blue. 
    The inset image shows a colour image of each galaxy generated using the \textit{g,r,} and \textit{Z} imaging. The pink outline shows the segment defined for the galaxy shown, stars are shown in blue, other GAMA galaxies in purple, and masked segments in yellow. 
    The right panel shows the best fit SFH in blue and the posterior sampling in pale blue with the scale given by the left axis. 
    The metallicity history for each galaxy is also presented in the right panel as the red line for the best fit solution and as the pink lines for the sampling of the posterior. The scale for the metallicity history is shown on the right axis.}
    }
    \label{fig:SEDExamples}
\end{figure*}

For this analysis we use the SED fitting results for the D10 and GAMA fields presented in \cite{ThorneDeepExtragalacticVIsible2022}.
Briefly, we use the \textsc{ProSpect} SED fitting code \citep{RobothamProSpectgeneratingspectral2020}, with the \cite{BruzualStellarpopulationsynthesis2003} stellar templates, \cite{ChabrierGalacticStellarSubstellar2003} IMF and the \cite{CharlotSimpleModelAbsorption2000} dust attenuation and \cite{DaleTwoParameterModelInfrared2014} dust re-emission models. 
In our analysis we use the \texttt{massfunc\_snorm\_trunc} parameterisation for the star formation history, which takes the form of a skewed Normal distribution, with the peak position (\texttt{mpeak}), peak SFR (\texttt{mSFR}), SFH width (\texttt{mperiod}), and SFH skewness (\texttt{mskew}) set as free parameters. 
The SFH is anchored to 0 at a lookback time of 13.4 Gyr, selected to be the age at which galaxies start forming (equivalent to $z=11$, \citealt{OeschREMARKABLYLUMINOUSGALAXY2016}).

Although a galaxy's SED shows imprints of the metallicity of its stellar content, we model metallicity in \textsc{ProSpect} by considering the evolution of a galaxy's gas phase metallicity across time. 
Our implementation maps the metallicity evolution of each galaxy linearly to the total stellar mass growth through the \texttt{Zfunc\_massmap\_lin} function where the final metallicity for each galaxy is a free parameter, \texttt{Zfinal}\footnote{
The metallicity at any given time is linked to the total stellar mass formed up to this point, as although some stars will no longer be present or have experienced mass loss, they will still have contributed to the enrichment of the interstellar medium.}. 
This ensures that chemical enrichment in a galaxy follows the assumed star formation rate, where increased star formation is associated with an increased rate of metal production.
In this implementation, the value of the metallicity at any point in the history of the galaxy represents the gas-phase metallicity and thus the metallicity of the next generation of stars.

In the context of this work the physical meaning of the \texttt{Zfinal} parameter differs from the typically measured $Z_\text{gas}$ and stellar metallicity ($Z_\star$) from spectral line measurements. 
In the implementation of the evolving metallicity in \textsc{ProSpect}, \texttt{Zfinal} represents the metallicity of the gas from which the final stars formed. 
In this sense it is closest to the $Z_\text{gas}$ measured from spectral lines. 
However, if the gas content of a galaxy changed since the last epoch of star formation, for example if it continued to accrete lower metallicity gas then the true $Z_\text{gas}$ of the galaxy would be lower than we measure with \textsc{ProSpect}.
Despite this, we will refer to our measured metallicity as $Z_\text{gas}$ for the remainder of this work. 

As outlined in \cite{BellstedtGalaxyMassAssembly2020b} allowing for chemical enrichment over the lifetime of a galaxy directly impacts the derived star formation histories for each galaxy. 
\add{Additionally, the metallicities recovered from SED fitting with \textsc{ProSpect} have been shown to agree well with measurements directly from spectra \citep[see figure 1 from][]{BellstedtGalaxyMassAssembly2021}, albeit with significant scatter.} 

In addition to the five free parameters specifying the star formation and metallicity histories, we include four free parameters to describe the contribution of dust to the SED. 
Within \textsc{ProSpect} the dust is assumed to exist in two forms; in birth clouds formed around young stars (age$<10^{7}\,$yr), or distributed as a screen in the ISM. 
For each of these components we include two free parameters, describing the dust opacity (\texttt{tau\_screen}, \texttt{tau\_birth}), and the dust radiation field intensity (\texttt{alpha\_screen}, \texttt{alpha\_birth}). 
Figure 3 of \cite{ThorneDeepExtragalacticVIsible2021} shows the impact of each parameter on a generated galaxy SED.

We also include an AGN component by incorporating the model outlined in \cite{FritzRevisitinginfraredspectra2006} and \cite{FeltreSmoothclumpydust2012}. 
This models the primary source as a composition of power-laws, with different spectral indices as a function of the wavelength. 
To model the contribution from the torus, the \cite{FritzRevisitinginfraredspectra2006} model uses a simple but realistic torus geometry, a flared disc, and a dust grain distribution function including a full range of grain sizes and assumes that the dust in the AGN torus is smoothly distributed. 
Within \textsc{ProSpect} we model the AGN contribution by fitting the luminosity of the central source (\texttt{AGNlum}), optical depth at 9.7$\mu$m (\texttt{AGNta}), and angle of observation (\texttt{AGNan}). 
We also re-attenuate the emission from the central source and dust torus through the general ISM screen (see figure 1 of \citealt{RobothamProSpectgeneratingspectral2020}).

\subsection{Sample Selection}\label{sec:SampleSelection}

\begin{figure*}
    \centering
    \includegraphics[width = \linewidth]{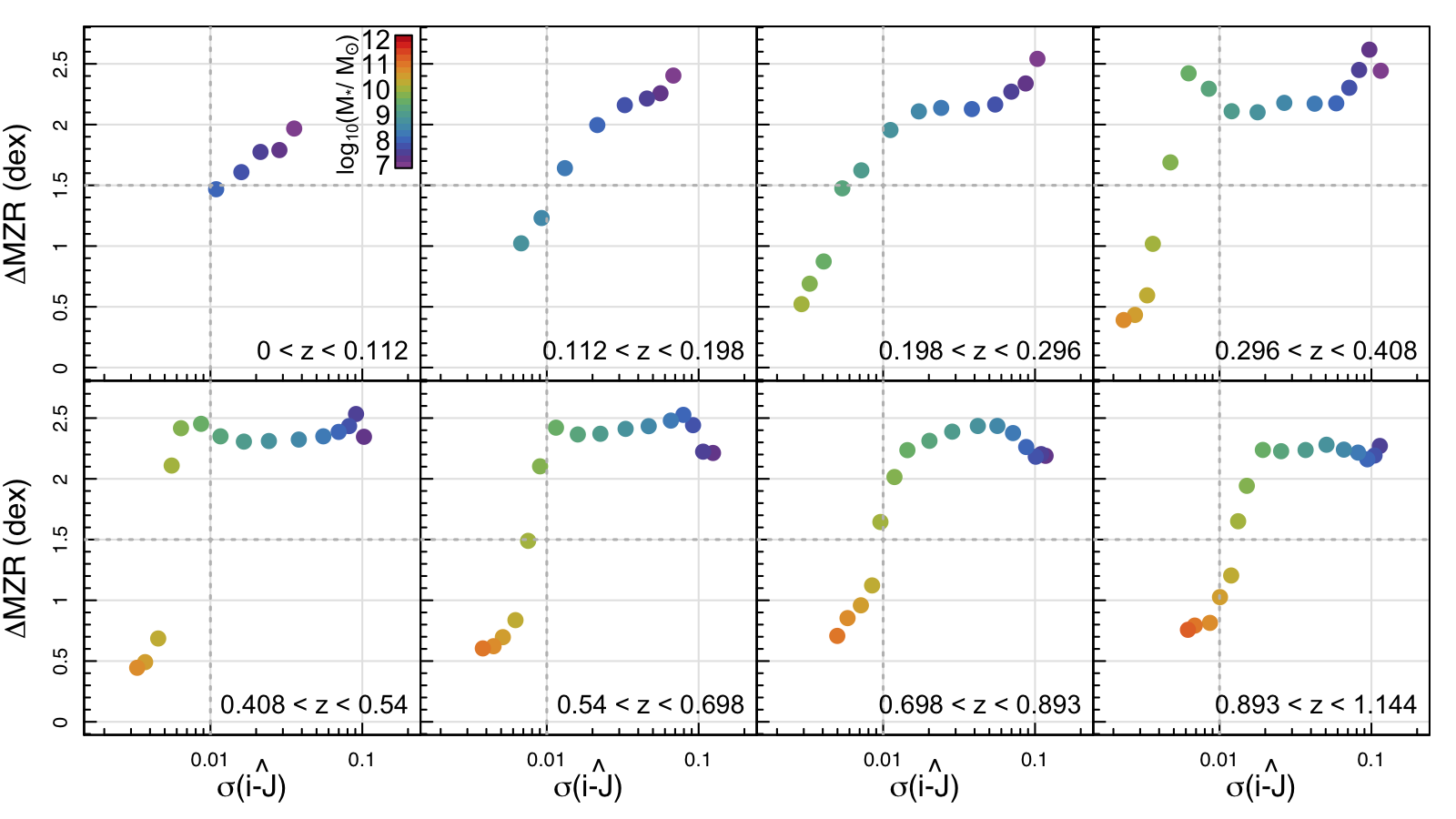}
    \caption{The scatter in the MZR as a function of average i-J colour error in bins of stellar mass (colour bar) and redshift (different panels).
    We show the $\sigma\text{MZR} < 1.5$ and $\sigma \hat{\text{i-J}}$ cuts as the dashed grey lines. 
    \add{
    The lower left region of each panel bounded by these cuts isolates the stellar mass regime where both the average i-J colour error and scatter in the MZR are low.
    This indicates that the metallicity estimates from \textsc{ProSpect} for these stellar masses at each redshift are reliable. 
    }
    }
    \label{fig:MZRScatter_ColError}
\end{figure*}

\begin{figure}
    \centering
    \includegraphics[width = \linewidth]{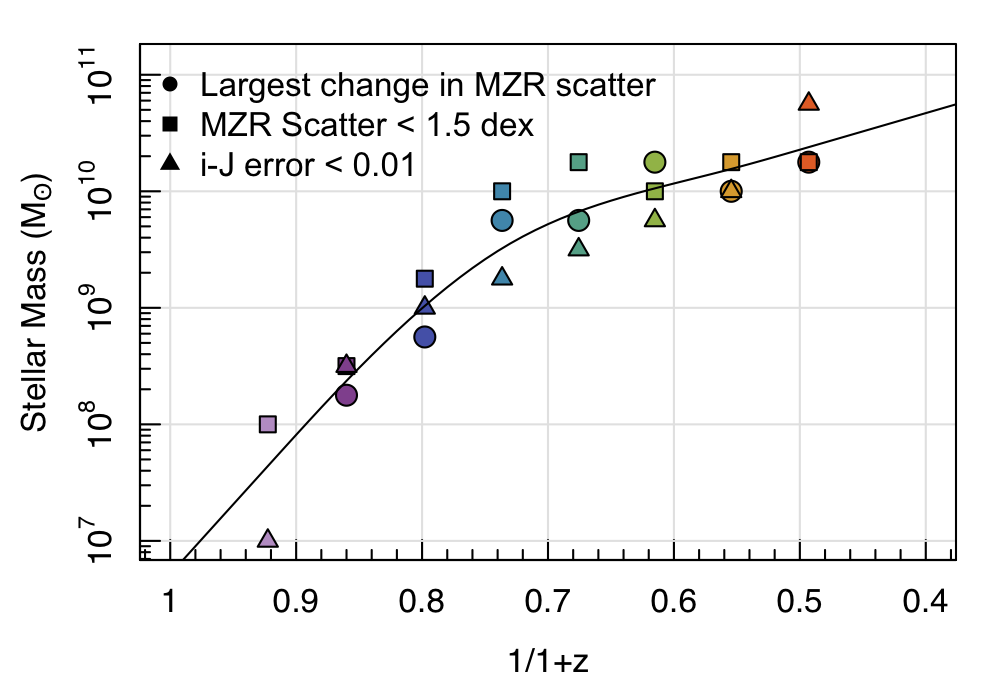}
    \caption{Stellar mass limits derived using (a) the largest change in the MZR scatter (circles), (b) the lowest stellar mass where the MZR scatter is less than 1.5\,dex (squares), and (c) the lowest stellar mass where the i-J error is less than 0.01 (triangles). 
    \add{The black line shows the smooth spline fit which is used to define the lower stellar mass limit for the D10 sample at each redshift.}
    }
    \label{fig:D10StellarMassCut}
\end{figure}

\begin{figure*}
    \centering
    \includegraphics[width = \linewidth]{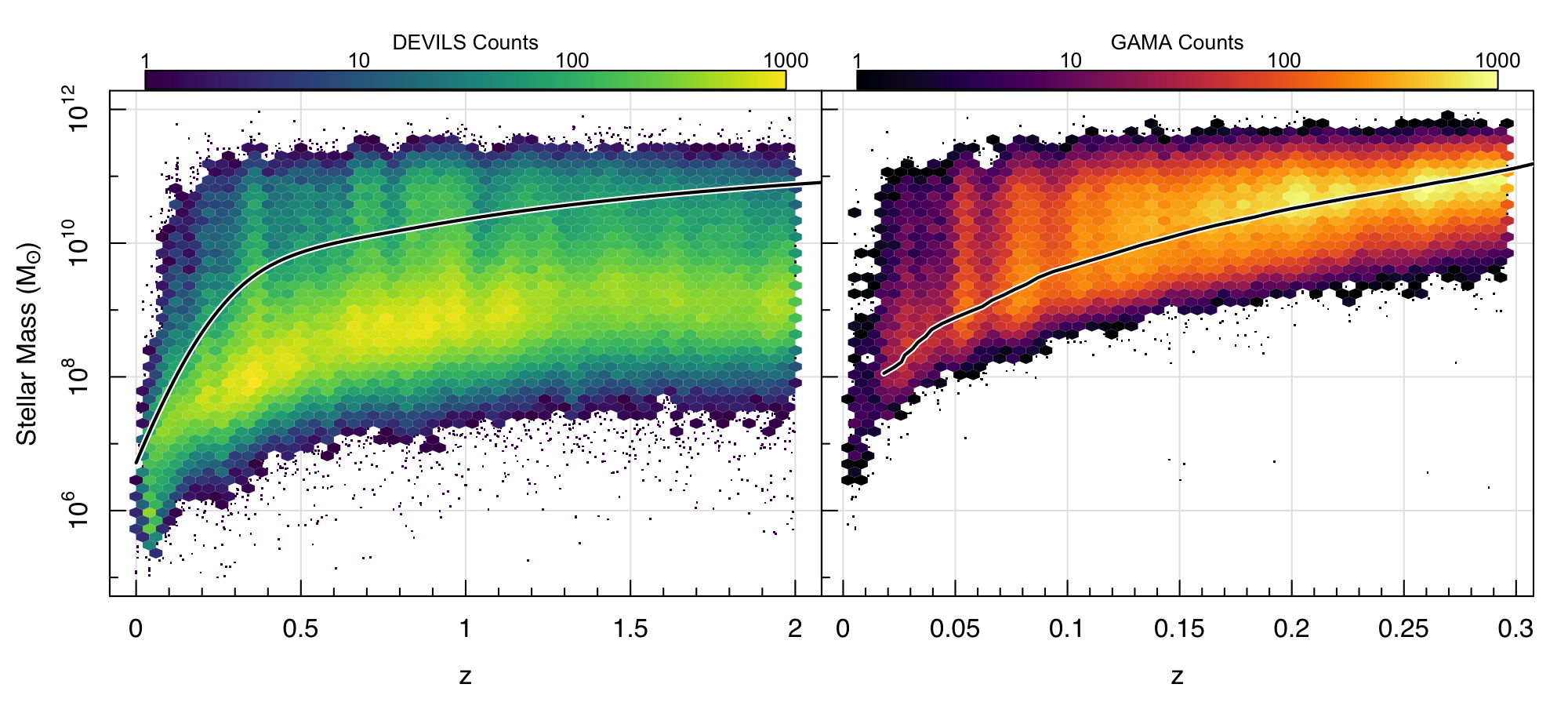}
    \caption{\add{The stellar mass distribution as a function of redshift for the DEVILS (left) and GAMA (right) samples. 
    In each panel the solid black line shows the mass selection function used in this work where, at each redshift, we select objects with masses above the black line. 
    For DEVILS this black line is as per Figure~\ref{fig:D10StellarMassCut} and for GAMA this is the mass completeness cut from \citet{RobothamGalaxyMassAssembly2014}.
    }}
    \label{fig:StellarMassCuts}
\end{figure*}

\begin{figure*}
    \centering
    \includegraphics[width = \linewidth]{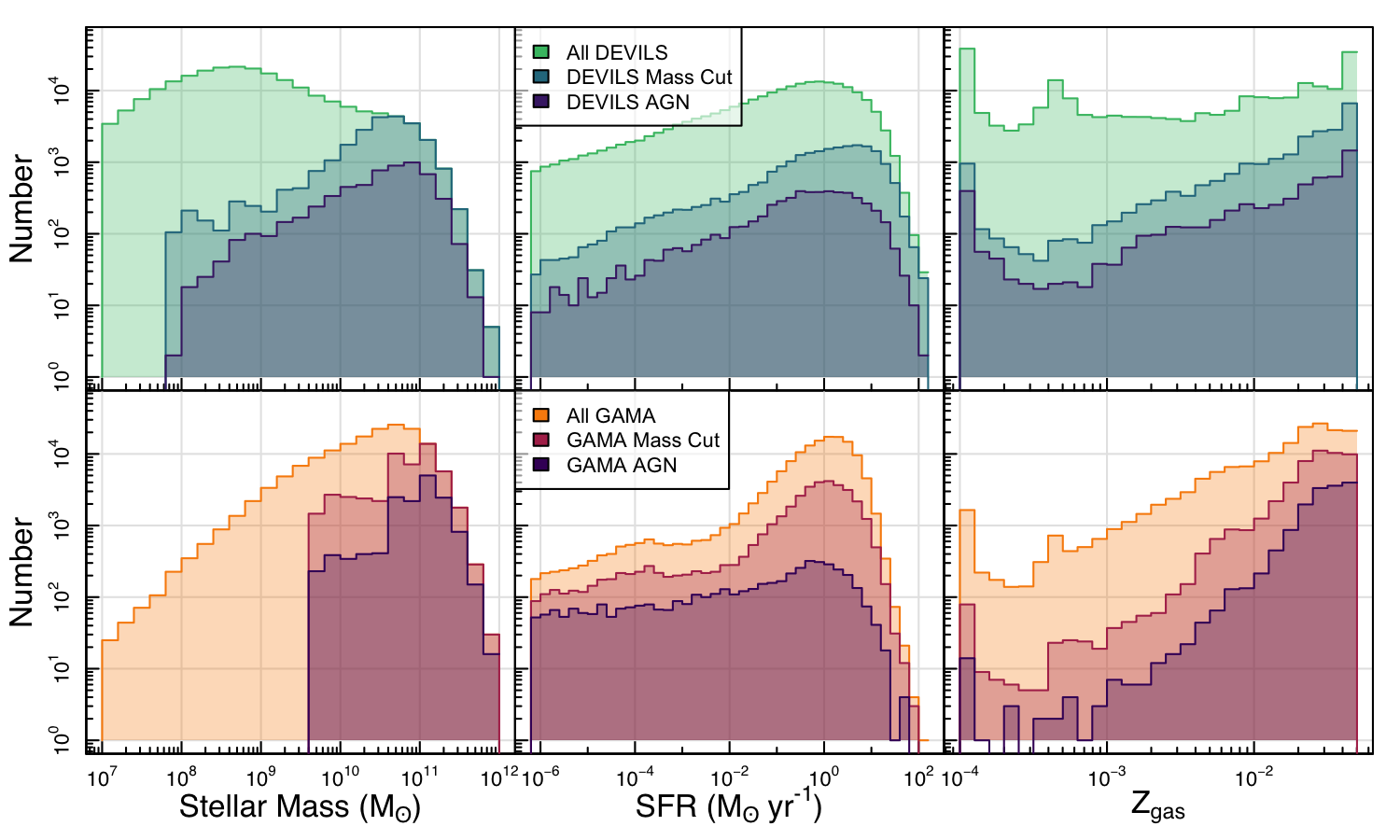}
    \caption{\add{The distribution of stellar masses (left), SFRs (middle), and metallicities (right) for DEVILS (top panels) and GAMA (bottom panels).
    In each panel we show the full sample (lightest colour), the sample after the mass cut has been applied (middle colour), and the sample of \textsc{ProSpect}-selected AGN (darkest colour).
    }}
    \label{fig:ParameterDistributions}
\end{figure*}

Due to the metallicity-dust-age degeneracy, and the subtle imprint left by metallicity on a galaxy's SED (Figure~\ref{fig:MetallicityInfluence}), measuring metallicities from broad-band photometry can be difficult. 
If the signal to noise ratio of the optical-NIR photometry is too low, then subtle differences in the SED caused by differences in metallicity can be hidden in the uncertainties and not recovered when SED fitting. 
This means that we are limited to brighter objects with higher signal to noise ratio, generally low-redshift or high-mass galaxies. 
But in contrast to typical metallicity measurements, we are not limited to star-forming galaxies by the need to have emission lines. 

In the context of this work, we are interested in understanding where SED-derived metallicity values are believable as a function of stellar mass and redshift.
This is to ensure that we are not making cuts that will bias the resulting sample to particular galaxy types and to ensure that at each redshift and stellar mass we have a complete galaxy sample. 

To determine the stellar mass cut for the D10 sample and how it evolves with redshift we investigate three techniques. 
The first technique identifies the stellar mass where the scatter in the MZR significantly increases, and cuts the sample below this mass. 
An artificial increase in the scatter of the MZR at low stellar masses can be caused by low signal to noise photometry being unable to constrain the metallicity of the galaxy.
When this happens, the optimization routines preferentially select lower metallicities, often the exact metallicity values of the \cite{BruzualStellarpopulationsynthesis2003} templates rather than metallicities that require interpolated spectra. 
To calculate where this artificial increase in scatter occurs, we bin the MZR in 0.25\,dex stellar mass bins and identify where the change in $1\sigma$ scatter between two adjacent mass bins is largest.
These are changes of $\sim0.6\,\text{dex}$ for the low redshift bins, and $\sim0.4\,\text{dex}$ for the two highest redshift bins.

The second and third techniques investigate the relationship between the scatter in the MZR and the average error in the $i-J$ colour as uncertainties in the optical-NIR photometry result in a lower constraint on the metallicity. 
In selecting the colour choice for these comparisons, we found that using short wavelength photometric bands (\textit{u,g,r}) introduced dependencies on SFH.
Additionally, we decided to avoid the longest wavelength NIR bands (\textit{H, Ks}) as in DEVILS these are shallower than the other NIR bands. 
We also found that using neighbouring bands provided little information on the metallicity, and the constraint on metallicity in SED fitting was driven by the overall slope of the optical-NIR SED. 
Because of this, we decided to use the $i-J$ colour as these bands are not adjacent, and not at too short or too long wavelengths.  

To have the greatest constraint on metallicity from an SED fit we require high signal to noise photometry with low uncertainties \add{($\le 0.1$\,mag; see \citealt{RobothamProSpectgeneratingspectral2020})}. 
At a given redshift the galaxies with the highest signal photometry will be the brightest and therefore most massive galaxies. 
If the photometry for a given galaxy has a larger associated uncertainty and therefore larger metallicity error, the scatter in the MZR will be artificially increased. 
In Figure~\ref{fig:MZRScatter_ColError} we show the scatter in the MZR as a function of the average $i-J$ colour error binned in stellar mass and redshift.
We find that there is a correlation between scatter and average colour error for the highest mass bins, where a higher colour error is associated with larger scatter in the MZR moving from high to low masses. 
We do find that in each redshift bin, there is a point in stellar mass where additional uncertainty in the colour does not correspond to an increase of scatter. 
Based on these relationships, a cut in $\Delta$\,MZR just below the flattening would be poorly motivated as it is apparent that the scatter has already reached the maximum value. 
We, therefore, use a cut at $\Delta\,\text{MZR}<1.5$ which lies more than $0.5$\,dex below the flattening at all redshifts. 
The third technique uses the same relationship but uses a cut in colour error of $i-J < 0.01$ as in most redshift bins, this value occurs at masses above the flattening in scatter.

Figure~\ref{fig:D10StellarMassCut} shows each of the stellar mass cuts from the three techniques as a function of the scale factor ($1/(1+z)$). 
Although these stellar mass limits were determined using three different methods, we find that they are in relatively good agreement in each redshift bin. 
To combine the three techniques together and determine a single stellar mass cut for the D10 sample we fit all the points with a smooth spline and use this to calculate a stellar mass cut at each redshift bin. 
We stress that this selection is unique to the combination and depth of photometry available as well as the quality and depth of the redshifts used.
Even in the context of the DEVILS survey, this selection will be unique to the D10 field due to differences in imaging depths and photometric redshift quality between fields. 

Although GAMA is a shallower survey, the limits of the photometry are not being pushed with the use of photometric redshifts. 
In the case of GAMA the imaging in the r-band is five magnitudes deeper than the spectroscopic completeness limit.\footnote{The photometry in the Y-band for DEVILS D10 is also five magnitudes below the spectroscopic completeness limit, however in the full D10 sample we push beyond the 5$\sigma$ depth by using all available photometric redshifts.}
For this reason, we use the spectroscopic completeness limit defined by \cite{RobothamGalaxyMassAssembly2014} as the lower stellar mass range for GAMA in this work. 
This results in a sample of 50\,147 galaxies.
Additionally, the stellar mass cut derived for the D10 sample is above the completeness cut derived by \cite{ThorneDeepExtragalacticVIsible2021} meaning that for both samples we can guarantee we are sampling all galaxy types present at a given stellar mass and are not biased to bright, star-forming galaxies. 

\add{Figure~\ref{fig:StellarMassCuts} shows the stellar mass distribution as a function of redshift of the GAMA and DEVILS samples. 
The stellar mass cut derived above for DEVILS is shown as the black line on the left panel while the completeness cut from \citet{RobothamGalaxyMassAssembly2014} for GAMA is shown as the black line on the right panel. 
In each case, we use only galaxies with stellar masses above this limit in our analysis. 
}

\subsubsection{AGN Selection}
As our implementation of \textsc{ProSpect} includes a flexible AGN component we can use our SED fits alone to identify galaxies hosting an AGN.
To identify and isolate AGN host galaxies we select objects that have $f_\text{AGN} > 0.1$ where $f_\text{AGN}$ is the fraction of flux contributed by the AGN component between 5-20$\mu$m (as per \citealt{ThorneDeepExtragalacticVIsible2022}). 
For DEVILS galaxies, to identify an AGN, we also require that the galaxy has $Y_\text{mag} < 21.2$ or was identified in the MIPS24 imaging and was therefore passed to the FIR photometry phase (as described in Section~\ref{sec:Data}, also see \citealt{DaviesDeepExtragalacticVIsible2021}). 
By ensuring that these objects have attempted FIR measurements we have a better constraint on the AGN component than without. 
This requirement is not necessary for GAMA as the redshift completeness limit is much shallower than DEVILS, and the FIR imaging in the GAMA fields has sufficient resolution and depth that FIR photometry can be extracted for all optically detected sources.
In \cite{ThorneDeepExtragalacticVIsible2022} we show that the inclusion of an AGN component in the SED fitting process has no systematic impact on the recovered masses or metallicities for galaxies with $f_\text{AGN}>0.1$ and should not impact these results.

\add{Selecting all GAMA galaxies with $f_\text{AGN}>0.1$ results in a sample of 14\,882/50\,147 galaxies identified as AGN. 
The additional FIR photometry requirement for DEVILS imposes a harsher cut and results in 5\,916/23\,784 galaxies identified as AGN. 
The resulting stellar mass, SFR, and metallicity distributions of the original sample, mass-selected sample, and sample of \textsc{ProSpect}-selected AGN are shown in Figure~\ref{fig:ParameterDistributions}.
Before the mass cut is applied, the DEVILS sample has a relatively uniform metallicity distribution with noticeable peaks at the values of the \citet{BruzualStellarpopulationsynthesis2003} metallicity templates.
However, applying the mass selection cut systematically removes more low-metallicity galaxies as is expected from the MZR. 
The same is true for GAMA. 
However, for both GAMA and DEVILS the distribution of metallicities are very similar for the mass-selected sample and AGN sample, however in GAMA we find that the AGN sample do not follow the same SFR-distribution as the mass-selected sample due to the lower SFRs associated with \textsc{ProSpect}-selected AGN.
}

\section{Mass -- Metallicity Relation}\label{sec:MZR}

\begin{figure*}
    \centering
    \includegraphics[width = \linewidth]{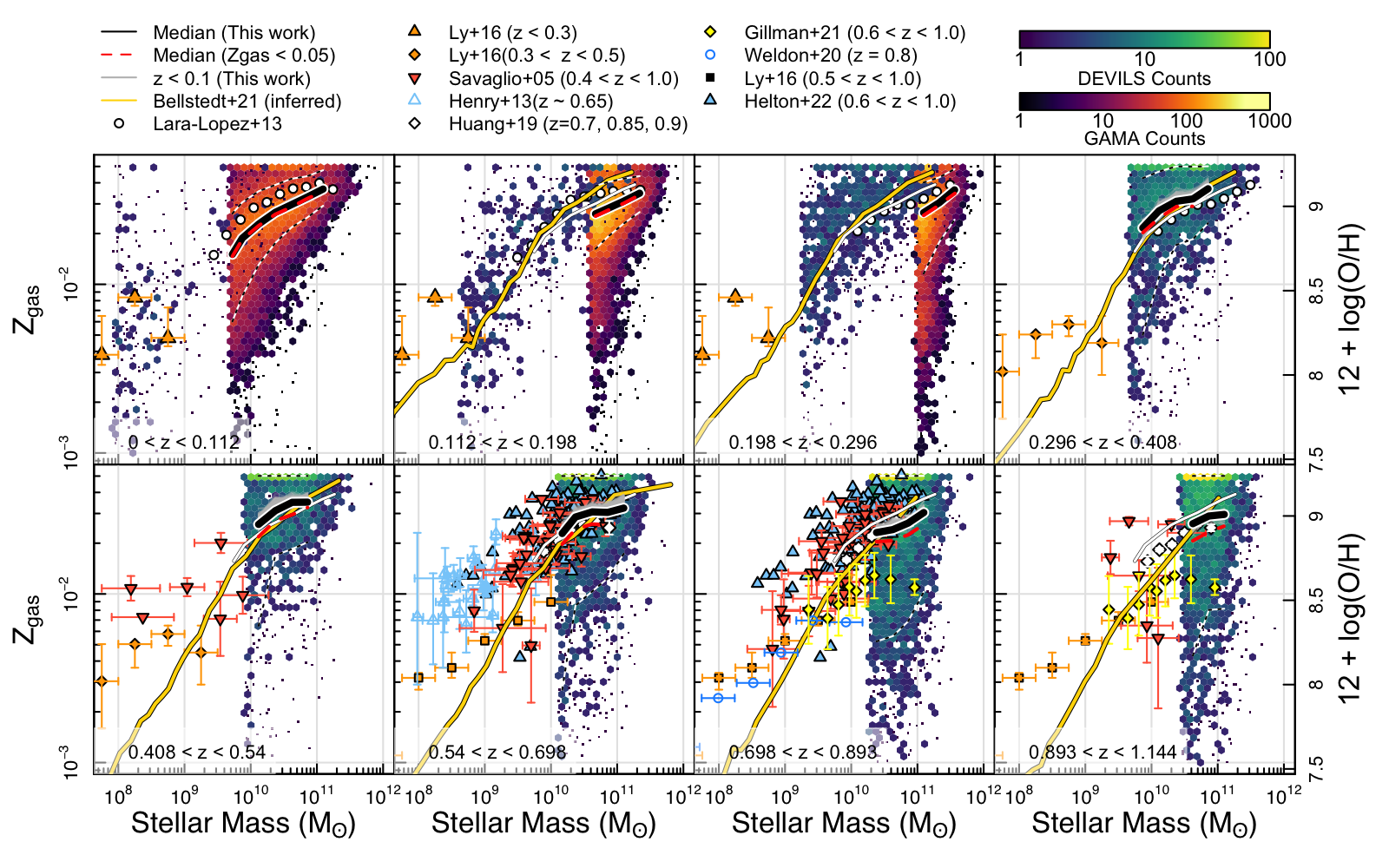}
    \caption{The mass--metallicity relation at 1\,Gyr intervals for $z=0$ to $z=1$ using the \textsc{ProSpect}-derived stellar masses and metallicities for galaxies with no significant AGN component from DEVILS and GAMA.
    We show the redshift range of each panel in the lower left. 
    At each interval, the running median is indicated with a solid black line, and the 1$\sigma$ range in the scatter is shown as the dashed lines. 
    \add{The dashed red line shows the running median re-calculated removing sources that have best-fitting metallicities equal to the highest metallicity in the \citet{BruzualStellarpopulationsynthesis2003} templates. }
    We also include the $z\approx0$ median in each panel as the white line, and the inferred mass-metallicity relation from \citet{BellstedtGalaxyMassAssembly2021} in yellow.
    Where possible we have included other observational measurements at the relevant epochs as a comparison. 
    These studies include those of \citet{SavaglioGeminiDeepDeep2005,HenryMetallicityEvolutionLowmass2013,Lara-LopezGalaxyMassAssembly2013,LyMetalAbundancesCosmic2016,HuangMassMetallicityRelationRedshift2019,Weldonstellarpopulationmetalpoor2020, Gillmanevolutiongasphasemetallicity2021,HeltonNebularPropertiesStarforming2022}. }
    \label{fig:MZR}
\end{figure*}

\citet{BellstedtGalaxyMassAssembly2021} previously demonstrated that the recovered MZR at $z\approx0$ from SED-fitting using \textsc{ProSpect} was consistent with the MZR obtained through spectral measurements. 
We extend this work to a larger sample of galaxies spanning a larger redshift range, as shown in Figure~\ref{fig:MZR}. 
The GAMA and DEVILS galaxies are shown in two 2D histograms with GAMA shown in warm colours (lower colour bar) and DEVILS shown in cool colours (upper colour bar).
We have removed SED-selected AGN from this Figure to ensure they are not contaminating the recovered MZR (the MZR derived for AGN host galaxies is discussed in Section~\ref{sec:AGNMZR}) but do not make any other cuts at this stage. 

The median trend of the MZR is calculated in stellar mass bins of width 0.25\,dex for bins with at least 200 galaxies and is shown as the solid black line.  
We show the error on the median by perturbing the metallicity of each galaxy by sampling from a normal distribution centered on the best fit value and using the metallicity uncertainties as the standard deviation. 
We do this 100 times and show each iteration as a grey line. 
This method does not account for the uncertainty in stellar mass, but typically this is much smaller than the uncertainty in metallicity \add{(the metallicity and stellar mass uncertainties are typically 0.4/0.2\,dex and 0.1/0.05\,dex for DEVILS/GAMA respectively).}
The $1\sigma$ scatter in the MZR is shown as the dashed lines. 

As is well known in the literature \citep{LequeuxChemicalCompositionEvolution1979,TremontiOriginMassMetallicity2004}, we find a correlation between mass and metallicity where higher mass objects have higher metallicities.
Some previous work noted a flattening or saturation of metallicity at the highest stellar masses \citep{Lara-LopezGalaxyMassAssembly2013,BellstedtGalaxyMassAssembly2021}, which we can see some evidence of in our lowest redshift bin. 
However, at higher redshifts, we do not see this as clearly as we are mostly limited to a $\sim1\,\text{dex}$ range in stellar masses. 

As mentioned in Section~\ref{sec:SampleSelection}, the scatter in the recovered MZR increases at low stellar mass at all redshifts, but this scatter reduces at higher stellar masses, particularly at $z\approx 0$. 
This is seen consistently across redshift bins despite the fact that we are significantly limited in mass range at higher redshifts due to the imposed mass cut. 

A clear artefact in these panels is the upper limit in the range of metallicities at $Z = 0.05$. 
This is particularly noticeable at $z>0.3$ where the upper region of our $1\sigma$ range is at this limit. 
This hard limit is the highest metallicity template in the \cite{BruzualStellarpopulationsynthesis2003} stellar population templates, and therefore the upper limit for the \texttt{Zfinal} parameter in \textsc{ProSpect}.
Hence, our application of \textsc{ProSpect} is not sensitive to gas-phase metallicity values larger than $Z_\text{gas} = 0.05$.
The pile-up of objects at the upper metallicity limit could represent cases where a higher final metallicity value would be preferred, or cases where a more rapid chemical enrichment needs to be accounted for to better produce realistic $Z_\text{gas}$ values. 
\add{We show the impact of removing sources with best fitting metallicities of $Z_\text{gas} = 0.05$ as the dashed red line in Figure~\ref{fig:MZR}.
In our three lowest redshift bins ($z< 0 .296$), we see no change in the recovered median MZR when removing these objects.
At these redshifts, the medians are driven by the sample of GAMA galaxies which do not hit the upper metallicity limit as frequently as the sample of DEVILS galaxies (see Figure~\ref{fig:ParameterDistributions}).
Removing galaxies at the upper limit results in an offset to lower metallicities which expands with increasing redshift. 
Despite this, the gradient and shape of the median MZR are consistent when these high metallicity sources are removed.  
However, we still expect that the majority of these sources truly have high metallicities and therefore the running median calculated including these sources is still a reasonable approximation for the whole sample. 
We present the maximal impact of these sources by showing the median when excluding them, however, we reiterate that this is an extreme solution.
}

To compare our results to previous measurements of the MZR, we include measurements made for GAMA galaxies by \cite{Lara-LopezGalaxyMassAssembly2013} as well as results from \cite{SavaglioGeminiDeepDeep2005,HenryMetallicityEvolutionLowmass2013,LyMetalAbundancesCosmic2016,HuangMassMetallicityRelationRedshift2019,Weldonstellarpopulationmetalpoor2020, Gillmanevolutiongasphasemetallicity2021,HeltonNebularPropertiesStarforming2022}.
\add{Comparing to spectroscopically derived values is difficult due to significant biases and implicit selection effects associated with spectroscopic measurements. 
Spectroscopic measurements rely on subsets of emission lines, strong line parameters, and calibration methods which all vary between data sets.
Two commonly used methods for deriving metallicities from spectroscopic data are strong line calibrations and electron temperature-based techniques.  
So-called T$_\text{e}$ methods are more directly linked to the physical processes governing ionised nebulae, and therefore more directly correlated with metallicity.
However, the required emission lines are weak, especially in metal-rich objects.
Strong-line derivations were proposed to measure abundances in faint, distant, and high metallicity galaxies where other techniques could not be used \citep{AlloinNitrogenoxygenabundances1979, PagelcompositionIIregions1979}. 
Strong-line derivations rely on ratios of collisionally excited lines and Balmer lines which have been shown to have a dependence on metallicity. 
However, these techniques must be calibrated either empirically, for samples in which T$_\text{e}$ based abundances have been derived \citep{PettiniOIIINIIabundance2004,PilyuginOxygenAbundanceDetermination2005,PilyuginCounterpartmethodabundance2012, PilyuginNewImprovedCalibration2010,MarinoO3N2N2abundance2013, PilyuginNewcalibrationsabundance2016}, or theoretically, using oxygen abundances that have been inferred via photoionisation models \citep[e.g.][]{KewleyUsingStrongLines2002,KobulnickyMetallicities2004,TremontiOriginMassMetallicity2004, DopitaNewStronglineAbundance2013,DopitaChemicalabundanceshighredshift2016,CurtiNewfullyempirical2017}.
Unfortunately, comparisons of metallicities estimated using different techniques and calibrations are highly discrepant, even for the same sample of galaxies.
\citet{KewleyMetallicityCalibrationsMassMetallicity2008} demonstrated that by using different calibrators, the resulting MZR can vary in normalization by up to 0.5\,dex but also have dramatically varying slopes. 
Additionally, spectroscopic metallicities are measured using different element bases -- stellar metallicities are often measured using iron, while gas phase metallicities are measured using oxygen (see \citealt{Fraser-McKelvieSAMIGalaxySurvey2022} for a comparison of stellar and gas metallicities).
This differs to the metallicity value from \textsc{ProSpect}, which represents the mass in all elements heavier than hydrogen or helium, and is not centered on a particular element.
The metallicity indicators and calibrations used for each of these literature measurements are outlined in table~2 of \cite{BellstedtGalaxyMassAssembly2021}.}

\add{
Additionally, spectroscopic measurements are derived from oxygen abundances ($12+\log(O/H)$) and require conversions to metallicity values using assumed oxygen and solar abundance values. }
To convert literature oxygen abundances to metallicities for comparison, we use the prescription from \citet{KobulnickyMetallicities2004}\footnote{This is equivalent to $Z_\text{gas} = Z_\odot \times 10^{[12+\log (\text{O/H}] - [12+\log (\text{O/H}]_\odot}$, where $Z_\odot = 0.0142$ and $[12+\log (\text{O/H}]_\odot = 8.69$ \citep{AsplundChemicalCompositionSun2009}}:

\begin{equation}
    Z = 29 \times 10^{[12+\log(\text{O/H})] - 12}.
\end{equation}
\add{However, additional uncertainty is associated with these conversions as the oxygen abundance and metallicty of the Sun are still uncertain. 
Changes in both the assumed oxygen abundance and metal mass fraction of the sun will change the normalisation of the conversion. 
Consequently, scatter between observed and our modelled metallicities could arise from the adopted scaling between oxygen abundances and total metallicities. 
}

Additionally, it is very difficult to measure gas-phase metallicities from spectra for galaxies with low-to-no star formation as the required emission lines are not detectable. 
Metallicities are also difficult to measure for AGN host galaxies as the associated broad and narrow emission lines can contaminate measurements of lines required for metallicity estimates.
Therefore, the previous measurements of the MZR using emission line techniques are predominantly based on star-forming, galaxies with no AGN component. 

At $z<0.3$, we recover an MZR systematically lower than that derived by \citet{Lara-LopezGalaxyMassAssembly2013}.
In our highest redshift bins ($z = 0.5 - 1.0$), we recover a mass-metallicity relation in close agreement with the MZR derived by \citet{HuangMassMetallicityRelationRedshift2019}. 
\add{
At these redshifts, our derived MZR is also consistent with the individual galaxy measurements from \cite{SavaglioGeminiDeepDeep2005} and \cite{HeltonNebularPropertiesStarforming2022}.
}
\add{However, we measure more low-metallicity values than the observations shown in Figure~\ref{fig:MZR}. 
This could be due to a number of factors. 
Recovering metallicty estimates through broad-band SED fitting allows us to work with larger samples and potentially trace the lower tail of metallicity values.
Additionally, using SED-derived metallicities allows us to probe different areas of the SFR-$M_\star$ parameter space where the emission lines might not be detectable.
Alternatively, emission line surveys may not be sensitive to low metallicities, or we may be recovering artificially low values in some cases.
}

\begin{figure*}
    \centering
    \includegraphics[width = \linewidth]{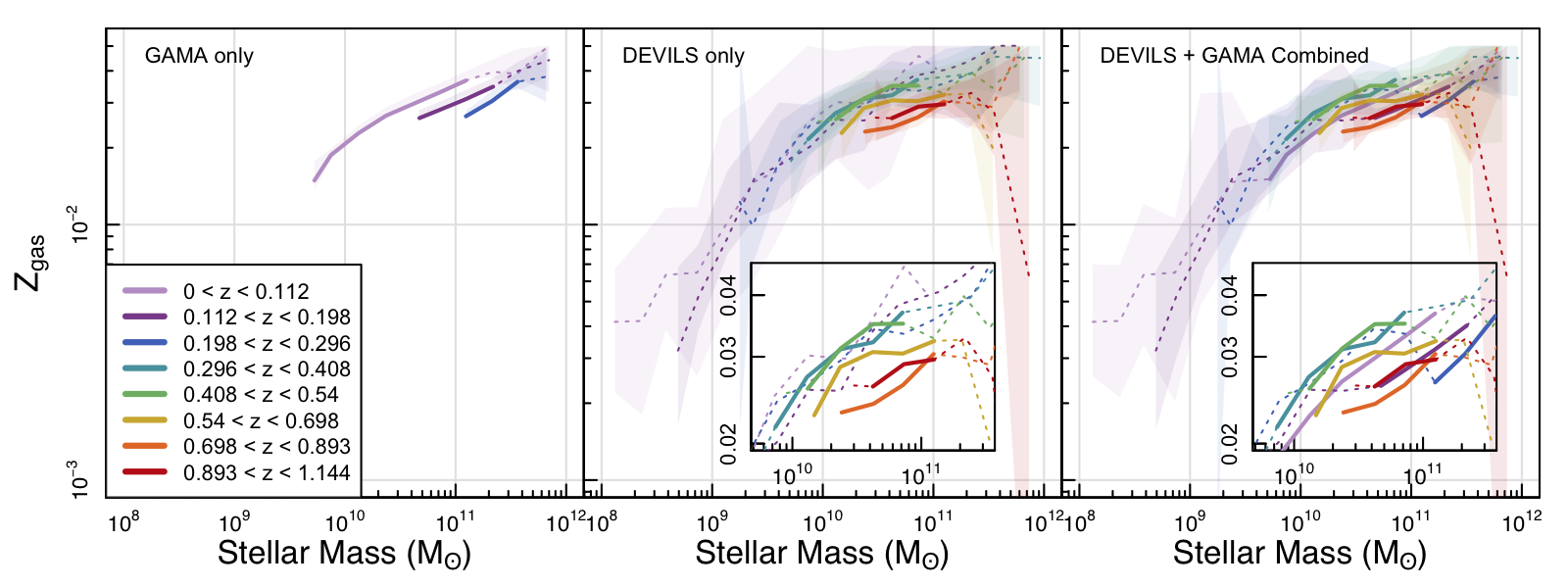}
    \caption{The evolution of the median MZR computed using only GAMA (left), only DEVILS (center), and the combination of GAMA and DEVILS (right). 
    Each line shows a different redshift bin from Figure~\ref{fig:MZR} and is shown as a solid (dashed) line if the mass bin contains more (less) than 200 galaxies. 
    The polygons show the error on the median. 
    In the inset panels, we show a zoom of the median relations without the polygons for clarity. 
    }
    \label{fig:MZR_evolve}
\end{figure*}

We also compare to results from \cite{BellstedtGalaxyMassAssembly2021} who used \textsc{ProSpect} to trace back in time the inferred MZR from the star formation and metallicity histories of a sample of $\sim$4,500 galaxies with $z<0.06$. 
\add{We refer to this as a forensic determination of the MZR and show the forensic MZR recovered in each redshift bin as the yellow line in each panel.} 
The masses and metallicities from \citet{BellstedtGalaxyMassAssembly2021} were derived using a very similar implementation of \textsc{ProSpect} as used in this work, with the major difference being the inclusion of the AGN model in this work and the use of a different optimization routine (Appendix~\ref{app:B21Diff} explores the impact of some of these differences). 
\cite{BellstedtGalaxyMassAssembly2021} also used GAMA galaxies for their analysis, but limited their analysis to 4,500 galaxies with $z<0.06$ and secure metallicity measurements in the G09, G12, and G15 fields (not including G23, which is included in this work). 
Interestingly, we find good agreement with the backwards modelled results from  \citet{BellstedtGalaxyMassAssembly2021} at higher redshifts ($z>0.3$) where our results are determined from the DEVILS galaxies while the  \citet{BellstedtGalaxyMassAssembly2021} results are extrapolated backwards from GAMA galaxies at $z\approx0$.

Despite the fact that the median MZR at $z=0$ from \cite{BellstedtGalaxyMassAssembly2021} is derived using the same SED fitting code on the same parent sample of galaxies, there is an offset in the recovered MZR at $z=0$. 
There are a number of differences between the implementations that could cause the offset including the difference in sample selection (due to field choice, metallicity uncertainty cuts, and redshift range used), inclusion of the AGN component in SED fitting, choice of optimization routines and selection of parameter values. 
See Appendix~\ref{app:B21Diff} for a demonstration of the impact of each implementation difference. 
Even after accounting for these differences, there are other reasons why we might expect the forensic MZR to differ from our measured MZR at higher redshifts. 
First, the forensic technique cannot take into account any merger activity, inflows or outflows which might change the metallicity after the final stars have formed.
Second, any agreement with the forensic technique is dependent on the implemented enrichment model being correct -- if it were not, we would expect a larger discrepancy.
Additionally, \cite{BellstedtGalaxyMassAssembly2020b} found rough agreement between the measured cosmic metal density using these forensic metallicity histories from \textsc{ProSpect} with those from observations suggesting that the evolving metallicity implementation in \textsc{ProSpect} does a reasonable job modelling the true chemical enrichment of galaxies. 

Typically the MZR is measured for star-forming galaxies, because gas-phase metallicity is typically measured from nebular emission lines. 
By extracting metallicities via SED fitting, we are not limited to star-forming objects and as such include passive systems in the MZR presented in Figure~\ref{fig:MZR}.
However, the concept of a gas-phase metallicity is unclear for a galaxy with \add{minimal gas} as is typical for high mass, passive galaxies. 
To test the impact of including passive galaxies, we trialled removing objects with a specific SFR ($\text{sSFR}) = \text{SFR}/M_\star < 10^{-13}\,\text{yr}^{-1}$ as motivated by the cut in \citet{ThorneDeepExtragalacticVIsible2021}. 
This cut had no impact on the median MZR, but slightly ($<0.05\,$dex) reduced the scatter in the highest mass bin of each redshift.

Figure~\ref{fig:MZR_evolve} shows the evolution of the median MZR in the redshift bins from Figure~\ref{fig:MZR}.
We calculate and show the MZR using just the GAMA sample (left), DEVILS sample (center), and the combination of the two (right) and show mass bins with less than 200 galaxies as the dashed lines. 
When examining the results from GAMA and DEVILS separately, there is a slight evolution present where, at fixed stellar mass, the normalisation of the MZR decreases with increasing redshift. 
When combining the two samples together however, there is a discontinuity in the evolution where we move from having GAMA and DEVILS to redshift bins with only DEVILS. 
Overall, this evolution is less than 0.1\,dex in the normalisation at $M_\star = 10^{10.75}\,M_\odot$ over the last 8\,Gyrs. 
The discontinuity between redshift bins demonstrates the potential impact of sample selection and highlights the importance of using cohesive data sets when exploring trends with redshift. 
Limited evolution in the MZR at high stellar masses ($M_\star > 10^{10}\,M_\odot$) since $z=1$ is also found by \cite{BellstedtGalaxyMassAssembly2021}.
They find that most of the evolution in the MZR at $M_\star > 10^{10}\,M_\odot$ occurs in the first $\sim5\,$Gyrs of cosmic time. 
These results are also supported by \cite{CrescimetallicitypropertieszCOSMOS2012} who find no evolution in the MZR over the last 7 billion years, however \cite{Lara-LopezStudystarforminggalaxies2010} find a $\sim0.2$\,dex increase in metallicity from $z\approx0.35$ to $z\approx0.07$. 

\subsection{Mass -- Metallicity Relation for AGN Host Galaxies}\label{sec:AGNMZR}
\begin{figure*}
    \centering
    \includegraphics[width = \linewidth]{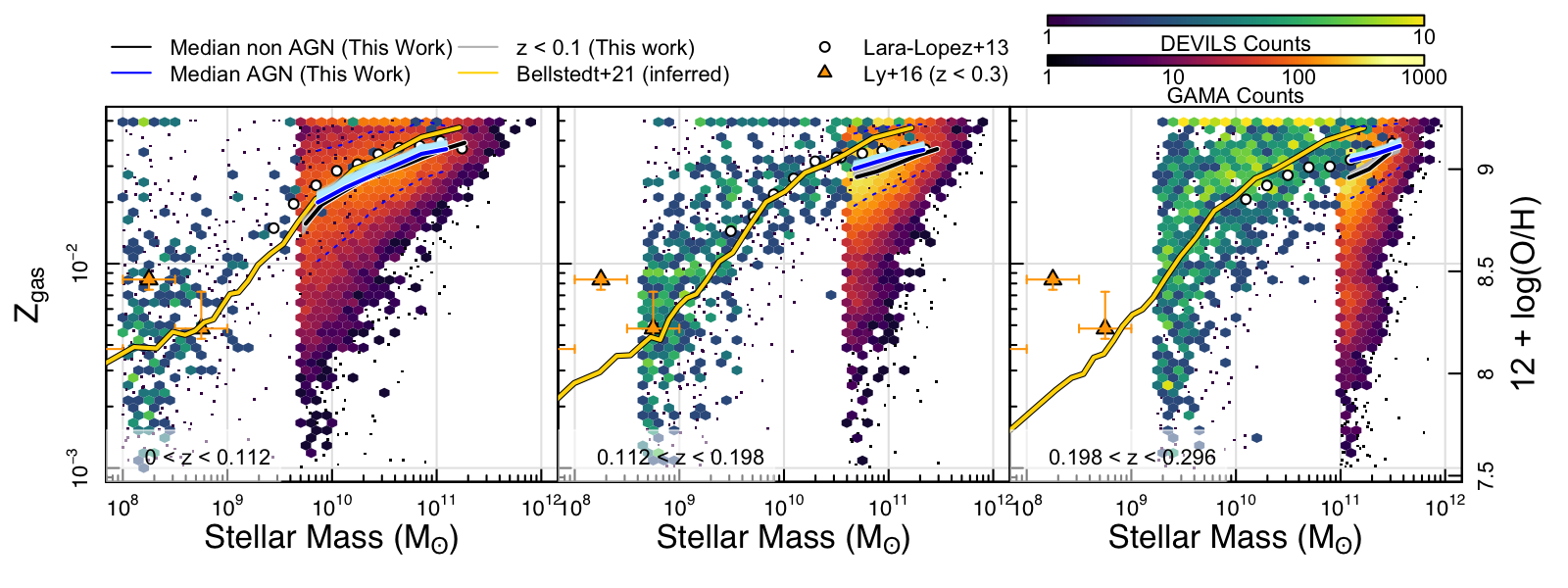}
    \caption{
    The mass-metallicity relation for galaxies with a significant AGN component ($f_\text{AGN} > 0.1$) at 1\,Gyr intervals for $z=0$ to $z=0.3$. 
    The running median and 1$\sigma$ range is shown in solid and dashed blue respectively. 
    We also show the error on the median as the light blue shaded region.
    We show the median MZR and error on the median from the equivalent panels of Figure~\ref{fig:MZR} (non AGN only) as the solid black line and grey shaded region for comparison. }
    \label{fig:MZR_AGN}
\end{figure*}

Typically, the MZR has been measured only for star-forming galaxies with no AGN.
This is primarily because the nebular line metallicity diagnostics have mostly been calibrated on star-forming galaxies classified as such based on the \cite*{BaldwinClassificationparametersemissionline1981} diagrams. 
These diagrams are a useful way to broadly divide galaxies by the excitation mechanism of the ISM into those with ongoing star formation, and the population of non-star-forming galaxies including LINERs and AGN. 
Extending the MZR to galaxies classified as LINERs and AGN is difficult due to a lack of suitable metallicity calibrators for estimating the gas-phase metallicity of such systems. 

The benefit of measuring galaxy metallicities through SED fitting is that we are not limited by existing metallicity calibrators or the presence of emission lines produced by star formation, allowing us to not only extend the MZR relation to quiescent or quenched galaxies, but also to galaxies with a significant AGN component.

Figure~\ref{fig:MZR_AGN} shows the recovered MZR for AGN galaxies. 
Due to the AGN selection criteria, we have significantly fewer galaxies at higher redshifts ($z>0.3$) and so we limit the redshift range here to $z<0.3$. 
For comparison we show the median MZR for non-AGN galaxies as the black line. 
We find that at $z=0$ the recovered MZRs for non-AGN and AGN galaxies are practically indistinguishable for $M_\star = 10^{10} - 10^{11} M_\odot$.
Between $z=0.2$ and $z=0.3$ we find that for galaxies with $M_\star > 10^{10} M_\odot$ galaxies that host an AGN have marginally higher metallicities than galaxies with no AGN. 
Below these masses we do not have a large enough sample to comment on the differences between selections.

These results are in agreement with \cite{CaiStellarPopulationsSample2020} who found that 136 dwarf galaxies with BPT-selected AGN follow a similar mass-metallicity relation to normal star-forming galaxies, indicating that AGN have little impact on the chemical evolution of the host galaxy. 
\cite{NetzerCosmicEvolutionMass2007} also find significant correlation between the normalised accretion rate ($L/L_\text{edd}$) and the Fe II / H$\beta$ line ratio. 
It is also well known that most AGN exist in galaxies with solar or supersolar metallicities \citep{Storchi-BergmannChemicalAbundanceCalibrations1998,HamannMetallicitiesAbundanceRatios2002}, although due to the high masses of AGN host galaxies this is not surprising.

\begin{figure*}
    \centering
    \includegraphics[width = \linewidth]{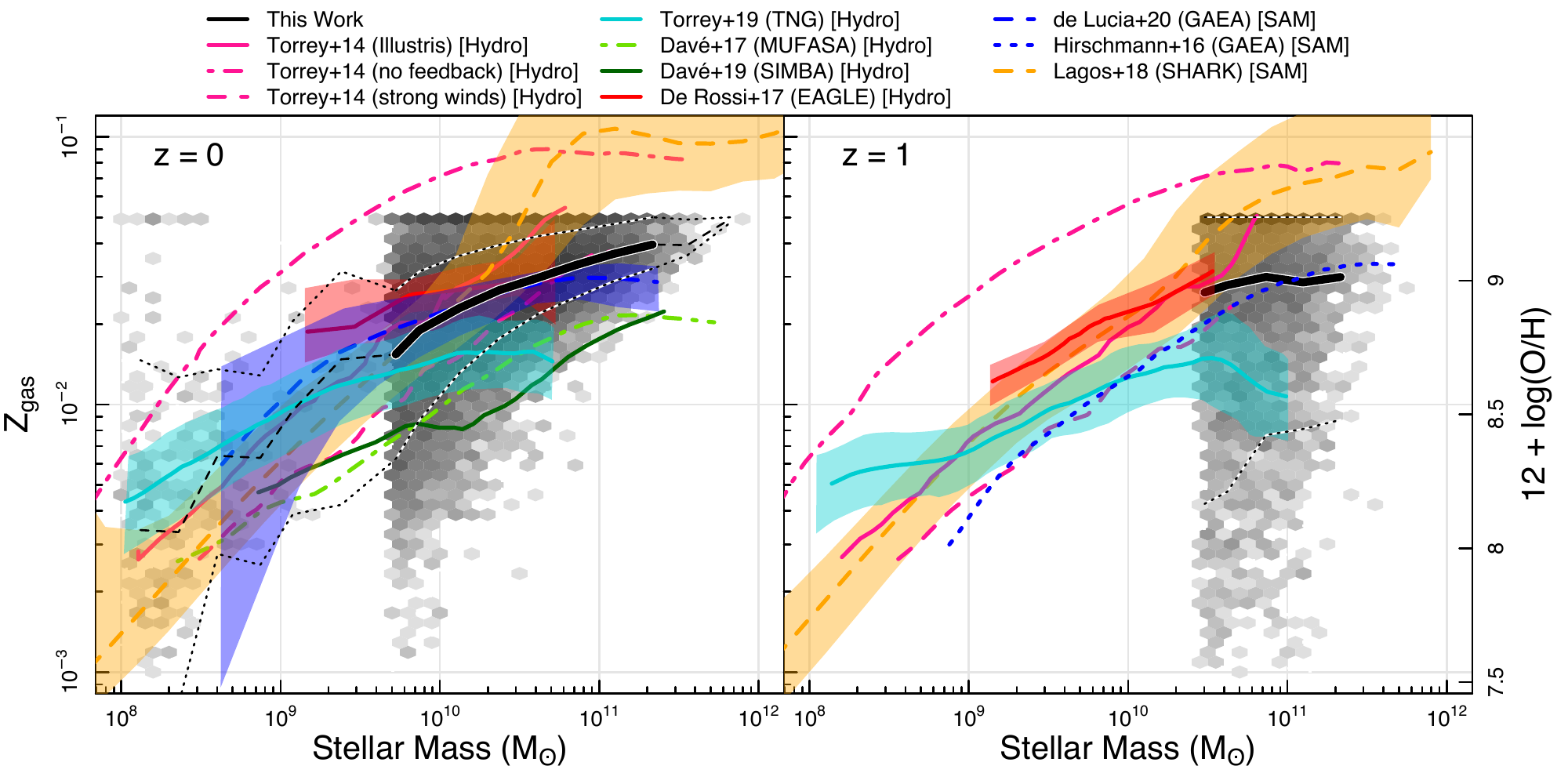}
    \caption{Comparisons of our MZR at $z=0$ and $z=1$ with simulations. 
    The underlying distribution from GAMA and DEVILS is shown as the grey scale 2D histogram. 
    Here, we include the cosmological hydrodynamic simulations Illustris (\citealt{Torreymodelcosmologicalsimulations2014}, showing the default model and two variations of the feedback model including no feedback, and strong winds), IllustrisTNG \citep{Torreyevolutionmassmetallicityrelation2019}, MUFASA \citep{DaveMUFASAGalaxystar2017}, SIMBA \citep{DaveSIMBACosmologicalsimulations2019}, EAGLE \citep{DeRossiGalaxymetallicityscaling2017} and the semi-analytic models GAEA \citep{HirschmannGalaxyassemblystellar2016,DeLuciaGasaccretionregulates2020a} and SHARK \citep{LagosSharkintroducingopen2018}.
    }
    \label{fig:SimulationsMZR}
\end{figure*}

\begin{table*}
    \centering
    \caption{\add{Summary of the simulations presented in Figure~\ref{fig:SimulationsMZR}. For each simulation we list the box length ($L_\text{box}$), number of dark matter particles and gas cells, and initial dark matter and baryonic particle mass. For the semi-analytic models (SHARK and GAEA), we include the details of the dark matter-only simulations used to generate the merger trees. 
    Note that the MZR derived from MUFASA combines the 12.5, 25, and 50\,Mpc boxes. 
    }}
    \label{tab:SimulationSummary}
    \begin{tabular}{l l l l l l l}
    \hline
Simulation &	Name	&	$L_\text{box}$	&	Cells/ particles	&	$m_\text{DM}$	&	$m_\text{baryon}$	&	Reference	\\
	&		&	(cMpc)	&		&	($\,M_\odot$)	&	($\,M_\odot$)	&		\\
 \hline
Illustris	&	L25n256	&	25	&	$ 2 \times 256^3$	&	$5.87 \times 10^7$	&	$1.18 \times 10^7$	&	\cite{VogelsbergerIntroducingIllustrisProject2014},\cite{Torreymodelcosmologicalsimulations2014}	\\
	&	No feedback	&	25	&	$ 2 \times 256^3$	&	$5.87 \times 10^7$	&	$1.18 \times 10^7$	&		\\
	&	Strong winds	&	25	&	$ 2 \times 256^3$	&	$5.87 \times 10^7$	&	$1.18 \times 10^7$	&		\\
IllustrisTNG	&	TNG100(-1)	&	110.7	&	$2 \times 1820^3$	&	$7.5 \times 10^6$	&	$1.4 \times 10^6$	&	\cite{PillepichFirstresultsIllustrisTNG2018}	\\
EAGLE	&	Recal-L025N0752	&	25	&	$2 \times 752^3$	&	$1.21 \times 10^6$	&	$2.26 \times 10^5$	&	\cite{SchayeEAGLEprojectsimulating2015}	\\
SIMBA	&	m100n1024	&	100	&	$2 \times 1024^3$	&	$9.6 \times 10^7$	&	$1.82 \times 10^7$	&	\cite{DaveSIMBACosmologicalsimulations2019}	\\
MUFASA	&	m50n512	&	50	&	$2 \times 512^3$	&	$9.6 \times 10^7$	&	$1.82 \times 10^7$	&	\citet{DaveMUFASAgalaxyformation2016}	\\
	&	m25n512	&	25	&	$2 \times 512^3$	&	$1.2 \times 10^7$	&	$2.28 \times 10^6$	&		\\
	&	m12.5n512	&	12.5	&	$2 \times 512^3$	&	$1.5 \times 10^6$	&	$2.85 \times 10^5$	&		\\
 \hline
GAEA	&		&	500	&	$2160^3$	&	$1.0078 \times 10^{10}$	&		&	\cite{HirschmannGalaxyassemblystellar2016}	\\
SHARK	&	L210N1536	&	210	&	$1536^3$	&	$2.21 \times 10^8$	&		&	\cite{LagosSharkintroducingopen2018}	\\
\hline
    \end{tabular}
\end{table*}

\subsection{Comparison with Simulations}
Historically, the MZR has been very difficult for simulations and semi-analytic models (SAMs) to reproduce. 
Recently, there has been an increased reporting of theoretical models producing MZR trends similar to observations at $z=0$.
In Figure~\ref{fig:SimulationsMZR} we compare our derived MZR at $z=0$ and $z=1$ to those produced by leading simulations/SAMs\footnote{\add{
Directly comparing observed metallicities to simulated metallicities is difficult as an assumption has to be made about stellar atmospheres either to turn simulations into observables (generate SEDs/spectra) or to invert measured SEDs/spectra to recover estimates for the metallicity. Figure~30 of \citet{RobothamProSpectgeneratingspectral2020} demonstrates that \textsc{ProSpect} can recover input stellar masses and metallicities for simulated galaxies. Additionally, \citet[appendix A]{BellstedtGalaxyMassAssembly2021} demonstrate the validity of proportionally evolving metallicity histories as implemented in \textsc{ProSpect} using simulated galaxies}}.
We include the MZRs derived by cosmological hydrodynamic simulations Illustris, IllustrisTNG, MUFASA, SIMBA, EAGLE and the semi-analytic models GAEA and SHARK. 

Using Illustris, \cite{Torreymodelcosmologicalsimulations2014} demonstrate that the adopted feedback models (both stellar and AGN) have a dramatic influence on the resulting MZR at all redshifts with no feedback resulting in much higher metallicities, whilst a strong wind implementation reduces the normalisation of the MZR. 
Note, that in the fiducial Illustris simulation, the normalisation of the MZR was used to set the metal content of ejected wind material. 
At both $z=0$ and $z=1$, the Illustris simulations do not find a flattening of the MZR at high stellar masses when including feedback, but the no-feedback run does recover this flattening.
However, the no-feedback run of Illustris recovers far too many galaxies at all stellar masses, and a higher cosmic star formation rate density at all times.
This suggests that the flattening at high stellar masses is driven by a saturation in metallicity through high SFRs and not driven by metal loss. 

Unlike Illustris, where little bending was observed, the MZR recovered from IllustrisTNG  \citep{Torreyevolutionmassmetallicityrelation2019} does recover a saturation in metallicity at $z=0$. 
However, both the stellar mass and metallicity where this bending occurs is much lower than that recovered by \textsc{ProSpect}. 
At $z=1$, the MZR recovered from IllustrisTNG has a negative slope for $M_\star > 10^{10.5} M_\odot$ which is not found in our results or other spectroscopic observations. 

SIMBA, and its predecessor MUFASA, recover systematically lower metallicities at all stellar masses than our results. 
Additionally, the shape of the MZR from SIMBA has a different shape to those from other simulations.
A dip in metallicity can be seen at $M_\star \sim 10^{10} M_\odot$ which is not evident in MUFASA. 
The normalisation and shape of the MZR recovered from MUFASA are closer to our derived MZR than SIMBA, despite the fact that \cite{DaveSIMBACosmologicalsimulations2019} describe the MZR from MUFASA as being too steep.

We also include the MZR from the recalibrated EAGLE simulations presented in \cite{DeRossiGalaxymetallicityscaling2017}. 
On average, EAGLE galaxies have slightly higher metallicities than other simulations but are in reasonable agreement with our results at $M_\star > 10^{10}M_\odot$.
However, EAGLE recovers a flatter MZR with little increase in metallicity with increasing stellar mass. 

In addition to results from cosmological hydrodynamic simulations, \add{Figure~\ref{fig:SimulationsMZR} also shows the MZR derived} from the GAEA \citep{DeLuciaGasaccretionregulates2020a} and SHARK \citep{LagosSharkintroducingopen2018} semi-analytic models. 
Of all the simulation results shown, the GAEA MZR agrees closest with our results in both normalisation and shape at $z=0$ over the full mass range covered by GAMA and the more sparsely sampled DEVILS mass range (dashed black line). 
The GAEA MZR recovers the expected lower metallicities for low-mass galaxies and a saturation of metallicity for high-mass galaxies. 
\cite{DeLuciaGasaccretionregulates2020a} describe the predicted scatter as larger than observed particularly for galaxies with $M_\star < 10^{10}\,M_\odot$, however their scatter is smaller than the 1$\sigma$ range that we derive.
At $z=1$ we show the MZR recovered from GAEA and presented in \cite{HirschmannGalaxyassemblystellar2016}. 
It matches the normalisation of our MZR at $M_\star = 10^{11}M_\odot$ and flattens at the same metallicity, but this flattening occurs at higher stellar masses than we find. 

The MZR from SHARK is relatively consistent with our results for $10^{9} < M_\star / M_\odot < 10^{10}$ but produces significantly higher metallicities for $M_\star > 10^{10.5}\,M_\odot$ at both $z=0$ and $z=1$.

\subsubsection{The impact of AGN feedback on the MZR}

One of the physical processes often used to explain the shape of the MZR is galactic winds and feedback removing metals from galaxy potential wells \citep{TremontiOriginMassMetallicity2004,KobayashiSimulationsCosmicChemical2007}.
For galaxies with low stellar masses, supernova explosions are effective at ejecting metals from the ISM via outflows \citep[e.g.][]{KobayashiSimulationsCosmicChemical2007,BlancCharacteristicMassScale2019}. 
Although dying stars produce metals, the force of a supernova explosion can be sufficient to eject this enriched gas from the ISM, especially for lower mass galaxies with a lower gravitational potential (see \citealt{DEugeniogasphasemetallicitiesstarforming2018} for a discussion on the relationship between metallicity and gravitational potential). 
Galaxies with a higher gravitational potential might be less affected by supernova feedback, but they are more likely to host AGN that can also remove gas from a galaxy through mechanical feedback \citep{FabianObservationalEvidenceActive2012}.
Mechanical feedback occurs when most of the energy of an AGN is released in kinetic form via radio-emitting jets, which can expel large amounts of gas from galaxies \citep{MorgantimanyroutesAGN2017}. 
This form of AGN feedback can readily eject enriched material but can also prevent accretion onto a galaxy, further impacting the metallicity of the ISM \citep{SilkQuasarsgalaxyformation1998}. 
These forms of AGN feedback are also implemented in simulations in order to match the bright end of the luminosity function, while stellar feedback drives the agreement at the faint end.
\add{However, the physical mechanisms and impact of AGN feedback on host galaxies are still poorly understood \citep{ShangguanAGNFeedbackStar2020,LahaIonizedoutflowsactive2021,KoudmaniTwocanplay2022,NobelsinterplayAGNfeedback2022}.}


Altering the physics implemented in theoretical simulations can allow us to understand the impact of AGN feedback on resulting galaxy properties. 
\cite{TayloreffectsAGNfeedback2015} find that including AGN feedback in their simulations has no impact on the derived stellar or gas-phase mass-metallicity relations. 
They attribute this to the fact that AGN feedback quenches star formation most efficiently at low redshift, after the peak of star formation and chemical enrichment.
However, in the context of the EAGLE simulations, \cite{DeRossiGalaxymetallicityscaling2017} find that the flattening of the MZR at $M_\star > 10^{10} M_\odot$ is mainly regulated by AGN feedback. 
By comparing a run with no AGN feedback to one with AGN feedback, while keeping the resolution and box size the same, they find a metallicity offset of $\approx 0.3\,$dex at $M_\star \sim 10^{11}M_\odot$. 
They comment that, despite the fact that galaxies with evidence of current AGN activity are removed from observed MZR studies, black hole feedback may have previously occurred influencing the chemical enrichment of high mass galaxies. 
Using three hydrodynamical simulations of a single galaxy, \cite{EisenreichActivegalacticnuclei2017} find that mechanical AGN feedback flattens the metallicity gradient across the galaxy, but the inclusion of radiation feedback decreases the overall metallicity. 

Our results suggest little impact on the MZR from galaxies with current signatures of AGN. 
However, as highlighted by \cite{DeRossiGalaxymetallicityscaling2017}, black hole activity could occur in cyclic episodes and galaxies which do not currently display signs of activity could have been impacted by past episodes. 

\begin{figure*}
    \centering
    \includegraphics[width = \linewidth]{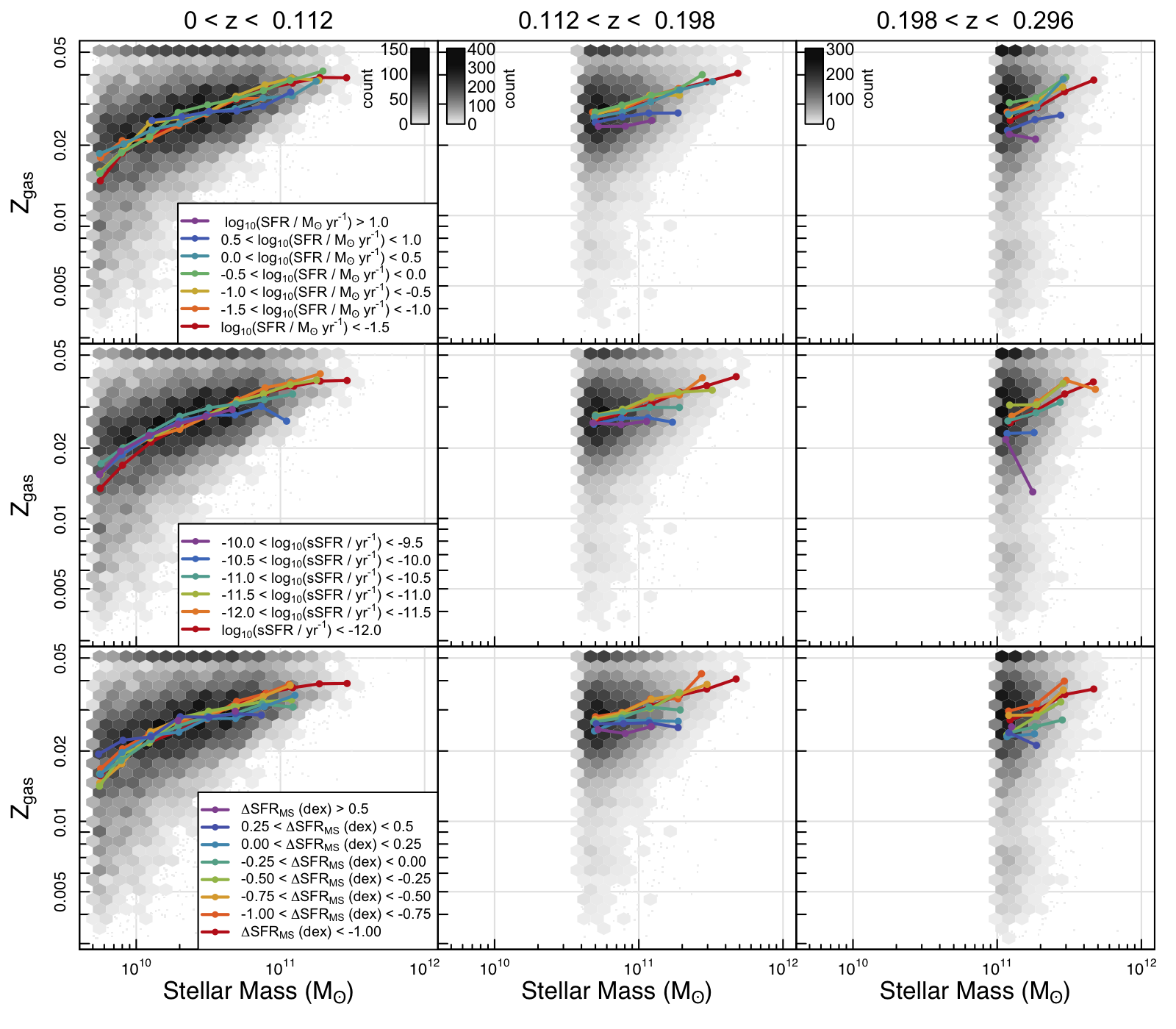}
    \caption{The dependence of the MZR on SFR (top), sSFR (middle) and location relative to the star-formation main sequence (bottom). 
    In all panels we show the median $Z_\text{gas}$ in $M_\star$ bins for each bin in SFR or sSFR, or the location relative to the main-sequence. 
    The underlying 2D histogram shows the total distribution of galaxies in each redshift bin.
    }
    \label{fig:MZR_SFRtrends}
\end{figure*}

\section{The influence of star formation history on the MZR}\label{sec:SFHs}
In this section, we explore additional higher dimensional correlations with the MZR to provide better constraints on the processes that regulate metallicity.  
Due to the nature of our SED fitting technique, we can unpack not only the relationship with current SFR or sSFR, but also with the overall star formation history. 

\add{For this section, we limit our sample to just galaxies from the GAMA survey at $z<0.3$ to maximize the number of galaxies in each stellar mass and metallicity bin. 
Additionally, the uncertainties on the stellar masses, metallicities, and SFRs estimates are lower for GAMA galaxies than for DEVILS galaxies. 
We also remove galaxies with $f_\text{AGN} > 0.1$ to limit to only galaxies with no sign of current AGN activity. }

\subsection{Star Formation Rate}\label{sec:SFRs}

Higher-order correlations of the MZR are most commonly explored using the SFR or sSFR as an indicator of the star formation history of the galaxy. 
The existence or not of a correlation of the MZR with SFR is still under debate \citep[e.g.][]{SanchezMassmetallicityrelationexplored2013,Sanchezmassmetallicityrelationrevisited2017,SanchezSAMIgalaxysurvey2019,SalimCriticalLookMassMetallicityStar2014,Barrera-BallesterosSeparateWaysMassMetallicity2017,CresciFundamentalmetallicityrelation2019}.

\add{Previous work by \cite{Mannuccifundamentalrelationmass2010} also showed an inverse correlation between SFR and metallicity at a given stellar mass for objects with $M_\star \lesssim 10^{11}\,M_\odot$ (see also \citealt{Dayalphysicsfundamentalmetallicity2013,Curtimassmetallicityfundamentalmetallicity2020a}).
\cite{SalimCriticalLookMassMetallicityStar2014} also found an anti-correlation between metallicity and sSFR but only for $M\star < 10^{10.5}\,M_\odot$. 
Above this mass, they find no correlation between metallicity and sSFR.
Similarly, at higher redshifts ($z\sim1.6$), \cite{ZahidFMOSCOSMOSSurveyStarforming2014} find an anti-correlation between metallicity and SFR for galaxies at fixed stellar mass (see also \citealt{Stottfundamentalmetallicityrelation2013,KashinoFMOSCOSMOSSurveyStarforming2017,KashinoFMOSCOSMOSSurveyStarforming2019, CresciFundamentalmetallicityrelation2019}).
This is opposite to the findings of \cite{Lara-LopezGalaxyMassAssembly2013} who find that at higher stellar masses ($M_\star > 10^{10}\,M_\odot$) a higher SFR/sSFR is associated with a higher metallicity at a given stellar mass. 
However, \cite{SanchezMassmetallicityrelationexplored2013,Sanchezmassmetallicityrelationrevisited2017,SanchezSAMIgalaxysurvey2019} find that there is no dependence of the MZR on SFR or sSFR and find no evidence for a SFR-$M_\star$-Z fundamental plane. 
}


Figure~\ref{fig:MZR_SFRtrends} shows the recovered MZR when binning by SFR, sSFR, and by location relative to the star-formation main sequence ($\Delta \text{SFR}_\text{MS}$, \citealt{NoeskeStarFormationAEGIS2007,SpeagleHighlyConsistentFramework2014}) for the GAMA sample. 
In each case, we only show bins with more than 20 galaxies. 

When considering the dependence of the MZR on SFR, we recover similar relations across all SFRs, especially at $z=0$.
At $z=0$ and $M_\star \approx 10^{11}\,M_\odot$, there is evidence of a trend within the highest SFR bins, where a higher SFR is associated with lower metallicity.
However, at all stellar masses, the range in median metallicities is $\sim0.13\,$dex while the SFR bins span 2.5 dex. 
At lower masses ($M_\star \approx 10^{10}\,M_\odot$), the trend in metallicity is reversed for galaxies with relatively high star formation rates (blue, teal, and green lines), where a higher metallicity is associated with a higher star formation rate. 
In the higher redshift bins, it is also evident that the highest star formation rates are offset to lower metallicity values, however the spread in median metallicities is greater than at $z=0$.
At all redshifts and mass ranges explored here, we do not find significant offsets to higher metallicities for galaxies with very low SFRs (yellow, orange, and red lines) which might be expected if a galaxy was undergoing strangulation (e.g. the stellar metallicity results from \citealt{PengStrangulationprimarymechanism2015}).

We also show the MZR binned in sSFR intervals to remove mass dependence in each bin.
There are similar trends between sSFR and metallicity as described above with SFR, where higher sSFRs are associated with lower metallicities. 
This is most extreme in our highest redshift bin where the spread in median metallicity is $\sim0.2$\,dex.
These trends are in agreement with the findings of \cite{Mannuccifundamentalrelationmass2010}, but opposite to the findings of \cite{Lara-LopezGalaxyMassAssembly2013}. 

We also present the MZR recovered when binned in distance from the star-formation main sequence in Figure~\ref{fig:MZR_SFRtrends}. 
The star-formation main sequence (MS) refers to the tight relation between stellar mass and star formation rate which holds over more than five orders of magnitude in stellar mass and has been shown to hold out to $z>5$ \citep{ThorneDeepExtragalacticVIsible2021}.
The reason for binning by distance from the MS is to isolate galaxies that have similar relative star formation activity across the full range of stellar masses (e.g. starburst, main-sequence, green-valley, or passive galaxies). 
By binning in distance from the MS we can more fairly compare population types than in the case of SFR or sSFR.

To calculate the distance from the MS for each galaxy in our sample we use the redshift dependent MS fit presented in \cite{ThorneDeepExtragalacticVIsible2021}.
The \cite{ThorneDeepExtragalacticVIsible2021} MS was derived using the same parent sample as this work and used a double power law to model the flattening of the MS at higher stellar masses.
For each galaxy, we calculate the vertical distance in dex from the MS ($\Delta \text{SFR}_\text{MS}$), where positive (negative) values represent galaxies that lie above (below) the main sequence at their given stellar mass. 
Binning by $\Delta \text{SFR}_\text{MS}$ results in similar trends across the MZR as with SFR and sSFR where high star formation activity correlates with lower metallicities at high masses ($M_\star > 10^{10.5}\,M_\odot$) at all redshifts.
However any trend between $\Delta \text{SFR}_\text{MS}$ and metallicity at lower masses ($M_\star < 10^{10.5}\,M_\odot$) can be excluded at a $\sim0.1\,$dex level, but not at levels below 0.1\,dex.

Overall, we find evidence of an anti-correlation between SFR, sSFR, and $\Delta \text{SFR}_\text{MS}$ with $Z_\text{gas}$ at $M_\star > 10^{11}\,M_\odot$ across all redshifts explored.
Of the three parameters, the association between $\Delta \text{SFR}_\text{MS}$ and metallicity is clearest especially in our two highest redshift bins. 
However, in some cases the range in median metallicities across different star formation bins is only $\sim 0.1\,$dex which could be within errors from total or bin selection effects, or could be within uncertainties due to the SED fitting technique.
Furthermore, due to the nature of SED fitting the recovered SFRs and metallicities are more intertwined than when measured from emission lines which could impact the direction and strength of the trends recovered.  

\add{However, these trends are consistent with results from \cite{SalimCriticalLookMassMetallicityStar2014}, who find an anti-correlation between metallicity and SFR, sSFR, \textit{and} $\Delta \text{sSFR}$ at fixed stellar mass using a sample of low redshift galaxies from the Sloan Digital Sky Survey (SDSS; \citealt{AbazajianSecondDataRelease2004,AbazajianSeventhDataRelease2009}) .}

\subsection{Star Formation History}\label{subsec:SFHs}
\begin{figure}
    \centering
    \includegraphics[width = \linewidth]{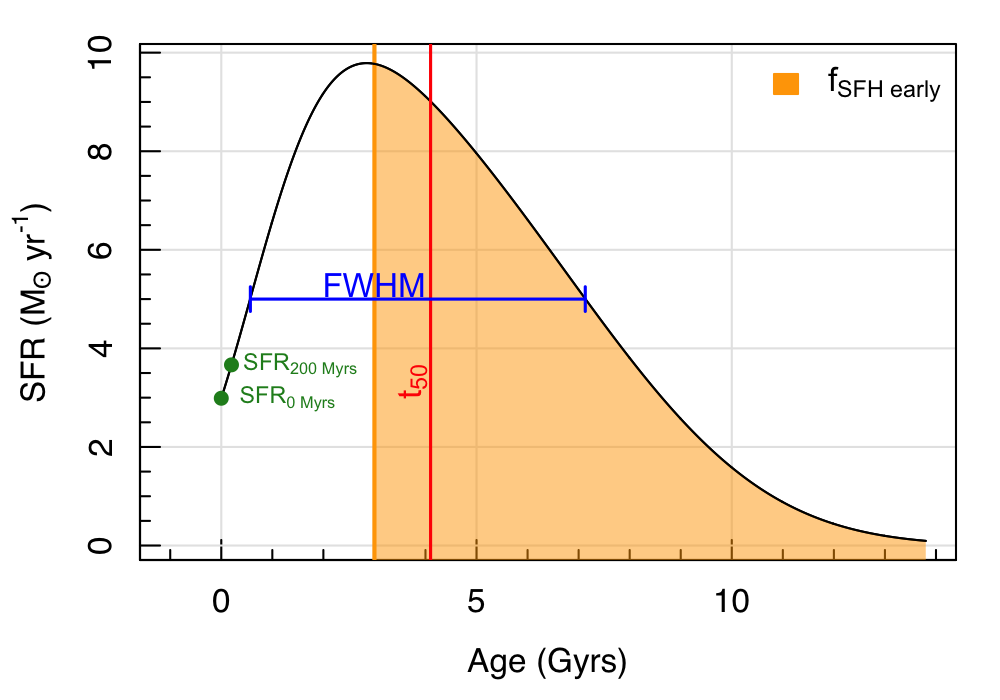}
    \caption{An example SFH with the descriptive parameters from  Section~\ref{subsec:SFHs} labelled.
    Note that the $\Delta\log\text{SFH}_\text{200}$ parameter is defined to be the difference of the $\log_{10} \text{SFR}$, but the SFR is shown here in linear space for ease.
    The $f_\text{SFH early}$ is defined as the integral of the orange shaded region divided by the integral of the whole SFH. 
    \add{$t_{50}$ is the half mass age or the age at which the galaxy has formed half of its total stellar mass and the FWHM is the width of the SFH in Gyrs at half of the maximum SFR. }
    }
    \label{fig:SFHparams}
\end{figure}

\begin{figure}
    \centering
    \includegraphics[width = \linewidth]{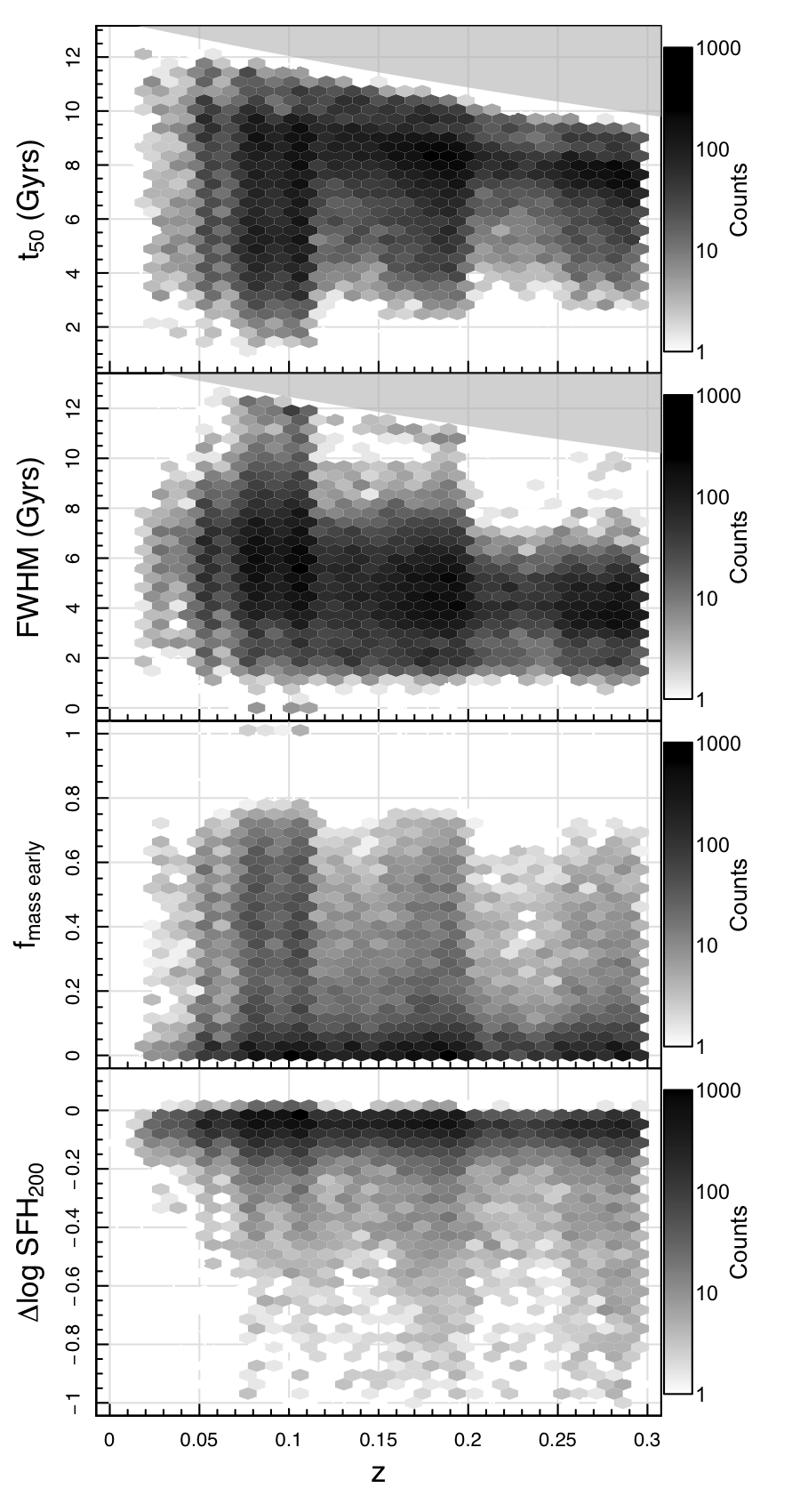}
    \caption{
    \add{The distribution of each of the parameters describing a galaxy's SFH from Figure~\ref{fig:SFHparams} as a function of redshift for all GAMA galaxies above the mass completeness limit and with $f_\text{AGN} < 0.1$.
    The grey shaded region in the top two panels indicates the part of parameter space that is not possible due to the dependence of both parameters on the age of the Universe at each redshift. 
    The visible vertical artefacts are the edges of the redshift bins used to calculate the completeness limit.}}
    \label{fig:NewSFHParmaDistributions}
\end{figure}

Although the SFR of galaxies provides insight into the current star formation activity, it is the overall star formation history (SFH) and chemical evolution history of the galaxy that influences the observed metallicity. 
In this section we use the SFHs derived using \textsc{ProSpect} for the sample of GAMA galaxies to explore higher-order correlations with the MZR. 
\add{As luminous AGN components can produce emission very similar to that generated by star formation, we remove galaxies with $f_\text{AGN} > 0.1$ to ensure that the star formation histories recovered by \textsc{ProSpect} are not biased by this degeneracy.}

Although we parameterise the SFH within \textsc{ProSpect} using four parameters to describe the skewed Normal shape, these can often be degenerate and produce similar SFHs despite having very different values. 
Because of this we define and use four different parameters that can better describe the features of a galaxy's SFH without being degenerate.
These are:
\begin{itemize}
    \item $t_{50}$ - the half mass age, i.e. the age at which the galaxy has formed half of its total stellar mass (Gyrs)
    \item FWHM - the full width half maximum, defined to be the width of the SFH at half the height of the maximum SFR (Gyrs)
    \item $f_\text{mass early}$ - the fraction of mass formed before the peak of star formation (e.g. $f_\text{mass early} = 0.5$ corresponds to a galaxy with a perfectly Normal SFH where half the mass is formed before the peak, and half the mass is formed after the peak.)
    \item $\Delta\log\text{SFH}_\text{200}= \log_{10} (\text{SFR}_\text{0\,Myrs}) - \log_{10} (\text{SFR}_\text{200\,Myrs})$, the change in SFR in dex over the last 200\,Myrs. Negative values correspond to declining SFHs and larger absolute values correspond to a steeper change over the last 200\,Myrs. 
\end{itemize}
For clarity we show an example SFH in Figure~\ref{fig:SFHparams} with each of the descriptive parameters labelled. 
\add{Figure~\ref{fig:NewSFHParmaDistributions} shows the distribution of each of these descriptive parameters as a function of redshift. 
We show only GAMA objects above the mass completeness limit and with $f_\text{AGN} < 0.1$. 
As the $t_{50}$ and FWHM parameters are dependent on the age of the Universe at the redshift of each object, the upper right corner of both parameter spaces is not a possible combination (indicated by the grey shading).
}

\begin{figure*}
    \centering
    \includegraphics[width = \linewidth]{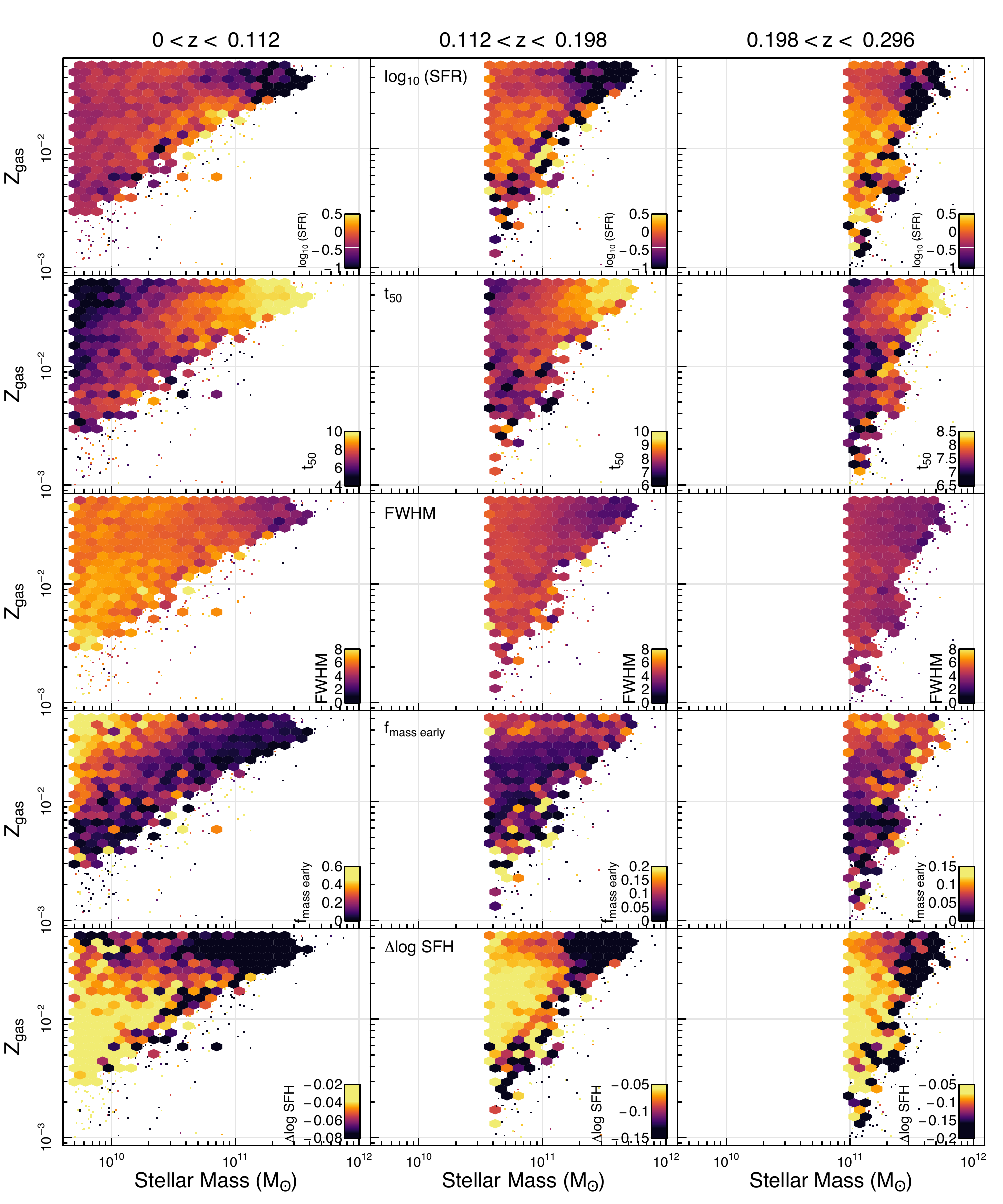}
    \caption{The MZR in three redshift bins using only the GAMA sample coloured by the median (first row) SFR, (second row) half mass age, (third row) the width of the star formation history at half the height of the peak, (fourth row) the fraction of the total mass formed before the peak of star formation, and (fifth row) the change in the SFR over the last 200\,Myrs in dex.}
    \label{fig:MZR_SFHs}
\end{figure*}

\begin{figure*}
    \centering
    \includegraphics[width=\linewidth]{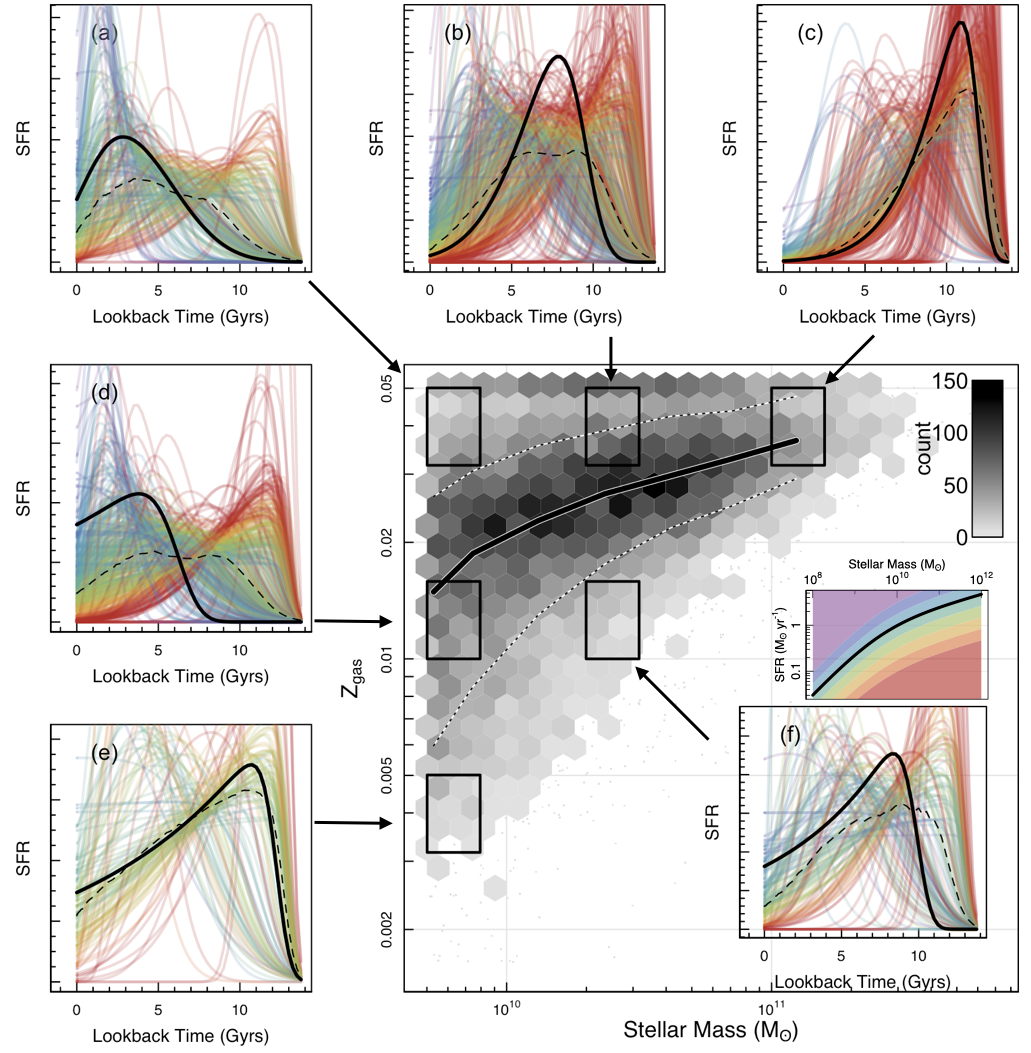}
    \caption{In the main panel we show the MZR at $z=0$ using the GAMA sample as the grey 2D histogram. 
    We also show the running median and $1\sigma$ range as the solid and dashed lines respectively. 
    We highlight six regions of the MZR as the  black rectangles and in the surrounding panels show the SFHs of all galaxies within that region of the MZR. 
    The colouring of each SFH corresponds to each galaxy's position relative to the star-formation main sequence from \citet{ThorneDeepExtragalacticVIsible2021}. 
    We show the $z\approx0$ main-sequence fit in the small inset panel as the solid black line and show each of the $\Delta \text{SFR}_\text{MS}$ bins as the shaded regions where purple corresponds to starburst galaxies while red represents passive galaxies. 
    \add{In each of the SFH panels, the dashed line shows the median inferred SFR of all SFHs at each age step. 
    The thick black line in each panel shows a toy model SFH informed by the distribution of parameters shown in Figure~\ref{fig:MZR_SFHs}.}
    As we are only concerned about the shape and not the normalisation of the SFH we do not show values on the SFR axis and every SFH in each box is normalised to the same total mass value.}
    \label{fig:SummaryPlot}
\end{figure*}

Figure~\ref{fig:MZR_SFHs} shows the MZR where each cell is coloured by the median value of each of these parameters. 
We also show the MZR coloured by the median SFR for comparison. 
As discussed in Section~\ref{sec:SFRs}, at a given stellar mass there is no clear trend of SFR with metallicity at $z\approx0$.
There is a slight trend with SFR at $z\approx0.25$ where higher star formation rates are associated with a lower metallicity at $M_\star = 10^{11}\,M_\odot$.
However, the most apparent trend is a manifestation of the star-forming main sequence \citep[see e.g.][]{NoeskeStarFormationAEGIS2007,LeeTurnoverGalaxyMain2015,ThorneDeepExtragalacticVIsible2021} where higher masses are associated with higher star formation rates, except at the highest stellar masses where we see a significant decrease in average SFR due to a large population of quenched/passive galaxies. 

When considering the dependence on half mass age ($t_{50}$), it is immediately evident that we are recovering downsizing in our results through the trend between stellar mass and age, where higher mass galaxies have older stellar populations than low mass galaxies. 
However, at lower masses ($M_\star \sim 10^{10}\,M_\odot$) we also find that galaxies with a higher metallicity at a given stellar mass are younger. 

The third row of Figure~\ref{fig:MZR_SFHs} colours the MZR by the median full-width half maximum of the SFH in a given mass and metallicity bin. 
This parameter is a proxy for how quickly the majority of star formation happened in a galaxy, where smaller FWHM values represent quicker star formation episodes. 
As with the median age, these panels are in agreement with cosmic downsizing where higher mass galaxies formed earlier with shorter star formation timescales while lower mass galaxies formed later on longer timescales. 
At $z=0$, there is also a slight vertical trend at the lowest stellar masses examined here ($M_\star < 10^{10} M_\odot$), where lower metallicities are associated with longer star formation timescales. 

The fourth row of Figure~\ref{fig:MZR_SFHs} shows each cell coloured by the median fraction of mass formed before the peak of star formation ($f_\text{SFH early}$). 
A higher value corresponds to a galaxy with a slow increase in star formation in the early Universe, while a value close to zero corresponds to a galaxy in which star formation ramped up very quickly. 
We find clear evidence that objects with low masses ($M_\star \approx 10^{10}\,M_\odot$) and high metallicities formed a larger fraction of mass before the peak of star formation than lower metallicity galaxies at the same stellar mass. 
At high stellar masses ($M_\star > 10^{11}\,M_\odot$), it is clear that, on average, the fraction of mass formed before the maximum star formation rate is low, consistent with a picture where these galaxies rapidly ramped up star formation in the early Universe, and formed quickly over short timescales. 

The final row shows the MZR coloured by the median change in SFR in dex of the galaxy over the last 200\,Myrs ($\Delta \log \text{SFH}_\text{200}$).
For the GAMA sample, most galaxies are experiencing declining SFHs over the last 200\,Myrs consistent with the overall star formation rate density trend. 
We find that higher mass galaxies are, on average, experiencing faster declines in SFR than low mass galaxies. 
However, at low masses we see that lower-metallicity galaxies have flatter recent SFHs than higher-metallicity galaxies at the same stellar mass which are declining at a faster rate.

Figure~\ref{fig:SummaryPlot} shows the range of SFHs found in different locations on the MZR at $z=0$. 
We show each individual SFH for galaxies in each region as the coloured lines, normalised to the same total stellar mass to highlight similarities/differences in shape. 
Each line is coloured by the galaxy's vertical distance from the MS ($\Delta \text{SFR}_\text{MS}$) as per Section~\ref{sec:SFRs} where purple represents starburst galaxies while red represents passive/quenched galaxies. 

\add{For each region, we show the median SFH, which is calculated by taking the median inferred SFR of all galaxies at each lookback time (dashed black line). 
As these median SFHs are independent at each lookback time they are not forced to take the form of a skewed-Normal.
We also show a toy-model SFH (solid black line) generated by transforming the median parameters in the left column of Figure~\ref{fig:MZR_SFHs} ($t_{50}$, FWHM, $f_\text{mass early}$, and $\Delta \log \text{SFH}$), to the required inputs of the \texttt{massfunc\_snorm\_trunc}.
These are generated with the following parameters:
\begin{enumerate}
\renewcommand{\theenumi}{(\alph{enumi})}
    \item \texttt{mpeak = 3}, \texttt{mperiod} = 3, \texttt{mskew} = -0.3
    \item \texttt{mpeak = 8}, \texttt{mperiod} = 2, \texttt{mskew} = 0.3
    \item \texttt{mpeak = 11}, \texttt{mperiod} = 1.5, \texttt{mskew} = 0.3
    \item \texttt{mpeak = 4}, \texttt{mperiod} = 3, \texttt{mskew} = 0.5
    \item \texttt{mpeak = 11}, \texttt{mperiod} = 3, \texttt{mskew} = 0.5
    \item \texttt{mpeak = 8.5}, \texttt{mperiod} = 2, \texttt{mskew} = 0.5
\end{enumerate}
}

The regions were selected to examine the SFHs in `extreme' locations on the MZR to explore how changes in SFH can explain the scatter in the MZR.
In each region there is a diversity in individual SFHs and cases where the individual SFHs are very different from the toy model. 
This is to be expected as the toy model is based on the median parameter distribution and there will always be outlier SFHs.
Additionally, there is nothing built into \textsc{ProSpect} to link a particular SFH to a final metallicity value, however some SFH and $Z_\text{gas}$ combinations could be the consequence of degeneracies, which we discuss below.

In the low mass, intermediate metallicity (panel d) and intermediate mass, high metallicity (panel b) regions, we find that two distinct shapes of SFH are equally prevalent - an old, and a young one. 
These two populations correlate well with the current position of each galaxy on the MS where the older population are currently passive while the younger population are still actively forming stars.
Meanwhile, in the high mass, high metallicity case (panel c) we recover SFH shapes that are consistently similar to the toy model.

The low mass, high metallicity region (a) also shows a diversity in SFHs with two dominant populations, the population that formed-early and have since quenched, and the, dominant, population of galaxies that have formed less than 4\,Gyrs ago. 
When comparing to the other high metallicity regions at both intermediate and high masses, there are significantly fewer galaxies that formed recently. 
If this population of low mass, high metallicity galaxies were actually older and more metal-poor, then due to the age-metallicity degeneracy it is possible for \textsc{ProSpect} to falsely characterise these galaxies as young and metal rich. 
In this scenario, the galaxies should actually be in the intermediate metallicity, low mass region (panel d) of the MZR (a more typically populated region), whereas instead they have been placed in the low mass, high metallicity region (panel a). 
Although it is unclear the extent to which this is occurring, we can use the \texttt{genbox} function included in \textsc{ProSpect} to explore potential situations in which low mass, high metallicity galaxies might exist. 
The \texttt{genbox} function allows the user to specify star formation functions, inflow and outflow rates, and metallicity arguments to explore the evolution of gas and stellar mass, and metallicity of a system over time.
Through investigation using these models\footnote{We tested using various star formation prescriptions including cases where the star formation is proportional to the stellar mass, gas mass, or total mass of the system. We also tested various inflow and outflow scenarios, including constant inflows and outflows, and more complicated models such as outflows coupled to supernova events with various mass loading factors, and inflows proportional to the total mass of the system. The trends reported are consistent across all physically motivated scenarios tested.}, it is clear that in order to form a galaxy with a very high $Z_\text{gas}$ ($Z \sim 0.05$) at any stellar mass, the galaxy has to reach low gas fractions ($M_\text{ISM} / M_\star < 0.1$). 
As the gas fraction of a galaxy decreases, supernova enrichment is more efficient at increasing the metallicity of the ISM due to the fact that there is much less gas to enrich. 
Observational measurements of the gas fraction of $z\approx0$ galaxies in the xGASS sample \citep{CatinellaxGASStotalcold2018} shows that there are populations of galaxies with $M_\star \sim 10^{10}\,M_\odot$ with the required gas fractions to reach high gas-phase metallicities as recovered in this work. 
Galaxies at $M_\star \sim 10^{10}\,M_\odot$ with high metallicities ($Z_\text{gas} > 0.04$) are also found in large spectral line samples  \citep[e.g.][]{TremontiOriginMassMetallicity2004,Lara-LopezStudystarforminggalaxies2010} and therefore the recovery of these properties in our work is not necessarily an artefact of measuring metallicities through SED fitting.
The consequence of requiring low gas fractions in galaxies is that there is very little gas left from which to form stars and would suggest that these galaxies are likely quenching and therefore have low current SFRs. 
The combination of these arguments suggests that the reason behind the existence of two populations in region (a) is that there are galaxies with physical SFHs and metallicities (the older, passive galaxies), and galaxies that are likely an artefact the age-metallicity degeneracy (the younger, star forming galaxies).

Despite these caveats, recent results suggest that these trends in SFH, and therefore gas mass, across the MZR are physical. 
For example, \cite{ZhouSemianalyticspectralfitting2022} use a semi-analytic spectral fitting technique to recover the star formation history and chemical evolution history of galaxies in the MANGA sample.
They find that both the SFHs and chemical evolution histories have strong mass dependence whereby massive galaxies accumulate their stellar masses and become enriched earlier. 
Their use of spectroscopic data means their results are less likely to be biased by the age-dust-metallicity degeneracy. 
Additionally, \cite{Brownroleatomichydrogen2018} and \cite{ChenRoleInnerMass2022} find anti-correlations between H{\sc\,i} gas mass and metallicity at a fixed stellar mass where at a fixed stellar mass, a higher metallicity corresponds to a lower gas mass \citep[see also][]{Bothwellfundamentalrelationmetallicity2013,Hughesrolecoldgas2013}. 

\section{Conclusion}\label{sec:conclusion}

It has been known for decades that broadband optical-NIR colors can be used to separate the age and metallicity of composite populations.
Despite this, metallicity is still treated as a nuisance parameter in most SED fitting codes. 
Almost all SED fitting codes assume that metallicity is constant over the lifetime of a galaxy, where the value is either set to the solar value, fitted as a free parameter, or constrained using the mass-metallicity as an informative prior.  
However, galaxies are known to evolve chemically over time and trend towards higher metallicities as the Universe evolves.

Using the SED fitting code \textsc{ProSpect} \citep{RobothamProSpectgeneratingspectral2020}, we have used a simple physically motivated prescription to model an evolving metallicity history for individual galaxies \citep[also used in][]{BellstedtGalaxyMassAssembly2020b,BellstedtGalaxyMassAssembly2021,ThorneDeepExtragalacticVIsible2021,ThorneDeepExtragalacticVIsible2022} to extract star formation and metallicity histories for $\sim90\,000$ galaxies from the GAMA and DEVILS surveys.

Measuring metallicities from SED fitting is difficult due to the age-dust-metallicity degeneracy even when making simple, physically motivated assumptions about the chemical enrichment history. 
It is even more challenging when trying to measure metallicities pushing near the depths of photometry or with poor quality redshifts. 
We stress that despite this, it is still important to model metallicity in a physically motivated way so that it does not impact other galaxy properties (\add{Appendix~\ref{App:Zimplementation}}). 
If the photometry quality is too poor to accurately measure the metallicity we suggest at least allowing the metallicity to evolve but using the MZR to provide an informative prior on the metallicity value. 

In this work we used the metallicity and stellar mass estimates from \textsc{ProSpect} to derive the mass-metallicity relation in eight redshift bins from $z=0$ to $z=1.1$ (Section~\ref{sec:MZR} and Figure~\ref{fig:MZR}).
While there was no fitting prior set on the resulting gas-phase metallicity values in our implementation of \textsc{ProSpect}, we recover a MZR that is consistent at all redshifts with previous literature measurements from both nebular emission lines and forensic predictions from \cite{BellstedtGalaxyMassAssembly2021}.
Our results suggest that the normalisation of the MZR has evolved by $0.1\,$dex since $z=1$, consistent with results from \cite{CrescimetallicitypropertieszCOSMOS2012} and \cite{BellstedtGalaxyMassAssembly2021} \add{who find little evolution in the normalisation of the MZR over the last 7 billion years} (Figure~\ref{fig:MZR_evolve}). 
However, as we find a discontinuity in the normalisation of the MZR when swapping from the GAMA and DEVILS combined sample to just DEVILS, the existence and strength of this evolution could be impacted by sample selection.

Using the \cite{FritzRevisitinginfraredspectra2006} AGN model included in \textsc{ProSpect} we also measure the MZR for AGN host galaxies for $z<0.3$ and can rule out differences in the normalisation or shape of the MZR to that measured for galaxies with no SED-detected AGN at a level of 0.1\,dex, but not at levels below 0.1\,dex (Section~\ref{sec:AGNMZR} and Figure~\ref{fig:MZR_AGN}).

When exploring higher order correlations of the MZR with star formation activity we find that for galaxies with $M_\star > 10^{10.5}\,M_\odot$, higher levels of star formation are associated with lower metallicities (Section~\ref{sec:SFRs}). 
This trend is most clear when binning by distance from the star-formation main-sequence ($\Delta\text{SFR}_\text{MS}$). 
\add{This trend is in agreement with results demonstrating the existence of an anti-correlation between between metallicity and star-formation activity (either SFR, sSFR, or $\Delta \text{sSFR}$) at fixed stellar mass \citep[see e.g.][]{Mannuccifundamentalrelationmass2010,SalimCriticalLookMassMetallicityStar2014,Curtimassmetallicityfundamentalmetallicity2020a}.}
We also find that the overall shape of a galaxy's SFH is correlated with its location on the MZR (Section~\ref{subsec:SFHs} and Figure~\ref{fig:MZR_SFHs}). 
Most apparent is the effect of galaxy downsizing, where we find that more massive galaxies formed at earlier times.
\add{This is consistent with the resolved galaxy study from \citet{ZhouSemianalyticspectralfitting2022} who found that chemical evolution histories have a strong dependence on stellar mass. }
We find that for galaxies with $M_\star \approx 10^{10}\,M_\odot$, lower metallicities are recovered for galaxies that formed half their mass earlier than galaxies with higher metallicities.

For galaxies at a given stellar mass, we also find that higher metallicity values are associated with shorter periods of star formation and star formation histories where the majority of stellar mass was formed before the peak star formation rate (Figure~\ref{fig:SummaryPlot}). 
These correlations between age and metallicity could be due to the age--metallicity--dust degeneracy, however recent results from integral field spectroscopy studies suggest these results could be physical.  

The analysis in this work, combined with the accurate derivation of the cosmic SFH \citep{BellstedtGalaxyMassAssembly2020b} and stellar mass functions and star-forming main sequence \citep{ThorneDeepExtragalacticVIsible2021}, highlights the importance of giving careful consideration to both the metallicity and star formation histories of galaxies and demonstrates that accurate galaxy properties can be derived using SED fitting on only broad-band photometry. 

\section*{Acknowledgements}
We thank the anonymous referee for their detailed reading and comments which improved this paper. 
JET is supported by the Australian
Government Research Training Program (RTP) Scholarship.
ASGR and LJMD acknowledge support from the \textit{Australian Research Council's} Future Fellowship scheme (FT200100375 and FT200100055 respectively). 
SB acknowledges support from the \textit{Australian Research Council’s} Discovery Project and Future Fellowship funding schemes (DP180103740, FT200100375).
LC acknowledges support from the \textit{Australian Research Council} Discovery Project and Future Fellowship funding schemes (DP210100337, FT180100066).

We acknowledge the traditional owners of the land on which this research was completed, the Whadjuk Noongar people, and the land on which the AAT stands, the Gamilaraay people, and pay our respects to elders past and present.

DEVILS is an Australian project based around a spectroscopic campaign using the Anglo-Australian Telescope. 
The DEVILS input catalogue is generated from data taken as part of the ESO VISTA-VIDEO \citep{JarvisVISTADeepExtragalactic2013} and UltraVISTA \citep{McCrackenUltraVISTAnewultradeep2012} surveys. DEVILS is part funded via Discovery Programs by the Australian Research Council and the participating institutions. The DEVILS website is \url{https://devilsurvey.org}. The DEVILS data is hosted and provided by AAO Data Central (\url{https://datacentral.org.au/}).

GAMA is a joint European-Australasian project based around a spectroscopic campaign using the Anglo- Australian Telescope. The GAMA input catalogue is based on data taken from the Sloan Digital Sky Survey and the UKIRT Infrared Deep Sky Survey. Complementary imaging of the GAMA regions is being obtained by a number of in-dependent survey programmes including GALEX MIS, VST KiDS, VISTA VIKING, WISE, Herschel-ATLAS, GMRT and ASKAP providing UV to radio coverage. GAMA is funded by the STFC (UK), the ARC (Australia), the AAO, and the participating institutions. The GAMA website is \url{http://www.gama-survey.org/}.

This work was supported by resources provided by the Pawsey Supercomputing Centre with funding from the Australian Government and the Government of Western Australia. We gratefully acknowledge DUG Technology for their support and HPC services.

All of the work presented here was made possible by the free and open R software environment \citep{RCoreTeamLanguageEnvironmentStatistical2020}. All figures in this paper were made using the R \textsc{magicaxis} package \citep{RobothammagicaxisPrettyscientific2016}. This work also makes use of the \textsc{celestial} package \citep{RobothamCelestialCommonastronomical2016}.

\section*{Data Availability}
The DEVILS data products used in this paper are from the internal DEVILS team data release and presented in \cite{DaviesDeepExtragalacticVIsible2021}, \cite{ThorneDeepExtragalacticVIsible2021} and \cite{ThorneDeepExtragalacticVIsible2022}. 
They will be made publicly available as part of the DEVILS first data release described in Davies et al. (in preparation). 
The GAMA data products used in this paper were released as part of GAMA DR4 \citep{DriverGalaxyMassAssembly2022} and are available on the GAMA website\footnote{\url{http://www.gama-survey.org/dr4/}}.




\bibliographystyle{mnras}
\bibliography{MyBib} 

\begin{thebibliography}{}
\makeatletter
\relax
\def\mn@urlcharsother{\let\do\@makeother \do\$\do\&\do\#\do\^\do\_\do\%\do\~}
\def\mn@doi{\begingroup\mn@urlcharsother \@ifnextchar [ {\mn@doi@}
  {\mn@doi@[]}}
\def\mn@doi@[#1]#2{\def\@tempa{#1}\ifx\@tempa\@empty \href
  {http://dx.doi.org/#2} {doi:#2}\else \href {http://dx.doi.org/#2} {#1}\fi
  \endgroup}
\def\mn@eprint#1#2{\mn@eprint@#1:#2::\@nil}
\def\mn@eprint@arXiv#1{\href {http://arxiv.org/abs/#1} {{\tt arXiv:#1}}}
\def\mn@eprint@dblp#1{\href {http://dblp.uni-trier.de/rec/bibtex/#1.xml}
  {dblp:#1}}
\def\mn@eprint@#1:#2:#3:#4\@nil{\def\@tempa {#1}\def\@tempb {#2}\def\@tempc
  {#3}\ifx \@tempc \@empty \let \@tempc \@tempb \let \@tempb \@tempa \fi \ifx
  \@tempb \@empty \def\@tempb {arXiv}\fi \@ifundefined
  {mn@eprint@\@tempb}{\@tempb:\@tempc}{\expandafter \expandafter \csname
  mn@eprint@\@tempb\endcsname \expandafter{\@tempc}}}

\bibitem[\protect\citeauthoryear{Abazajian et~al.,}{Abazajian
  et~al.}{2004}]{AbazajianSecondDataRelease2004}
Abazajian K.,  et~al., 2004, \mn@doi [AJ] {10.1086/421365}, 128, 502

\bibitem[\protect\citeauthoryear{Abazajian et~al.,}{Abazajian
  et~al.}{2009}]{AbazajianSeventhDataRelease2009}
Abazajian K.~N.,  et~al., 2009, \mn@doi [ApJS] {10.1088/0067-0049/182/2/543},
  182, 543

\bibitem[\protect\citeauthoryear{Aihara et~al.,}{Aihara
  et~al.}{2019}]{AiharaSeconddatarelease2019}
Aihara H.,  et~al., 2019, \mn@doi [PASJ] {10.1093/pasj/psz103}, 71, 114

\bibitem[\protect\citeauthoryear{Alloin, {Collin-Souffrin}, Joly  \&
  Vigroux}{Alloin et~al.}{1979}]{AlloinNitrogenoxygenabundances1979}
Alloin D.,  {Collin-Souffrin} S.,  Joly M.,   Vigroux L.,  1979, A\&A, 78, 200

\bibitem[\protect\citeauthoryear{Asplund, Grevesse, Sauval  \& Scott}{Asplund
  et~al.}{2009}]{AsplundChemicalCompositionSun2009}
Asplund M.,  Grevesse N.,  Sauval A.~J.,   Scott P.,  2009, \mn@doi [ARA\&A]
  {10.1146/annurev.astro.46.060407.145222}, 47, 481

\bibitem[\protect\citeauthoryear{Baldwin, Phillips  \& Terlevich}{Baldwin
  et~al.}{1981}]{BaldwinClassificationparametersemissionline1981}
Baldwin J.~A.,  Phillips M.~M.,   Terlevich R.,  1981, \mn@doi [PASP]
  {10.1086/130766}, 93, 5

\bibitem[\protect\citeauthoryear{{Barrera-Ballesteros}, S{\'a}nchez, Heckman,
  Blanc  \& {MaNGA Team}}{{Barrera-Ballesteros}
  et~al.}{2017}]{Barrera-BallesterosSeparateWaysMassMetallicity2017}
{Barrera-Ballesteros} J.~K.,  S{\'a}nchez S.~F.,  Heckman T.,  Blanc G.~A.,
  {MaNGA Team} 2017, \mn@doi [ApJ] {10.3847/1538-4357/aa7aa9}, 844, 80

\bibitem[\protect\citeauthoryear{Bellstedt et~al.,}{Bellstedt
  et~al.}{2020a}]{BellstedtGalaxyMassAssembly2020a}
Bellstedt S.,  et~al., 2020a, \mn@doi [MNRAS] {10.1093/mnras/staa1466}, 496,
  3235

\bibitem[\protect\citeauthoryear{Bellstedt et~al.,}{Bellstedt
  et~al.}{2020b}]{BellstedtGalaxyMassAssembly2020b}
Bellstedt S.,  et~al., 2020b, \mn@doi [MNRAS] {10.1093/mnras/staa2620}, 498,
  5581

\bibitem[\protect\citeauthoryear{Bellstedt et~al.,}{Bellstedt
  et~al.}{2021}]{BellstedtGalaxyMassAssembly2021}
Bellstedt S.,  et~al., 2021, \mn@doi [MNRAS] {10.1093/mnras/stab550}, 503, 3309

\bibitem[\protect\citeauthoryear{Blanc, Lu, Benson, Katsianis  \&
  Barraza}{Blanc et~al.}{2019}]{BlancCharacteristicMassScale2019}
Blanc G.~A.,  Lu Y.,  Benson A.,  Katsianis A.,   Barraza M.,  2019, \mn@doi
  [ApJ] {10.3847/1538-4357/ab16ec}, 877, 6

\bibitem[\protect\citeauthoryear{Boquien, Burgarella, Roehlly, Buat, Ciesla,
  Corre, Inoue  \& Salas}{Boquien et~al.}{2019}]{BoquienCIGALEpythonCode2019}
Boquien M.,  Burgarella D.,  Roehlly Y.,  Buat V.,  Ciesla L.,  Corre D.,
  Inoue A.~K.,   Salas H.,  2019, \mn@doi [A\&A] {10.1051/0004-6361/201834156},
  622, A103

\bibitem[\protect\citeauthoryear{Bothwell, Maiolino, Kennicutt, Cresci,
  Mannucci, Marconi  \& Cicone}{Bothwell
  et~al.}{2013}]{Bothwellfundamentalrelationmetallicity2013}
Bothwell M.~S.,  Maiolino R.,  Kennicutt R.,  Cresci G.,  Mannucci F.,  Marconi
  A.,   Cicone C.,  2013, \mn@doi [MNRAS] {10.1093/mnras/stt817}, 433, 1425

\bibitem[\protect\citeauthoryear{Brooks, Governato, Booth, Willman, Gardner,
  Wadsley, Stinson  \& Quinn}{Brooks
  et~al.}{2007}]{BrooksOriginEvolutionMassMetallicity2007}
Brooks A.~M.,  Governato F.,  Booth C.~M.,  Willman B.,  Gardner J.~P.,
  Wadsley J.,  Stinson G.,   Quinn T.,  2007, \mn@doi [ApJ] {10.1086/511765},
  655, L17

\bibitem[\protect\citeauthoryear{Brown, Cortese, Catinella  \& Kilborn}{Brown
  et~al.}{2018}]{Brownroleatomichydrogen2018}
Brown T.,  Cortese L.,  Catinella B.,   Kilborn V.,  2018, \mn@doi [MNRAS]
  {10.1093/mnras/stx2452}, 473, 1868

\bibitem[\protect\citeauthoryear{Bruzual \& Charlot}{Bruzual \&
  Charlot}{2003}]{BruzualStellarpopulationsynthesis2003}
Bruzual G.,  Charlot S.,  2003, \mn@doi [MNRAS]
  {10.1046/j.1365-8711.2003.06897.x}, 344, 1000

\bibitem[\protect\citeauthoryear{Cai, Zhao, Zhang, Bai  \& Liu}{Cai
  et~al.}{2020}]{CaiStellarPopulationsSample2020}
Cai W.,  Zhao Y.,  Zhang H.-X.,  Bai J.-M.,   Liu H.-T.,  2020, \mn@doi [ApJ]
  {10.3847/1538-4357/abb81c}, 903, 58

\bibitem[\protect\citeauthoryear{Calura, Pipino, Chiappini, Matteucci  \&
  Maiolino}{Calura et~al.}{2009}]{Caluraevolutionmassmetallicityrelation2009}
Calura F.,  Pipino A.,  Chiappini C.,  Matteucci F.,   Maiolino R.,  2009,
  \mn@doi [A\&A] {10.1051/0004-6361/200911756}, 504, 373

\bibitem[\protect\citeauthoryear{{Camps-Fari{\~n}a}, S{\'a}nchez, Carigi,
  Lacerda, {Garc{\'i}a-Benito}, Mast, Galbany  \&
  {Barrera-Ballesteros}}{{Camps-Fari{\~n}a}
  et~al.}{2021}]{Camps-FarinaSignaturesAGNinducedMetal2021}
{Camps-Fari{\~n}a} A.,  S{\'a}nchez S.~F.,  Carigi L.,  Lacerda E. A.~D.,
  {Garc{\'i}a-Benito} R.,  Mast D.,  Galbany L.,   {Barrera-Ballesteros} J.~K.,
   2021, \mn@doi [ApJL] {10.3847/2041-8213/ac37c1}, 922, L20

\bibitem[\protect\citeauthoryear{Capak et~al.,}{Capak
  et~al.}{2007}]{CapakFirstReleaseCOSMOS2007}
Capak P.,  et~al., 2007, \mn@doi [ApJS] {10.1086/519081}, 172, 99

\bibitem[\protect\citeauthoryear{Carnall, McLure, Dunlop  \& Dav{\'e}}{Carnall
  et~al.}{2018}]{CarnallInferringstarformation2018}
Carnall A.~C.,  McLure R.~J.,  Dunlop J.~S.,   Dav{\'e} R.,  2018, \mn@doi
  [MNRAS] {10.1093/mnras/sty2169}, 480, 4379

\bibitem[\protect\citeauthoryear{Catinella et~al.,}{Catinella
  et~al.}{2018}]{CatinellaxGASStotalcold2018}
Catinella B.,  et~al., 2018, \mn@doi [MNRAS] {10.1093/mnras/sty089}, 476, 875

\bibitem[\protect\citeauthoryear{Chabrier}{Chabrier}{2003}]{ChabrierGalacticStellarSubstellar2003}
Chabrier G.,  2003, \mn@doi [PASP] {10.1086/376392}, 115, 763

\bibitem[\protect\citeauthoryear{Charlot \& Fall}{Charlot \&
  Fall}{2000}]{CharlotSimpleModelAbsorption2000}
Charlot S.,  Fall S.~M.,  2000, \mn@doi [ApJ] {10.1086/309250}, 539, 718

\bibitem[\protect\citeauthoryear{Chartab et~al.,}{Chartab
  et~al.}{2021}]{ChartabMOSDEFSurveyEnvironmental2021b}
Chartab N.,  et~al., 2021, \mn@doi [ApJ] {10.3847/1538-4357/abd71f}, 908, 120

\bibitem[\protect\citeauthoryear{Chen, Wang  \& Kong}{Chen
  et~al.}{2022}]{ChenRoleInnerMass2022}
Chen X.,  Wang J.,   Kong X.,  2022, \mn@doi [ApJ] {10.3847/1538-4357/ac70d0},
  933, 39

\bibitem[\protect\citeauthoryear{Chisholm, Tremonti  \& Leitherer}{Chisholm
  et~al.}{2018}]{ChisholmMetalenrichedgalacticoutflows2018}
Chisholm J.,  Tremonti C.,   Leitherer C.,  2018, \mn@doi [MNRAS]
  {10.1093/mnras/sty2380}, 481, 1690

\bibitem[\protect\citeauthoryear{Conroy}{Conroy}{2013}]{ConroyModelingPanchromaticSpectral2013}
Conroy C.,  2013, \mn@doi [ARA\&A] {10.1146/annurev-astro-082812-141017}, 51,
  393

\bibitem[\protect\citeauthoryear{Cresci, Mannucci, Sommariva, Maiolino, Marconi
   \& Brusa}{Cresci et~al.}{2012}]{CrescimetallicitypropertieszCOSMOS2012}
Cresci G.,  Mannucci F.,  Sommariva V.,  Maiolino R.,  Marconi A.,   Brusa M.,
  2012, \mn@doi [MNRAS] {10.1111/j.1365-2966.2011.20299.x}, 421, 262

\bibitem[\protect\citeauthoryear{Cresci, Mannucci  \& Curti}{Cresci
  et~al.}{2019}]{CresciFundamentalmetallicityrelation2019}
Cresci G.,  Mannucci F.,   Curti M.,  2019, \mn@doi [A\&A]
  {10.1051/0004-6361/201834637}, 627, A42

\bibitem[\protect\citeauthoryear{Cullen et~al.,}{Cullen
  et~al.}{2021}]{CullenNIRVANDELSSurveyrobust2021}
Cullen F.,  et~al., 2021, \mn@doi [MNRAS] {10.1093/mnras/stab1340}, 505, 903

\bibitem[\protect\citeauthoryear{Curti, Cresci, Mannucci, Marconi, Maiolino  \&
  Esposito}{Curti et~al.}{2017}]{CurtiNewfullyempirical2017}
Curti M.,  Cresci G.,  Mannucci F.,  Marconi A.,  Maiolino R.,   Esposito S.,
  2017, \mn@doi [MNRAS] {10.1093/mnras/stw2766}, 465, 1384

\bibitem[\protect\citeauthoryear{Curti, Mannucci, Cresci  \& Maiolino}{Curti
  et~al.}{2020}]{Curtimassmetallicityfundamentalmetallicity2020a}
Curti M.,  Mannucci F.,  Cresci G.,   Maiolino R.,  2020, \mn@doi [MNRAS]
  {10.1093/mnras/stz2910}, 491, 944

\bibitem[\protect\citeauthoryear{D'Eugenio, Colless, Groves, Bian  \&
  Barone}{D'Eugenio
  et~al.}{2018}]{DEugeniogasphasemetallicitiesstarforming2018}
D'Eugenio F.,  Colless M.,  Groves B.,  Bian F.,   Barone T.~M.,  2018, \mn@doi
  [MNRAS] {10.1093/mnras/sty1424}, 479, 1807

\bibitem[\protect\citeauthoryear{Dalcanton}{Dalcanton}{2007}]{DalcantonMetallicityGalaxyDisks2007}
Dalcanton J.~J.,  2007, \mn@doi [ApJ] {10.1086/508913}, 658, 941

\bibitem[\protect\citeauthoryear{Dale, Helou, Magdis, Armus, {D{\'i}az-Santos}
  \& Shi}{Dale et~al.}{2014}]{DaleTwoParameterModelInfrared2014}
Dale D.~A.,  Helou G.,  Magdis G.~E.,  Armus L.,  {D{\'i}az-Santos} T.,   Shi
  Y.,  2014, \mn@doi [ApJ] {10.1088/0004-637X/784/1/83}, 784, 83

\bibitem[\protect\citeauthoryear{Dav{\'e}, Finlator  \& Oppenheimer}{Dav{\'e}
  et~al.}{2011}]{DaveGalaxyevolutioncosmological2011}
Dav{\'e} R.,  Finlator K.,   Oppenheimer B.~D.,  2011, \mn@doi [MNRAS]
  {10.1111/j.1365-2966.2011.19132.x}, 416, 1354

\bibitem[\protect\citeauthoryear{Dav{\'e}, Thompson  \& Hopkins}{Dav{\'e}
  et~al.}{2016}]{DaveMUFASAgalaxyformation2016}
Dav{\'e} R.,  Thompson R.,   Hopkins P.~F.,  2016, \mn@doi [MNRAS]
  {10.1093/mnras/stw1862}, 462, 3265

\bibitem[\protect\citeauthoryear{Dav{\'e}, Rafieferantsoa, Thompson  \&
  Hopkins}{Dav{\'e} et~al.}{2017}]{DaveMUFASAGalaxystar2017}
Dav{\'e} R.,  Rafieferantsoa M.~H.,  Thompson R.~J.,   Hopkins P.~F.,  2017,
  \mn@doi [MNRAS] {10.1093/mnras/stx108}, 467, 115

\bibitem[\protect\citeauthoryear{Dav{\'e}, {Angl{\'e}s-Alc{\'a}zar}, Narayanan,
  Li, Rafieferantsoa  \& Appleby}{Dav{\'e}
  et~al.}{2019}]{DaveSIMBACosmologicalsimulations2019}
Dav{\'e} R.,  {Angl{\'e}s-Alc{\'a}zar} D.,  Narayanan D.,  Li Q.,
  Rafieferantsoa M.~H.,   Appleby S.,  2019, \mn@doi [MNRAS]
  {10.1093/mnras/stz937}, 486, 2827

\bibitem[\protect\citeauthoryear{Davies et~al.,}{Davies
  et~al.}{2018}]{DaviesDeepExtragalacticVIsible2018}
Davies L. J.~M.,  et~al., 2018, \mn@doi [MNRAS] {10.1093/mnras/sty1553}, 480,
  768

\bibitem[\protect\citeauthoryear{Davies et~al.,}{Davies
  et~al.}{2021}]{DaviesDeepExtragalacticVIsible2021}
Davies L. J.~M.,  et~al., 2021, \mn@doi [MNRAS] {10.1093/mnras/stab1601}, 506,
  256

\bibitem[\protect\citeauthoryear{Dayal, Ferrara  \& Dunlop}{Dayal
  et~al.}{2013}]{Dayalphysicsfundamentalmetallicity2013}
Dayal P.,  Ferrara A.,   Dunlop J.~S.,  2013, \mn@doi [MNRAS]
  {10.1093/mnras/stt083}, 430, 2891

\bibitem[\protect\citeauthoryear{De~Lucia, Xie, Fontanot  \&
  Hirschmann}{De~Lucia et~al.}{2020}]{DeLuciaGasaccretionregulates2020a}
De~Lucia G.,  Xie L.,  Fontanot F.,   Hirschmann M.,  2020, \mn@doi [MNRAS]
  {10.1093/mnras/staa2556}, 498, 3215

\bibitem[\protect\citeauthoryear{De~Rossi, Bower, Font, Schaye  \&
  Theuns}{De~Rossi et~al.}{2017}]{DeRossiGalaxymetallicityscaling2017}
De~Rossi M.~E.,  Bower R.~G.,  Font A.~S.,  Schaye J.,   Theuns T.,  2017,
  \mn@doi [MNRAS] {10.1093/mnras/stx2158}, 472, 3354

\bibitem[\protect\citeauthoryear{Dopita, Sutherland, Nicholls, Kewley  \&
  Vogt}{Dopita et~al.}{2013}]{DopitaNewStronglineAbundance2013}
Dopita M.~A.,  Sutherland R.~S.,  Nicholls D.~C.,  Kewley L.~J.,   Vogt F.
  P.~A.,  2013, \mn@doi [ApJS] {10.1088/0067-0049/208/1/10}, 208, 10

\bibitem[\protect\citeauthoryear{Dopita, Kewley, Sutherland  \&
  Nicholls}{Dopita et~al.}{2016}]{DopitaChemicalabundanceshighredshift2016}
Dopita M.~A.,  Kewley L.~J.,  Sutherland R.~S.,   Nicholls D.~C.,  2016,
  \mn@doi [Astrophys. Space Sci.] {10.1007/s10509-016-2657-8}, 361, 61

\bibitem[\protect\citeauthoryear{Driver et~al.,}{Driver
  et~al.}{2022}]{DriverGalaxyMassAssembly2022}
Driver S.~P.,  et~al., 2022, \mn@doi [MNRAS] {10.1093/mnras/stac472}, 513, 439

\bibitem[\protect\citeauthoryear{Eales et~al.,}{Eales
  et~al.}{2010}]{EalesHerschelATLAS2010}
Eales S.,  et~al., 2010, \mn@doi [PASP] {10.1086/653086}, 122, 499

\bibitem[\protect\citeauthoryear{Edge, Sutherland, Kuijken, Driver, McMahon,
  Eales  \& Emerson}{Edge et~al.}{2013}]{EdgeVISTAKilodegreeInfrared2013}
Edge A.,  Sutherland W.,  Kuijken K.,  Driver S.,  McMahon R.,  Eales S.,
  Emerson J.~P.,  2013, The Messenger, 154, 32

\bibitem[\protect\citeauthoryear{Eisenreich, Naab, Choi, Ostriker  \&
  Emsellem}{Eisenreich et~al.}{2017}]{EisenreichActivegalacticnuclei2017}
Eisenreich M.,  Naab T.,  Choi E.,  Ostriker J.~P.,   Emsellem E.,  2017,
  \mn@doi [MNRAS] {10.1093/mnras/stx473}, 468, 751

\bibitem[\protect\citeauthoryear{Eldridge \& Stanway}{Eldridge \&
  Stanway}{2012}]{Eldridgeeffectstellarevolution2012}
Eldridge J.~J.,  Stanway E.~R.,  2012, \mn@doi [MNRAS]
  {10.1111/j.1365-2966.2011.19713.x}, 419, 479

\bibitem[\protect\citeauthoryear{Eldridge, Stanway, Xiao, McClelland, Taylor,
  Ng, Greis  \& Bray}{Eldridge
  et~al.}{2017}]{EldridgeBinaryPopulationSpectral2017}
Eldridge J.~J.,  Stanway E.~R.,  Xiao L.,  McClelland L. A.~S.,  Taylor G.,  Ng
  M.,  Greis S. M.~L.,   Bray J.~C.,  2017, \mn@doi [PASA]
  {10.1017/pasa.2017.51}, 34, e058

\bibitem[\protect\citeauthoryear{Ellison, Patton, Simard  \&
  McConnachie}{Ellison et~al.}{2008}]{EllisonCluesOriginMassMetallicity2008}
Ellison S.~L.,  Patton D.~R.,  Simard L.,   McConnachie A.~W.,  2008, \mn@doi
  [ApJ] {10.1086/527296}, 672, L107

\bibitem[\protect\citeauthoryear{Eminian, Kauffmann, Charlot, Wild, Bruzual,
  Rettura  \& Loveday}{Eminian
  et~al.}{2008}]{EminianPhysicalinterpretationnearinfrared2008}
Eminian C.,  Kauffmann G.,  Charlot S.,  Wild V.,  Bruzual G.,  Rettura A.,
  Loveday J.,  2008, \mn@doi [MNRAS] {10.1111/j.1365-2966.2007.12742.x}, 384,
  930

\bibitem[\protect\citeauthoryear{Fabian}{Fabian}{2012}]{FabianObservationalEvidenceActive2012}
Fabian A.~C.,  2012, \mn@doi [ARA\&A] {10.1146/annurev-astro-081811-125521},
  50, 455

\bibitem[\protect\citeauthoryear{Feltre, Hatziminaoglou, Fritz  \&
  Franceschini}{Feltre et~al.}{2012}]{FeltreSmoothclumpydust2012}
Feltre A.,  Hatziminaoglou E.,  Fritz J.,   Franceschini A.,  2012, \mn@doi
  [MNRAS] {10.1111/j.1365-2966.2012.21695.x}, 426, 120

\bibitem[\protect\citeauthoryear{{Fraser-McKelvie} et~al.,}{{Fraser-McKelvie}
  et~al.}{2022}]{Fraser-McKelvieSAMIGalaxySurvey2022}
{Fraser-McKelvie} A.,  et~al., 2022, \mn@doi [MNRAS] {10.1093/mnras/stab3430},
  510, 320

\bibitem[\protect\citeauthoryear{Fritz, Franceschini  \& Hatziminaoglou}{Fritz
  et~al.}{2006}]{FritzRevisitinginfraredspectra2006}
Fritz J.,  Franceschini A.,   Hatziminaoglou E.,  2006, \mn@doi [MNRAS]
  {10.1111/j.1365-2966.2006.09866.x}, 366, 767

\bibitem[\protect\citeauthoryear{Garnett}{Garnett}{2002}]{GarnettLuminosityMetallicityRelationEffective2002}
Garnett D.~R.,  2002, \mn@doi [ApJ] {10.1086/344301}, 581, 1019

\bibitem[\protect\citeauthoryear{Gillman et~al.,}{Gillman
  et~al.}{2021}]{Gillmanevolutiongasphasemetallicity2021}
Gillman S.,  et~al., 2021, \mn@doi [MNRAS] {10.1093/mnras/staa3400}, 500, 4229

\bibitem[\protect\citeauthoryear{Gillman et~al.,}{Gillman
  et~al.}{2022}]{Gillmanresolvedchemicalabundance2022}
Gillman S.,  et~al., 2022, \mn@doi [MNRAS] {10.1093/mnras/stac580}, 512, 3480

\bibitem[\protect\citeauthoryear{Hamann, Korista, Ferland, Warner  \&
  Baldwin}{Hamann et~al.}{2002}]{HamannMetallicitiesAbundanceRatios2002}
Hamann F.,  Korista K.~T.,  Ferland G.~J.,  Warner C.,   Baldwin J.,  2002,
  \mn@doi [ApJ] {10.1086/324289}, 564, 592

\bibitem[\protect\citeauthoryear{Helton, Strom, Greene, Bezanson  \&
  Beaton}{Helton et~al.}{2022}]{HeltonNebularPropertiesStarforming2022}
Helton J.~M.,  Strom A.~L.,  Greene J.~E.,  Bezanson R.,   Beaton R.,  2022,
  \mn@doi [ApJ] {10.3847/1538-4357/ac78e5}, 934, 81

\bibitem[\protect\citeauthoryear{Henry, Martin, Finlator  \& Dressler}{Henry
  et~al.}{2013}]{HenryMetallicityEvolutionLowmass2013}
Henry A.,  Martin C.~L.,  Finlator K.,   Dressler A.,  2013, \mn@doi [ApJ]
  {10.1088/0004-637X/769/2/148}, 769, 148

\bibitem[\protect\citeauthoryear{Hirschauer, Salzer, Janowiecki  \&
  Wegner}{Hirschauer et~al.}{2018}]{HirschauerMetalAbundancesKISS2018}
Hirschauer A.~S.,  Salzer J.~J.,  Janowiecki S.,   Wegner G.~A.,  2018, \mn@doi
  [AJ] {10.3847/1538-3881/aaa4ba}, 155, 82

\bibitem[\protect\citeauthoryear{Hirschmann, De~Lucia  \& Fontanot}{Hirschmann
  et~al.}{2016}]{HirschmannGalaxyassemblystellar2016}
Hirschmann M.,  De~Lucia G.,   Fontanot F.,  2016, \mn@doi [MNRAS]
  {10.1093/mnras/stw1318}, 461, 1760

\bibitem[\protect\citeauthoryear{Huang et~al.,}{Huang
  et~al.}{2019}]{HuangMassMetallicityRelationRedshift2019}
Huang C.,  et~al., 2019, \mn@doi [ApJ] {10.3847/1538-4357/ab4902}, 886, 31

\bibitem[\protect\citeauthoryear{Hughes, Cortese, Boselli, Gavazzi  \&
  Davies}{Hughes et~al.}{2013}]{Hughesrolecoldgas2013}
Hughes T.~M.,  Cortese L.,  Boselli A.,  Gavazzi G.,   Davies J.~I.,  2013,
  \mn@doi [A\&A] {10.1051/0004-6361/201218822}, 550, A115

\bibitem[\protect\citeauthoryear{Hunt, Dayal, Magrini  \& Ferrara}{Hunt
  et~al.}{2016}]{HuntCoevolutionmetallicitystar2016}
Hunt L.,  Dayal P.,  Magrini L.,   Ferrara A.,  2016, \mn@doi [MNRAS]
  {10.1093/mnras/stw1993}, 463, 2002

\bibitem[\protect\citeauthoryear{Jarvis et~al.,}{Jarvis
  et~al.}{2013}]{JarvisVISTADeepExtragalactic2013}
Jarvis M.~J.,  et~al., 2013, \mn@doi [MNRAS] {10.1093/mnras/sts118}, 428, 1281

\bibitem[\protect\citeauthoryear{Johnson, Leja, Conroy  \& Speagle}{Johnson
  et~al.}{2021}]{JohnsonStellarPopulationInference2021}
Johnson B.~D.,  Leja J.,  Conroy C.,   Speagle J.~S.,  2021, \mn@doi [ApJS]
  {10.3847/1538-4365/abef67}, 254, 22

\bibitem[\protect\citeauthoryear{Kashino et~al.,}{Kashino
  et~al.}{2017}]{KashinoFMOSCOSMOSSurveyStarforming2017}
Kashino D.,  et~al., 2017, \mn@doi [ApJ] {10.3847/1538-4357/835/1/88}, 835, 88

\bibitem[\protect\citeauthoryear{Kashino et~al.,}{Kashino
  et~al.}{2019}]{KashinoFMOSCOSMOSSurveyStarforming2019}
Kashino D.,  et~al., 2019, \mn@doi [ApJS] {10.3847/1538-4365/ab06c4}, 241, 10

\bibitem[\protect\citeauthoryear{Kewley \& Dopita}{Kewley \&
  Dopita}{2002}]{KewleyUsingStrongLines2002}
Kewley L.~J.,  Dopita M.~A.,  2002, \mn@doi [ApJS] {10.1086/341326}, 142, 35

\bibitem[\protect\citeauthoryear{Kewley \& Ellison}{Kewley \&
  Ellison}{2008}]{KewleyMetallicityCalibrationsMassMetallicity2008}
Kewley L.~J.,  Ellison S.~L.,  2008, \mn@doi [ApJ] {10.1086/587500}, 681, 1183

\bibitem[\protect\citeauthoryear{Kobayashi, Springel  \& White}{Kobayashi
  et~al.}{2007}]{KobayashiSimulationsCosmicChemical2007}
Kobayashi C.,  Springel V.,   White S. D.~M.,  2007, \mn@doi [MNRAS]
  {10.1111/j.1365-2966.2007.11555.x}, 376, 1465

\bibitem[\protect\citeauthoryear{Kobulnicky \& Kewley}{Kobulnicky \&
  Kewley}{2004}]{KobulnickyMetallicities2004}
Kobulnicky H.~A.,  Kewley L.~J.,  2004, \mn@doi [ApJ] {10.1086/425299}, 617,
  240

\bibitem[\protect\citeauthoryear{Koudmani, Sijacki  \& Smith}{Koudmani
  et~al.}{2022}]{KoudmaniTwocanplay2022}
Koudmani S.,  Sijacki D.,   Smith M.~C.,  2022, \mn@doi [MNRAS]
  {10.1093/mnras/stac2252}

\bibitem[\protect\citeauthoryear{Lagos, Tobar, Robotham, Obreschkow, Mitchell,
  Power  \& Elahi}{Lagos et~al.}{2018}]{LagosSharkintroducingopen2018}
Lagos C. d.~P.,  Tobar R.~J.,  Robotham A. S.~G.,  Obreschkow D.,  Mitchell
  P.~D.,  Power C.,   Elahi P.~J.,  2018, \mn@doi [MNRAS]
  {10.1093/mnras/sty2440}, 481, 3573

\bibitem[\protect\citeauthoryear{Laha, Reynolds, Reeves, Kriss, Guainazzi,
  Smith, Veilleux  \& Proga}{Laha et~al.}{2021}]{LahaIonizedoutflowsactive2021}
Laha S.,  Reynolds C.~S.,  Reeves J.,  Kriss G.,  Guainazzi M.,  Smith R.,
  Veilleux S.,   Proga D.,  2021, \mn@doi [Nat. Astron.]
  {10.1038/s41550-020-01255-2}, 5, 13

\bibitem[\protect\citeauthoryear{Laigle et~al.,}{Laigle
  et~al.}{2016}]{LaigleCOSMOS2015CATALOGEXPLORING2016}
Laigle C.,  et~al., 2016, \mn@doi [ApJS] {10.3847/0067-0049/224/2/24}, 224, 24

\bibitem[\protect\citeauthoryear{{Lara-L{\'o}pez}, Bongiovanni, Cepa,
  P{\'e}rez~Garc{\'i}a, {S{\'a}nchez-Portal}, Casta{\~n}eda,
  Fern{\'a}ndez~Lorenzo  \& Povi{\'c}}{{Lara-L{\'o}pez}
  et~al.}{2010}]{Lara-LopezStudystarforminggalaxies2010}
{Lara-L{\'o}pez} M.~A.,  Bongiovanni A.,  Cepa J.,  P{\'e}rez~Garc{\'i}a A.~M.,
   {S{\'a}nchez-Portal} M.,  Casta{\~n}eda H.~O.,  Fern{\'a}ndez~Lorenzo M.,
  Povi{\'c} M.,  2010, \mn@doi [A\&A] {10.1051/0004-6361/200913886}, 519, A31

\bibitem[\protect\citeauthoryear{{Lara-L{\'o}pez} et~al.,}{{Lara-L{\'o}pez}
  et~al.}{2013}]{Lara-LopezGalaxyMassAssembly2013}
{Lara-L{\'o}pez} M.~A.,  et~al., 2013, \mn@doi [MNRAS] {10.1093/mnras/stt1031},
  434, 451

\bibitem[\protect\citeauthoryear{Larson \& Dinerstein}{Larson \&
  Dinerstein}{1975}]{LarsonGaslossgroups1975}
Larson R.~B.,  Dinerstein H.~L.,  1975, \mn@doi [PASP] {10.1086/129870}, 87,
  911

\bibitem[\protect\citeauthoryear{Lee, Worthey, Trager  \& Faber}{Lee
  et~al.}{2007}]{LeeAgeMetallicityEstimation2007}
Lee H.-c.,  Worthey G.,  Trager S.~C.,   Faber S.~M.,  2007, \mn@doi [AJ]
  {10.1086/518855}, 664, 215

\bibitem[\protect\citeauthoryear{Lee et~al.,}{Lee
  et~al.}{2015}]{LeeTurnoverGalaxyMain2015}
Lee N.,  et~al., 2015, \mn@doi [ApJ] {10.1088/0004-637X/801/2/80}, 801, 80

\bibitem[\protect\citeauthoryear{Leja, Johnson, Conroy  \& {van Dokkum}}{Leja
  et~al.}{2018}]{LejaHotDustPanchromatic2018}
Leja J.,  Johnson B.~D.,  Conroy C.,   {van Dokkum} P.,  2018, \mn@doi [ApJ]
  {10.3847/1538-4357/aaa8db}, 854, 62

\bibitem[\protect\citeauthoryear{Lequeux, Peimbert, Rayo, Serrano  \&
  {Torres-Peimbert}}{Lequeux
  et~al.}{1979}]{LequeuxChemicalCompositionEvolution1979}
Lequeux J.,  Peimbert M.,  Rayo J.~F.,  Serrano A.,   {Torres-Peimbert} S.,
  1979, A\&A, 80, 155

\bibitem[\protect\citeauthoryear{Lutz et~al.,}{Lutz
  et~al.}{2011}]{LutzPACSEvolutionaryProbe2011}
Lutz D.,  et~al., 2011, \mn@doi [A\&A] {10.1051/0004-6361/201117107}, 532, A90

\bibitem[\protect\citeauthoryear{Ly, Malkan, Rigby  \& Nagao}{Ly
  et~al.}{2016}]{LyMetalAbundancesCosmic2016}
Ly C.,  Malkan M.~A.,  Rigby J.~R.,   Nagao T.,  2016, \mn@doi [ApJ]
  {10.3847/0004-637X/828/2/67}, 828, 67

\bibitem[\protect\citeauthoryear{Maiolino et~al.,}{Maiolino
  et~al.}{2008}]{MaiolinoAMAZEevolutionmassmetallicity2008}
Maiolino R.,  et~al., 2008, \mn@doi [A\&A] {10.1051/0004-6361:200809678}, 488,
  463

\bibitem[\protect\citeauthoryear{Mannucci et~al.,}{Mannucci
  et~al.}{2009}]{MannucciLSDLymanbreakgalaxies2009}
Mannucci F.,  et~al., 2009, \mn@doi [MNRAS] {10.1111/j.1365-2966.2009.15185.x},
  398, 1915

\bibitem[\protect\citeauthoryear{Mannucci, Cresci, Maiolino, Marconi  \&
  Gnerucci}{Mannucci et~al.}{2010}]{Mannuccifundamentalrelationmass2010}
Mannucci F.,  Cresci G.,  Maiolino R.,  Marconi A.,   Gnerucci A.,  2010,
  \mn@doi [MNRAS] {10.1111/j.1365-2966.2010.17291.x}, 408, 2115

\bibitem[\protect\citeauthoryear{Maraston, Daddi, Renzini, Cimatti, Dickinson,
  Papovich, Pasquali  \& Pirzkal}{Maraston
  et~al.}{2006}]{MarastonEvidenceTPAGBStars2006}
Maraston C.,  Daddi E.,  Renzini A.,  Cimatti A.,  Dickinson M.,  Papovich C.,
  Pasquali A.,   Pirzkal N.,  2006, \mn@doi [ApJ] {10.1086/508143}, 652, 85

\bibitem[\protect\citeauthoryear{Marchesini, {van Dokkum},
  F{\"o}rster~Schreiber, Franx, Labb{\'e}  \& Wuyts}{Marchesini
  et~al.}{2009}]{MarchesiniEvolutionStellarMass2009}
Marchesini D.,  {van Dokkum} P.~G.,  F{\"o}rster~Schreiber N.~M.,  Franx M.,
  Labb{\'e} I.,   Wuyts S.,  2009, \mn@doi [ApJ]
  {10.1088/0004-637X/701/2/1765}, 701, 1765

\bibitem[\protect\citeauthoryear{Marino et~al.,}{Marino
  et~al.}{2013}]{MarinoO3N2N2abundance2013}
Marino R.~A.,  et~al., 2013, \mn@doi [A\&A] {10.1051/0004-6361/201321956}, 559,
  A114

\bibitem[\protect\citeauthoryear{Martin et~al.,}{Martin
  et~al.}{2005}]{MartinGalaxyEvolutionExplorer2005}
Martin D.~C.,  et~al., 2005, \mn@doi [ApJ] {10.1086/426387}, 619, L1

\bibitem[\protect\citeauthoryear{McCracken et~al.,}{McCracken
  et~al.}{2012}]{McCrackenUltraVISTAnewultradeep2012}
McCracken H.~J.,  et~al., 2012, \mn@doi [A\&A] {10.1051/0004-6361/201219507},
  544, A156

\bibitem[\protect\citeauthoryear{Morganti}{Morganti}{2017}]{MorgantimanyroutesAGN2017}
Morganti R.,  2017, \mn@doi [Front. Astron. Space Sci.]
  {10.3389/fspas.2017.00042}, 4, 42

\bibitem[\protect\citeauthoryear{Netzer \& Trakhtenbrot}{Netzer \&
  Trakhtenbrot}{2007}]{NetzerCosmicEvolutionMass2007}
Netzer H.,  Trakhtenbrot B.,  2007, \mn@doi [ApJ] {10.1086/509650}, 654, 754

\bibitem[\protect\citeauthoryear{Nobels, Schaye, Schaller, Bah{\'e}  \&
  Chaikin}{Nobels et~al.}{2022}]{NobelsinterplayAGNfeedback2022}
Nobels F. S.~J.,  Schaye J.,  Schaller M.,  Bah{\'e} Y.~M.,   Chaikin E.,
  2022, \mn@doi [MNRAS] {10.1093/mnras/stac2061}, 515, 4838

\bibitem[\protect\citeauthoryear{Noeske et~al.,}{Noeske
  et~al.}{2007}]{NoeskeStarFormationAEGIS2007}
Noeske K.~G.,  et~al., 2007, \mn@doi [ApJL] {10.1086/517926}, 660, L43

\bibitem[\protect\citeauthoryear{Oesch et~al.,}{Oesch
  et~al.}{2016}]{OeschREMARKABLYLUMINOUSGALAXY2016}
Oesch P.~A.,  et~al., 2016, \mn@doi [ApJ] {10.3847/0004-637X/819/2/129}, 819,
  129

\bibitem[\protect\citeauthoryear{Oliver et~al.,}{Oliver
  et~al.}{2012}]{OliverHerschelMultitieredExtragalactic2012}
Oliver S.~J.,  et~al., 2012, \mn@doi [MNRAS]
  {10.1111/j.1365-2966.2012.20912.x}, 424, 1614

\bibitem[\protect\citeauthoryear{Pagel, Edmunds, Blackwell, Chun  \&
  Smith}{Pagel et~al.}{1979}]{PagelcompositionIIregions1979}
Pagel B. E.~J.,  Edmunds M.~G.,  Blackwell D.~E.,  Chun M.~S.,   Smith G.,
  1979, \mn@doi [MNRAS] {10.1093/mnras/189.1.95}, 189, 95

\bibitem[\protect\citeauthoryear{Papovich, Dickinson  \& Ferguson}{Papovich
  et~al.}{2001}]{PapovichStellarPopulationsEvolution2001}
Papovich C.,  Dickinson M.,   Ferguson H.~C.,  2001, \mn@doi [ApJ]
  {10.1086/322412}, 559, 620

\bibitem[\protect\citeauthoryear{Peeples \& Shankar}{Peeples \&
  Shankar}{2011}]{PeeplesConstraintsstarformation2011}
Peeples M.~S.,  Shankar F.,  2011, \mn@doi [MNRAS]
  {10.1111/j.1365-2966.2011.19456.x}, 417, 2962

\bibitem[\protect\citeauthoryear{Peng, Maiolino  \& Cochrane}{Peng
  et~al.}{2015}]{PengStrangulationprimarymechanism2015}
Peng Y.,  Maiolino R.,   Cochrane R.,  2015, \mn@doi [Nature]
  {10.1038/nature14439}, 521, 192

\bibitem[\protect\citeauthoryear{Pettini \& Pagel}{Pettini \&
  Pagel}{2004}]{PettiniOIIINIIabundance2004}
Pettini M.,  Pagel B. E.~J.,  2004, \mn@doi [MNRAS]
  {10.1111/j.1365-2966.2004.07591.x}, 348, L59

\bibitem[\protect\citeauthoryear{Pforr, Maraston  \& Tonini}{Pforr
  et~al.}{2012}]{PforrRecoveringgalaxystellar2012}
Pforr J.,  Maraston C.,   Tonini C.,  2012, \mn@doi [MNRAS]
  {10.1111/j.1365-2966.2012.20848.x}, 422, 3285

\bibitem[\protect\citeauthoryear{Pillepich et~al.,}{Pillepich
  et~al.}{2018}]{PillepichFirstresultsIllustrisTNG2018}
Pillepich A.,  et~al., 2018, \mn@doi [MNRAS] {10.1093/mnras/stx3112}, 475, 648

\bibitem[\protect\citeauthoryear{Pilyugin \& Grebel}{Pilyugin \&
  Grebel}{2016}]{PilyuginNewcalibrationsabundance2016}
Pilyugin L.~S.,  Grebel E.~K.,  2016, \mn@doi [MNRAS] {10.1093/mnras/stw238},
  457, 3678

\bibitem[\protect\citeauthoryear{Pilyugin \& Thuan}{Pilyugin \&
  Thuan}{2005}]{PilyuginOxygenAbundanceDetermination2005}
Pilyugin L.~S.,  Thuan T.~X.,  2005, \mn@doi [ApJ] {10.1086/432408}, 631, 231

\bibitem[\protect\citeauthoryear{Pilyugin, V{\'i}lchez  \& Thuan}{Pilyugin
  et~al.}{2010}]{PilyuginNewImprovedCalibration2010}
Pilyugin L.~S.,  V{\'i}lchez J.~M.,   Thuan T.~X.,  2010, \mn@doi [ApJ]
  {10.1088/0004-637X/720/2/1738}, 720, 1738

\bibitem[\protect\citeauthoryear{Pilyugin, Grebel  \& Mattsson}{Pilyugin
  et~al.}{2012}]{PilyuginCounterpartmethodabundance2012}
Pilyugin L.~S.,  Grebel E.~K.,   Mattsson L.,  2012, \mn@doi [MNRAS]
  {10.1111/j.1365-2966.2012.21398.x}, 424, 2316

\bibitem[\protect\citeauthoryear{{Planck Collaboration} et~al.,}{{Planck
  Collaboration} et~al.}{2016}]{PlanckCollaborationPlanck2015results2016}
{Planck Collaboration} et~al., 2016, \mn@doi [A\&A]
  {10.1051/0004-6361/201525830}, 594, A13

\bibitem[\protect\citeauthoryear{{R Core Team}}{{R Core
  Team}}{2020}]{RCoreTeamLanguageEnvironmentStatistical2020}
{R Core Team} 2020, R: {{A Language}} and {{Environment}} for {{Statistical}}
  {{Computing}}, R Foundation for Statistical Computing

\bibitem[\protect\citeauthoryear{Robotham}{Robotham}{2016a}]{RobothamCelestialCommonastronomical2016}
Robotham A. S.~G.,  2016a, Astrophysics Source Code Library, p. ascl:1602.011

\bibitem[\protect\citeauthoryear{Robotham}{Robotham}{2016b}]{RobothammagicaxisPrettyscientific2016}
Robotham A. S.~G.,  2016b, Astrophysics Source Code Library, p. ascl:1604.004

\bibitem[\protect\citeauthoryear{Robotham et~al.,}{Robotham
  et~al.}{2014}]{RobothamGalaxyMassAssembly2014}
Robotham A. S.~G.,  et~al., 2014, \mn@doi [MNRAS] {10.1093/mnras/stu1604}, 444,
  3986

\bibitem[\protect\citeauthoryear{Robotham, Davies, Driver, Koushan, Taranu,
  Casura  \& Liske}{Robotham
  et~al.}{2018}]{RobothamProFoundSourceExtraction2018}
Robotham A. S.~G.,  Davies L. J.~M.,  Driver S.~P.,  Koushan S.,  Taranu D.~S.,
   Casura S.,   Liske J.,  2018, \mn@doi [MNRAS] {10.1093/mnras/sty440}, 476,
  3137

\bibitem[\protect\citeauthoryear{Robotham, Bellstedt, Lagos, Thorne, Davies,
  Driver  \& Bravo}{Robotham
  et~al.}{2020}]{RobothamProSpectgeneratingspectral2020}
Robotham A. S.~G.,  Bellstedt S.,  Lagos C. d.~P.,  Thorne J.~E.,  Davies
  L.~J.,  Driver S.~P.,   Bravo M.,  2020, \mn@doi [MNRAS]
  {10.1093/mnras/staa1116}, 495, 905

\bibitem[\protect\citeauthoryear{Salim, Lee, Ly, Brinchmann, Dav{\'e},
  Dickinson, Salzer  \& Charlot}{Salim
  et~al.}{2014}]{SalimCriticalLookMassMetallicityStar2014}
Salim S.,  Lee J.~C.,  Ly C.,  Brinchmann J.,  Dav{\'e} R.,  Dickinson M.,
  Salzer J.~J.,   Charlot S.,  2014, \mn@doi [ApJ]
  {10.1088/0004-637X/797/2/126}, 797, 126

\bibitem[\protect\citeauthoryear{S{\'a}nchez et~al.,}{S{\'a}nchez
  et~al.}{2013}]{SanchezMassmetallicityrelationexplored2013}
S{\'a}nchez S.~F.,  et~al., 2013, \mn@doi [A\&A] {10.1051/0004-6361/201220669},
  554, A58

\bibitem[\protect\citeauthoryear{S{\'a}nchez et~al.,}{S{\'a}nchez
  et~al.}{2017}]{Sanchezmassmetallicityrelationrevisited2017}
S{\'a}nchez S.~F.,  et~al., 2017, \mn@doi [MNRAS] {10.1093/mnras/stx808}, 469,
  2121

\bibitem[\protect\citeauthoryear{S{\'a}nchez et~al.,}{S{\'a}nchez
  et~al.}{2019}]{SanchezSAMIgalaxysurvey2019}
S{\'a}nchez S.~F.,  et~al., 2019, \mn@doi [MNRAS] {10.1093/mnras/stz019}, 484,
  3042

\bibitem[\protect\citeauthoryear{Sanders et~al.,}{Sanders
  et~al.}{2007}]{SandersSCOSMOSSpitzerLegacy2007}
Sanders D.~B.,  et~al., 2007, \mn@doi [ApJS] {10.1086/517885}, 172, 86

\bibitem[\protect\citeauthoryear{Sanders et~al.,}{Sanders
  et~al.}{2015}]{SandersMOSDEFSurveyMass2015}
Sanders R.~L.,  et~al., 2015, \mn@doi [ApJ] {10.1088/0004-637X/799/2/138}, 799,
  138

\bibitem[\protect\citeauthoryear{Sanders et~al.,}{Sanders
  et~al.}{2018}]{SandersMOSDEFSurveyStellar2018}
Sanders R.~L.,  et~al., 2018, \mn@doi [ApJ] {10.3847/1538-4357/aabcbd}, 858, 99

\bibitem[\protect\citeauthoryear{Sanders et~al.,}{Sanders
  et~al.}{2020}]{SandersMOSDEFsurveydirectmethod2020}
Sanders R.~L.,  et~al., 2020, \mn@doi [MNRAS] {10.1093/mnras/stz3032}, 491,
  1427

\bibitem[\protect\citeauthoryear{Sanders et~al.,}{Sanders
  et~al.}{2021}]{SandersMOSDEFSurveyEvolution2021}
Sanders R.~L.,  et~al., 2021, \mn@doi [ApJ] {10.3847/1538-4357/abf4c1}, 914, 19

\bibitem[\protect\citeauthoryear{Sanders et~al.,}{Sanders
  et~al.}{2022}]{SandersCOEmissionMolecular2022}
Sanders R.~L.,  et~al., 2022, {{CO Emission}}, {{Molecular Gas}}, and
  {{Metallicity}} in {{Main-Sequence Star-Forming Galaxies}} at
  \$z\textbackslash sim2.3\$

\bibitem[\protect\citeauthoryear{Savaglio et~al.,}{Savaglio
  et~al.}{2005}]{SavaglioGeminiDeepDeep2005}
Savaglio S.,  et~al., 2005, \mn@doi [ApJ] {10.1086/497331}, 635, 260

\bibitem[\protect\citeauthoryear{Schaye et~al.,}{Schaye
  et~al.}{2015}]{SchayeEAGLEprojectsimulating2015}
Schaye J.,  et~al., 2015, \mn@doi [MNRAS] {10.1093/mnras/stu2058}, 446, 521

\bibitem[\protect\citeauthoryear{Scoville et~al.,}{Scoville
  et~al.}{2017}]{ScovilleEvolutionInterstellarMedium2017}
Scoville N.,  et~al., 2017, \mn@doi [ApJ] {10.3847/1538-4357/aa61a0}, 837, 150

\bibitem[\protect\citeauthoryear{Shangguan, Ho, Bauer, Wang  \&
  Treister}{Shangguan et~al.}{2020}]{ShangguanAGNFeedbackStar2020}
Shangguan J.,  Ho L.~C.,  Bauer F.~E.,  Wang R.,   Treister E.,  2020, \mn@doi
  [ApJ] {10.3847/1538-4357/aba8a1}, 899, 112

\bibitem[\protect\citeauthoryear{Silk \& Rees}{Silk \&
  Rees}{1998}]{SilkQuasarsgalaxyformation1998}
Silk J.,  Rees M.~J.,  1998, A\&A, 331, L1

\bibitem[\protect\citeauthoryear{Somerville \& Dav{\'e}}{Somerville \&
  Dav{\'e}}{2015}]{SomervillePhysicalModelsGalaxy2015}
Somerville R.~S.,  Dav{\'e} R.,  2015, \mn@doi [ARA\&A]
  {10.1146/annurev-astro-082812-140951}, 53, 51

\bibitem[\protect\citeauthoryear{Speagle, Steinhardt, Capak  \&
  Silverman}{Speagle et~al.}{2014}]{SpeagleHighlyConsistentFramework2014}
Speagle J.~S.,  Steinhardt C.~L.,  Capak P.~L.,   Silverman J.~D.,  2014,
  \mn@doi [ApJS] {10.1088/0067-0049/214/2/15}, 214, 15

\bibitem[\protect\citeauthoryear{Spitoni, Calura, Mignoli, Gilli, Aguirre  \&
  Gallazzi}{Spitoni et~al.}{2020}]{SpitoniConnectiongalacticdownsizing2020}
Spitoni E.,  Calura F.,  Mignoli M.,  Gilli R.,  Aguirre V.~S.,   Gallazzi A.,
  2020, \mn@doi [A\&A] {10.1051/0004-6361/202037879}, 642, A113

\bibitem[\protect\citeauthoryear{{Storchi-Bergmann}, Schmitt, Calzetti  \&
  Kinney}{{Storchi-Bergmann}
  et~al.}{1998}]{Storchi-BergmannChemicalAbundanceCalibrations1998}
{Storchi-Bergmann} T.,  Schmitt H.~R.,  Calzetti D.,   Kinney A.~L.,  1998,
  \mn@doi [AJ] {10.1086/300242}, 115, 909

\bibitem[\protect\citeauthoryear{Stott et~al.,}{Stott
  et~al.}{2013}]{Stottfundamentalmetallicityrelation2013}
Stott J.~P.,  et~al., 2013, \mn@doi [MNRAS] {10.1093/mnras/stt1641}, 436, 1130

\bibitem[\protect\citeauthoryear{Taylor \& Kobayashi}{Taylor \&
  Kobayashi}{2015}]{TayloreffectsAGNfeedback2015}
Taylor P.,  Kobayashi C.,  2015, \mn@doi [MNRAS] {10.1093/mnras/stv139}, 448,
  1835

\bibitem[\protect\citeauthoryear{Thorne et~al.,}{Thorne
  et~al.}{2021}]{ThorneDeepExtragalacticVIsible2021}
Thorne J.~E.,  et~al., 2021, \mn@doi [MNRAS] {10.1093/mnras/stab1294}, 505, 540

\bibitem[\protect\citeauthoryear{Thorne et~al.,}{Thorne
  et~al.}{2022}]{ThorneDeepExtragalacticVIsible2022}
Thorne J.~E.,  et~al., 2022, \mn@doi [MNRAS] {10.1093/mnras/stab3208}, 509,
  4940

\bibitem[\protect\citeauthoryear{Tinsley}{Tinsley}{1980}]{TinsleyEvolutionStarsGas1980}
Tinsley B.~M.,  1980, \mn@doi [Fundam. Cosm. Phys.]
  {10.48550/arXiv.2203.02041}, 5, 287

\bibitem[\protect\citeauthoryear{Topping et~al.,}{Topping
  et~al.}{2021}]{ToppingMOSDEFsurveymassmetallicity2021}
Topping M.~W.,  et~al., 2021, \mn@doi [MNRAS] {10.1093/mnras/stab1793}, 506,
  1237

\bibitem[\protect\citeauthoryear{Torrey, Vogelsberger, Genel, Sijacki, Springel
   \& Hernquist}{Torrey et~al.}{2014}]{Torreymodelcosmologicalsimulations2014}
Torrey P.,  Vogelsberger M.,  Genel S.,  Sijacki D.,  Springel V.,   Hernquist
  L.,  2014, \mn@doi [MNRAS] {10.1093/mnras/stt2295}, 438, 1985

\bibitem[\protect\citeauthoryear{Torrey et~al.,}{Torrey
  et~al.}{2019}]{Torreyevolutionmassmetallicityrelation2019}
Torrey P.,  et~al., 2019, \mn@doi [MNRAS] {10.1093/mnras/stz243}, 484, 5587

\bibitem[\protect\citeauthoryear{Tremonti et~al.,}{Tremonti
  et~al.}{2004}]{TremontiOriginMassMetallicity2004}
Tremonti C.~A.,  et~al., 2004, \mn@doi [ApJ] {10.1086/423264}, 613, 898

\bibitem[\protect\citeauthoryear{Tumlinson, Peeples  \& Werk}{Tumlinson
  et~al.}{2017}]{TumlinsonCircumgalacticMedium2017}
Tumlinson J.,  Peeples M.~S.,   Werk J.~K.,  2017, \mn@doi [ARA\&A]
  {10.1146/annurev-astro-091916-055240}, 55, 389

\bibitem[\protect\citeauthoryear{Vale~Asari, Stasi{\'n}ska, Cid~Fernandes,
  Gomes, Schlickmann, Mateus  \& Schoenell}{Vale~Asari
  et~al.}{2009}]{ValeAsarievolutionmassmetallicityrelation2009}
Vale~Asari N.,  Stasi{\'n}ska G.,  Cid~Fernandes R.,  Gomes J.~M.,  Schlickmann
  M.,  Mateus A.,   Schoenell W.,  2009, \mn@doi [MNRAS]
  {10.1111/j.1745-3933.2009.00664.x}, 396, L71

\bibitem[\protect\citeauthoryear{Vazdekis, Koleva, Ricciardelli, R{\"o}ck  \&
  {Falc{\'o}n-Barroso}}{Vazdekis
  et~al.}{2016}]{VazdekisUVextendedEMILESstellar2016}
Vazdekis A.,  Koleva M.,  Ricciardelli E.,  R{\"o}ck B.,   {Falc{\'o}n-Barroso}
  J.,  2016, \mn@doi [MNRAS] {10.1093/mnras/stw2231}, 463, 3409

\bibitem[\protect\citeauthoryear{Vogelsberger et~al.,}{Vogelsberger
  et~al.}{2014}]{VogelsbergerIntroducingIllustrisProject2014}
Vogelsberger M.,  et~al., 2014, \mn@doi [MNRAS] {10.1093/mnras/stu1536}, 444,
  1518

\bibitem[\protect\citeauthoryear{Weldon, Ly  \& Cooper}{Weldon
  et~al.}{2020}]{Weldonstellarpopulationmetalpoor2020}
Weldon A.,  Ly C.,   Cooper M.,  2020, \mn@doi [MNRAS] {10.1093/mnras/stz3047},
  491, 2254

\bibitem[\protect\citeauthoryear{Worthey}{Worthey}{1994}]{WortheyComprehensivestellarpopulation1994}
Worthey G.,  1994, \mn@doi [ApJS] {10.1086/192096}, 95, 107

\bibitem[\protect\citeauthoryear{Wright et~al.,}{Wright
  et~al.}{2010}]{WrightWidefieldInfraredSurvey2010}
Wright E.~L.,  et~al., 2010, \mn@doi [AJ] {10.1088/0004-6256/140/6/1868}, 140,
  1868

\bibitem[\protect\citeauthoryear{Wuyts, Franx, Cox, Hernquist, Hopkins,
  Robertson  \& {van Dokkum}}{Wuyts
  et~al.}{2009}]{WuytsRecoveringStellarPopulation2009}
Wuyts S.,  Franx M.,  Cox T.~J.,  Hernquist L.,  Hopkins P.~F.,  Robertson
  B.~E.,   {van Dokkum} P.~G.,  2009, \mn@doi [ApJ]
  {10.1088/0004-637X/696/1/348}, 696, 348

\bibitem[\protect\citeauthoryear{Yates, Kauffmann  \& Guo}{Yates
  et~al.}{2012}]{Yatesrelationmetallicitystellar2012}
Yates R.~M.,  Kauffmann G.,   Guo Q.,  2012, \mn@doi [MNRAS]
  {10.1111/j.1365-2966.2012.20595.x}, 422, 215

\bibitem[\protect\citeauthoryear{Zahid, Kewley  \& Bresolin}{Zahid
  et~al.}{2011}]{ZahidMassMetallicityLuminosityMetallicityRelations2011}
Zahid H.~J.,  Kewley L.~J.,   Bresolin F.,  2011, \mn@doi [ApJ]
  {10.1088/0004-637X/730/2/137}, 730, 137

\bibitem[\protect\citeauthoryear{Zahid et~al.,}{Zahid
  et~al.}{2014}]{ZahidFMOSCOSMOSSurveyStarforming2014}
Zahid H.~J.,  et~al., 2014, \mn@doi [Astrophys. J.]
  {10.1088/0004-637X/792/1/75}, 792, 75

\bibitem[\protect\citeauthoryear{Zamojski et~al.,}{Zamojski
  et~al.}{2007}]{ZamojskiDeepGALEXImaging2007}
Zamojski M.~A.,  et~al., 2007, \mn@doi [ApJSS] {10.1086/516593}, 172, 468

\bibitem[\protect\citeauthoryear{Zhou, Merrifield  \&
  {Arag{\'o}n-Salamanca}}{Zhou
  et~al.}{2022}]{ZhouSemianalyticspectralfitting2022}
Zhou S.,  Merrifield M.,   {Arag{\'o}n-Salamanca} A.,  2022, \mn@doi [MNRAS]
  {10.1093/mnras/stac1279}, 513, 5446

\bibitem[\protect\citeauthoryear{{de Jong}, Verdoes~Kleijn, Kuijken  \&
  Valentijn}{{de Jong} et~al.}{2013a}]{deJongKiloDegreeSurvey2013}
{de Jong} J. T.~A.,  Verdoes~Kleijn G.~A.,  Kuijken K.~H.,   Valentijn E.~A.,
  2013a, \mn@doi [Exp. Astron.] {10.1007/s10686-012-9306-1}, 35, 25

\bibitem[\protect\citeauthoryear{{de Jong} et~al.,}{{de Jong}
  et~al.}{2013b}]{deJongKiloDegreeSurvey2013a}
{de Jong} J. T.~A.,  et~al., 2013b, The Messenger, 154, 44

\bibitem[\protect\citeauthoryear{{de Rossi}, Tissera  \& Scannapieco}{{de
  Rossi} et~al.}{2007}]{deRossiCluesoriginfundamental2007}
{de Rossi} M.~E.,  Tissera P.~B.,   Scannapieco C.,  2007, \mn@doi [MNRAS]
  {10.1111/j.1365-2966.2006.11150.x}, 374, 323

\makeatother
\end{thebibliography}



\appendix

\section{Impact of metallicity implementation on derived galaxy properties}\label{App:Zimplementation}
\begin{figure*}
    \centering
    \includegraphics[width = 0.47 \linewidth]{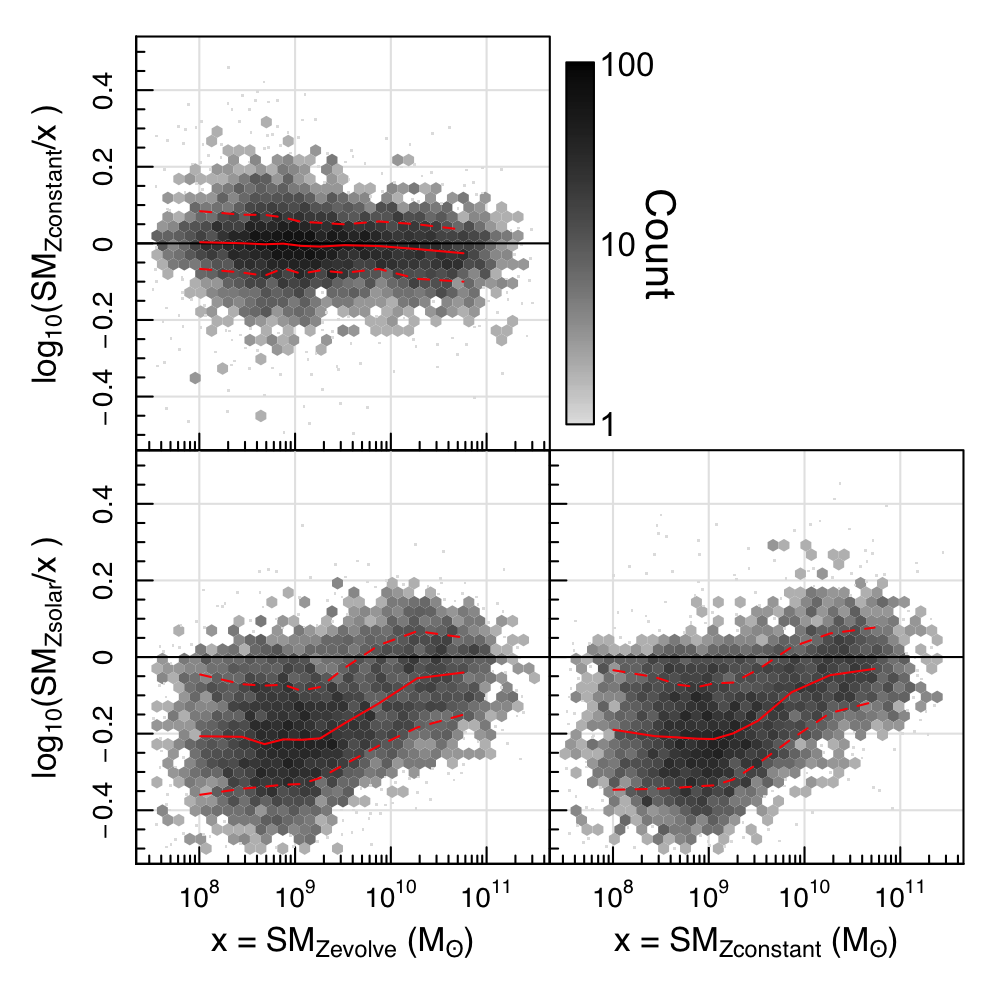}
    \includegraphics[width = 0.47 \linewidth]{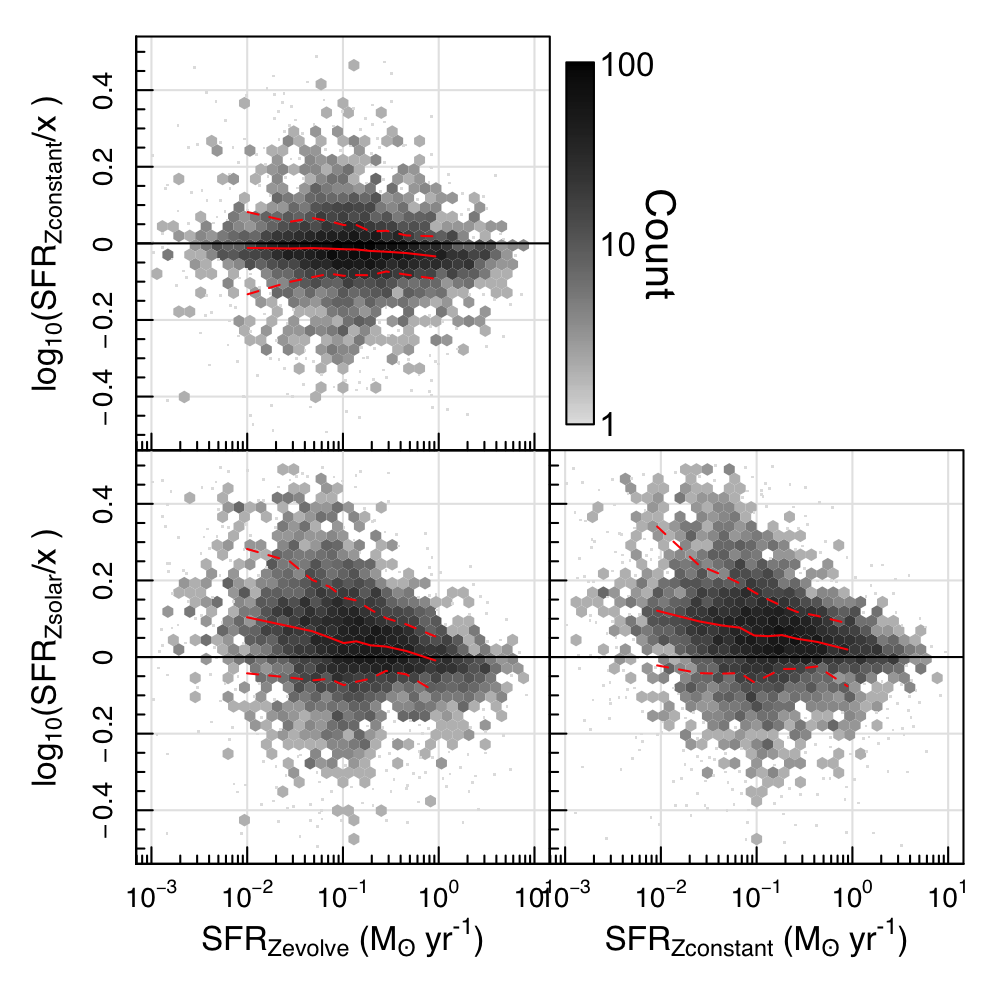}\\
    \includegraphics[width = 0.47 \linewidth]{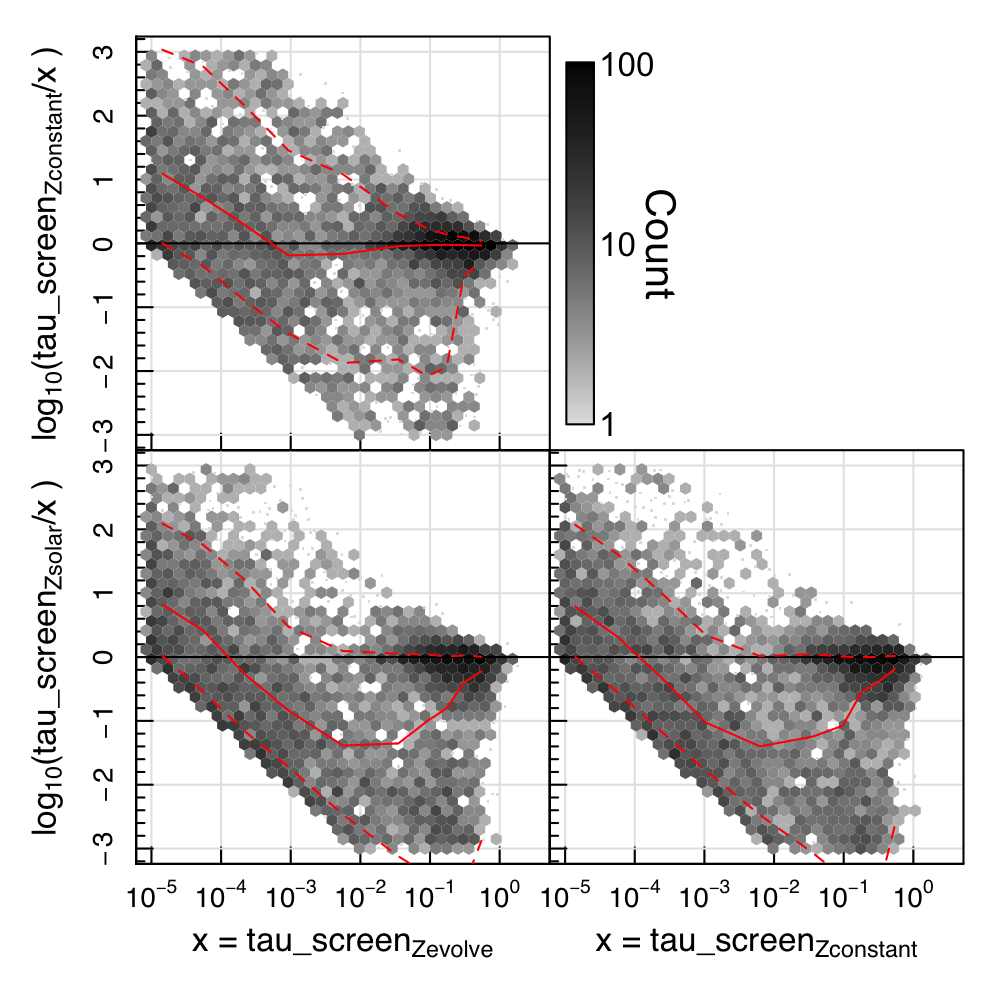}
    \includegraphics[width = 0.47 \linewidth]{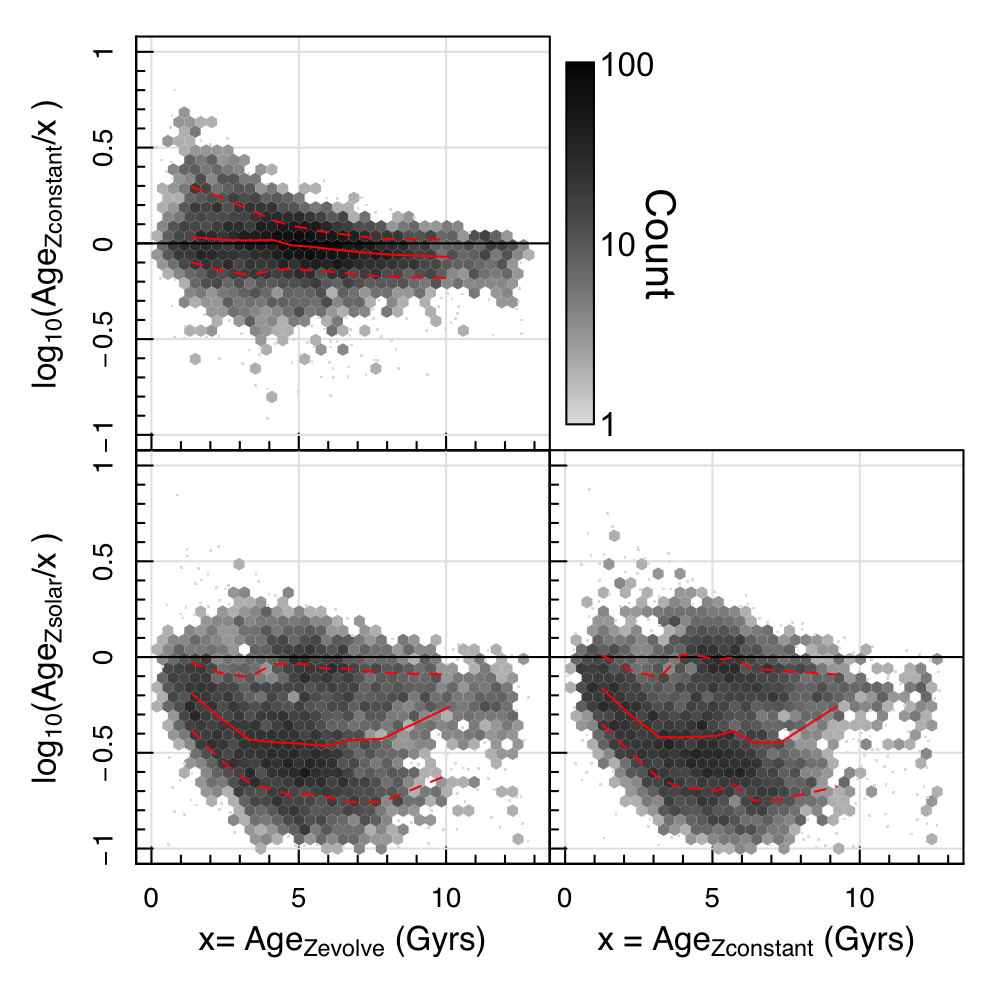}\\
    \caption[Comparison of derived properties when changing metallicity implementation]{
    \add{Impact of changing metallicity implementation on derived stellar masses (top left), SFRs (topright), ISM dust opaciticies (tau\_screen, lower left), and half-mass ages (lower right).
    The different metallicity implementations are assuming constant solar ($Z=0.02$; Zsolar), constant but the value of the metallicity is free (Zconstant), and using the evolving metallicity implementation described in Section~\ref{sec:ProSpect} (Zevolve). 
    The running median and $1\sigma$ ranges are shown as the red solid and dashed lines respectively. 
    }}
    \label{fig:Zchange_params}
\end{figure*}

\add{To demonstrate the impact of different metallicty implementations on derived galaxy properties, we use the sample of $\sim7,000$ low-redshift ($z<0.06$) GAMA galaxies studied in \cite{BellstedtGalaxyMassAssembly2020b}. 
We fit this sample using same implementation of \textsc{ProSpect} as described in \cite{BellstedtGalaxyMassAssembly2020b} including SFH (\texttt{massfunc\_snorm\_trunc}), band dependent error floors, and dust priors. 
However, we fit the sample using three different metallicity prescriptions; 
\begin{enumerate}
    \item using the linear mapping of metallicity growth to mass growth used in this work (\texttt{Zfunc\_massmap\_lin} - `Zevolve'),
    \item assuming that all stars in a galaxy formed with the same metallicity but allowing this metallicity to be modelled as a free parameter (`Zconstant'), 
    \item or assuming that all stars formed at solar metallicity ($Z=0.02$, as adopted by \citealt{BoquienCIGALEpythonCode2019} - 'Zsolar').
\end{enumerate}
}

\add{
Figure~\ref{fig:Zchange_params} shows comparisons of the stellar mass, SFR, half-mass age, and dust opacity (tau\_screen) when changing the metallicity prescription. 
It is immediately obvious that fixing the metallicity to $Z=0.02$ results in systematic offsets in all properties when compared to the `Zevolve' and `Zconstant' cases. 
Fixing the metallicity to $Z=0.02$ has the largest impact on the derived dust opacity, which can be offset by up to $\sim3$\,dex, however, the derived age is also significantly impacted with systematic offsets of $\sim0.5$\,dex at intermediate ages. 
The stellar masses, and SFRs are not systematically offset when allowing the metallicity value to change with either a constant or evolving metallicity history. 
However, age can be underestimated by $\sim0.1\,$dex at high ages when assuming a constant metallicity history as opposed to an evolving one due to the age-metallicty degeneracy \citep{WortheyComprehensivestellarpopulation1994,PapovichStellarPopulationsEvolution2001}. 
}

\section{Impact of chosen stellar population models on derived galaxy properties}\label{App:BPASSBC03}
\begin{figure*}
    \centering
    \includegraphics[width = \linewidth]{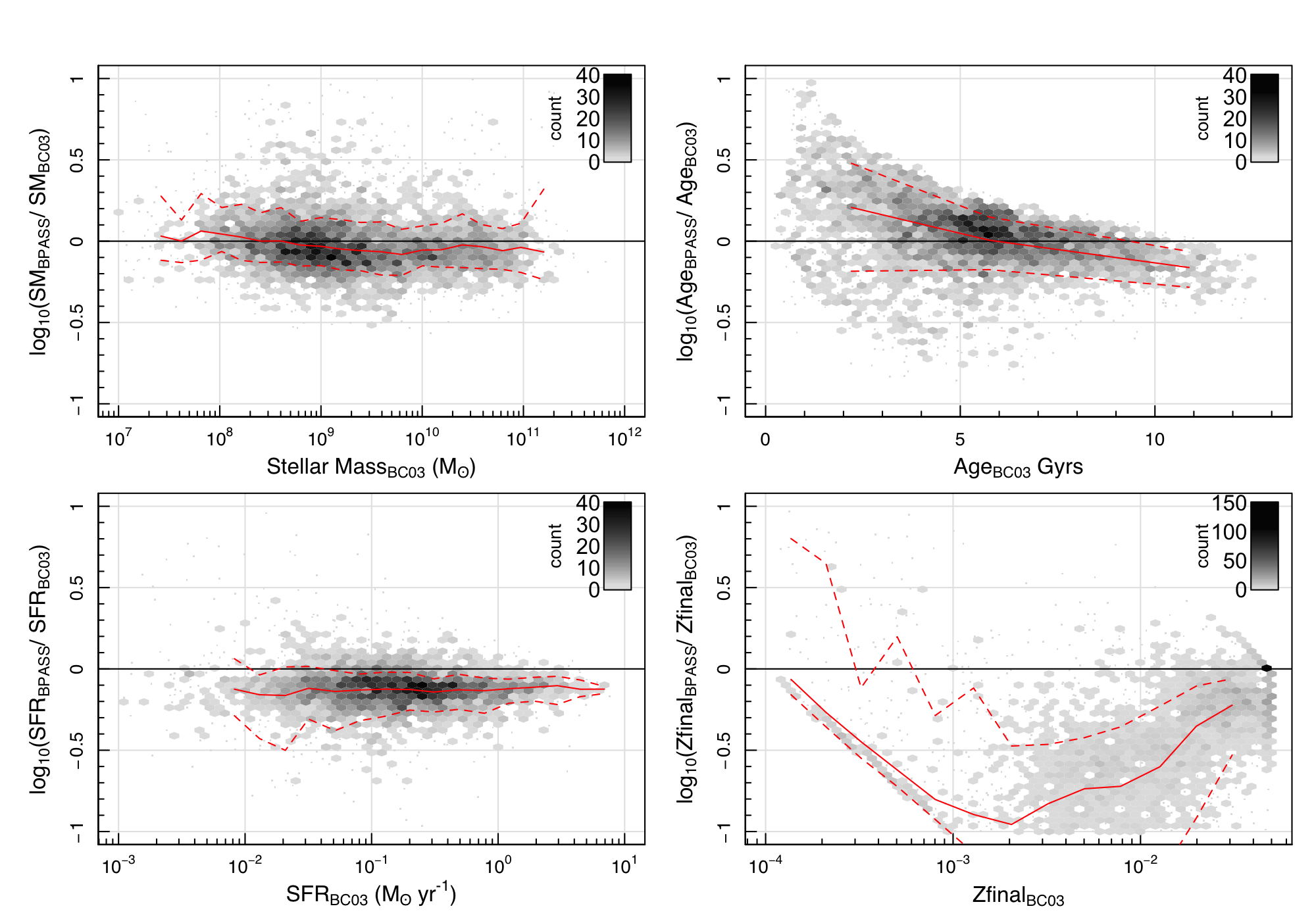}
    \caption[Comparison of galaxy properties derived using BC03 and BPASS.]{Comparison of galaxy properties derived using BPASS to those derived using BC03. 
    We show the difference in stellar mass, star formation rate, half mass age, and metallicity.
    In each panel the running median and 1$\sigma$ range are shown as the red solid and dashed lines respectively.
    }
    \label{fig:BPASS_BC03_ParameterComparisons}
\end{figure*}

\begin{figure}
    \centering
    \includegraphics[width = \linewidth]{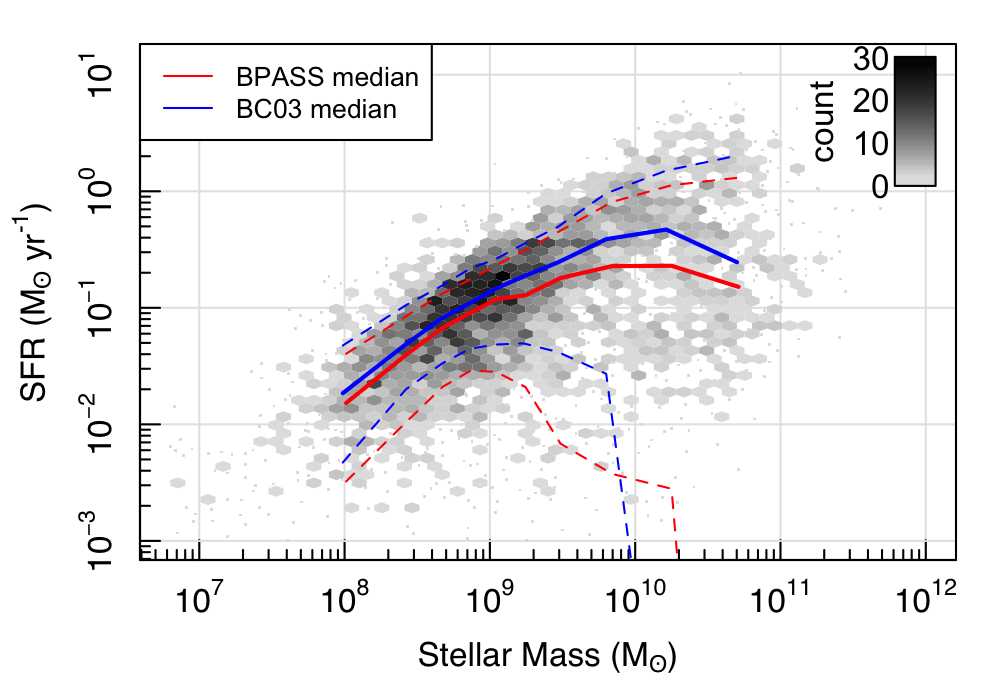}
    \caption[Comparison of SFMS derived using BC03 and BPASS]{Comparison of the SFMS derived using BC03 and BPASS at $z=0$.
    We show the underlying distribution of galaxies using the BPASS SPS models in the gay 2D histogram with the running median and 1$\sigma$ ranges in solid and dashed red respectively.
    We also show the running median and 1$\sigma$ range from the BC03 version in blue. 
    }
    \label{fig:BPASS_BC03_SFMS}
\end{figure}

\begin{figure}
    \centering
    \includegraphics[width = \linewidth]{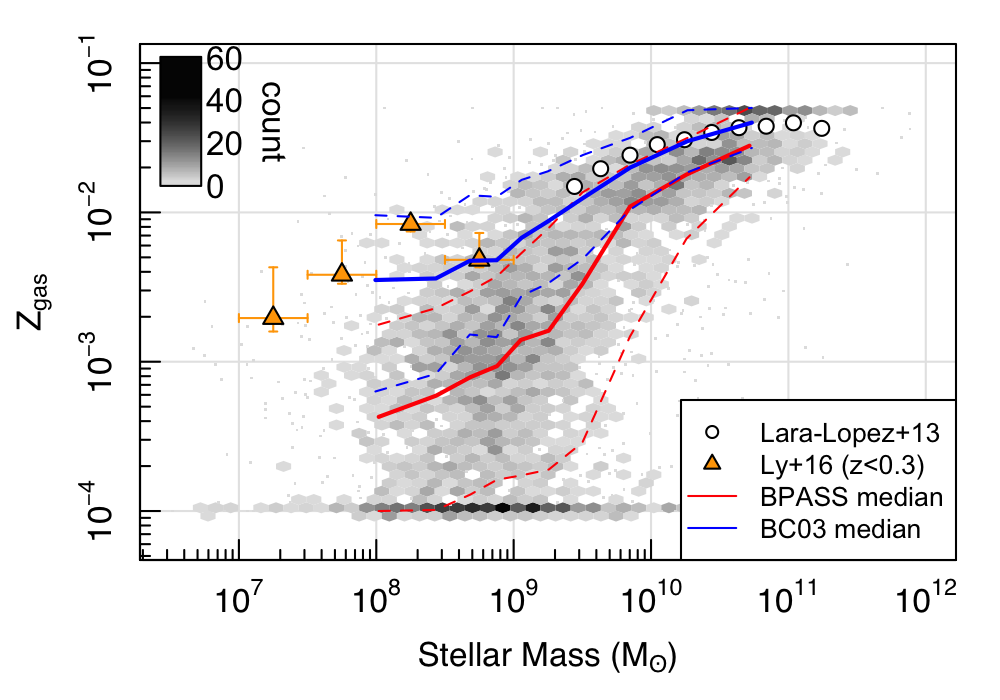}
    \caption[Comparison of the MZR derived using BC03 and BPASS]{Comparison of the MZR derived using BC03 (blue) and BPASS (red).
    As per Figure~\ref{fig:BPASS_BC03_SFMS}, the underlying 2D histogram shows the distribution of galaxies when using BPASS. 
    For comparison we also show values from \citet{Lara-LopezGalaxyMassAssembly2013} and \cite{LyMetalAbundancesCosmic2016} both of which are measured through spectra.}
    \label{fig:BPASS_BC03_MZR}
\end{figure}

One of the choices required when fitting the SEDs of galaxies is the set of templates, known as stellar population synthesis (SPS) models, used to model the expected emission from stars of certain ages and metallicities.  
The \cite{BruzualStellarpopulationsynthesis2003} SPS model is the most commonly employed SPS model in broad-band SED fitting (see Figure~1 or Table~A1 of \citealt{ThorneDeepExtragalacticVIsible2021} for an outline of the SPS models employed by different SED fitting codes), while the E-MILES SPS models are typically used for full spectral fitting techniques due to the finer wavelength resolution.
However, E-MILES has limited wavelength coverage and does not cover the far ultraviolet for all age templates, and does not cover the near-infrared for the youngest age models (age $< 450\,$Myr).

The creation of SPS models is conceptually straightforward, but current SPS templates are limited by incomplete isochrone tables, incomplete stellar libraries, and poorly calibrated physics (see \citealt{ConroyModelingPanchromaticSpectral2013} for more details). 
Some of the challenges of SPS models include the potential importance of thermally pulsating asymptotic giant branch stars \citep{MarastonEvidenceTPAGBStars2006}, and the importance of binary evolution \citep{Eldridgeeffectstellarevolution2012,EldridgeBinaryPopulationSpectral2017}. 

As \citet[BC03]{BruzualStellarpopulationsynthesis2003} and \citet[BPASS]{EldridgeBinaryPopulationSpectral2017} are both implemented in \textsc{ProSpect}, we can compare the impact on derived galaxy properties when changing the employed SPS model\footnote{\citet[E-MILES]{VazdekisUVextendedEMILESstellar2016} is also incorporated in \textsc{ProSpect}, but due to the limited wavelength coverage, there are many additional caveats that would make these comparisons far more difficult.}.
Using the sample of $\sim7,000$ low redshift ($z<0.06$) GAMA galaxies used in \cite{BellstedtGalaxyMassAssembly2020b, BellstedtGalaxyMassAssembly2021} and the methods described in \cite{BellstedtGalaxyMassAssembly2020b} we re-fit the galaxies using the BPASS. 
The BPASS models include 13 metallicities spanning $Z = 10^{-5} - 0.04$, while BC03 only has 5 metallicity templates spanning $Z = 10^{-4} - 0.05$. 
However, BC03 includes a significantly finer sampling of ages with 196 age bins with ages less than 14\,Gyrs while BPASS only has 42 bins with ages less than 14\,Gyrs. 
BPASS has very high time resolution for young stars ($<100\,$Myrs) with age bins of millions of years, and very coarse resolution for older populations with some bins separated by 3 Gyrs. 

Figure~\ref{fig:BPASS_BC03_ParameterComparisons} shows comparisons between general galaxy properties derived using BPASS to those derived using BC03. 
The stellar masses show good agreement, however the stellar masses derived using BPASS are offset at high stellar masses at a level of approximately 0.1\,dex.  
The SFRs show a systematic offset where BPASS recovers systematically lower SFRs at the level of $\sim0.15\,$dex across the entire SFR range. 
To quantify the ages of galaxies we use the half mass age, which is the time at which the galaxy had formed half of its stellar mass. 
We find that BPASS recovers older ages for young galaxies, and younger ages for old galaxies than BC03.
We also show the difference in recovered metallicity when using BPASS, where the different choice of SPS model results in significantly different metallicity values than when using BC03. 
The straight line artefacts in both panels are a result of galaxies hitting the upper and lower limits of the metallicity templates.

Figure~\ref{fig:BPASS_BC03_SFMS} shows the impact of the offset between the BC03, and BPASS fits on the SFMS. 
As BPASS predicts systematically lower SFRs, the SFMS recovered using BPASS is below that recovered using BC03. 
Despite the difference in normalisation, both SPS models recover the same overall bending trend.

Figure~\ref{fig:BPASS_BC03_MZR} presents the recovered MZR using BC03 and BPASS. 
Due to the large offset in metallicities discussed previously, the MZR recovered using BPASS lies significantly below the MZR recovered using BC03.
Despite the offset in metallicity values between the two SPS codes, the derived MZR follow the same trend with low mass galaxies having lower gas metallicities than high mass galaxies. 
The scatter on the MZR relation is higher for the BPASS derived sample however. 
As the same trends exist for the SFMS and MZR when fitting with both BC03 and BPASS, we do not expect that changing SPS model will have an impact on the general trends we find in this work.
However, there would be differences in metallicity normalisation due to the differences discussed above. 

\section{Differences between the z=0 MZR from this work and Bellstedt et al. 2021}\label{app:B21Diff}
\begin{figure}
    \centering
    \includegraphics[width = \linewidth]{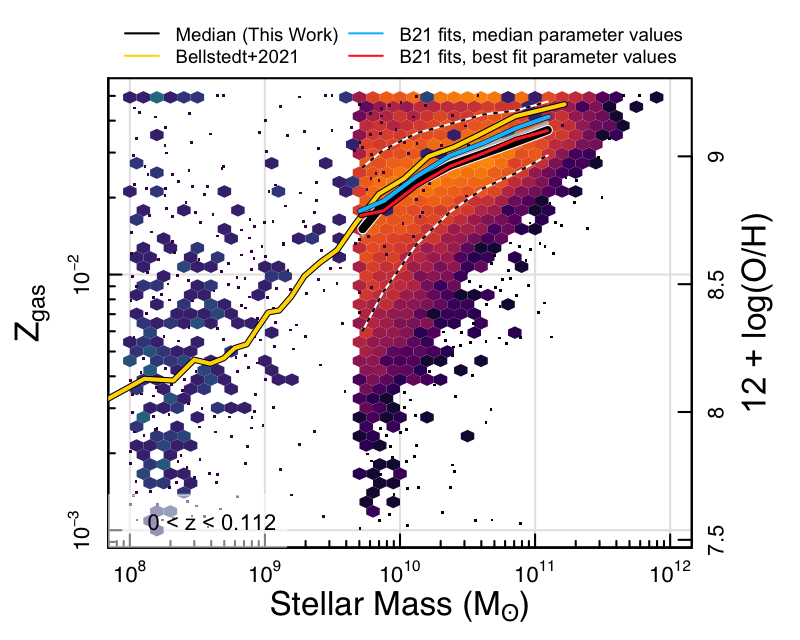}
    \caption{The lowest redshift panel from Figure~\ref{fig:MZR} with the addition of the median MZR derived using the SED fits presented in \citet[B21]{BellstedtGalaxyMassAssembly2021} using the same sample of galaxies used in this work. 
    We recalculate the median MZR using the median and the maximum-likelihood (best fit) parameter values provided by \citet{BellstedtGalaxyMassAssembly2021} which are shown as the blue and red lines respectively.}
    \label{fig:BellstedtMZR}
\end{figure}

Figure~\ref{fig:BellstedtMZR} reproduces the $z\approx0$ panel of Figure~\ref{fig:MZR} but also includes a recalculation of the MZR using the catalogue of SED fits from \citet{BellstedtGalaxyMassAssembly2021} (released with GAMA DR4\footnote{\url{http://www.gama-survey.org/dr4/}}) but with various changes. 
First, we extend the sample from \citet{BellstedtGalaxyMassAssembly2021} to match the sample used in this work. 
Specifically, this includes the addition of the G23 field, not using a $\Delta Z$ cut, and extending the redshift range from $z<0.06$ to $z<0.112$. 
The \citet{BellstedtGalaxyMassAssembly2021} results were derived using the median parameter value of both the mass and metallicity, whereas our parameter values are based on the maximum-likelihood (best fit) step of the posterior. 
As both the median and best fit parameter values are provided by \citet{BellstedtGalaxyMassAssembly2021} we recalculate the MZR using both to highlight the difference.
These are shown as the blue and red lines respectively. 
The difference between the blue and yellow lines highlights the impact from changing the sample selection to encompass a larger sample of galaxies as these are both extracted using the median parameter values. 
The difference between the blue and red lines highlights the effect of how the galaxy parameter values were chosen, where the best fit values recover a MZR relation with systematically lower metallicities at a given stellar mass. 
And finally, the difference between the red and black lines demonstrates that the incorporation of the AGN component in the SED fitting, and the change from a full MCMC chain to the \textsc{highlander} optimization routine does not significantly impact the derived MZR. 
Additionally the \cite{BellstedtGalaxyMassAssembly2021} MZR shown in the first panel of Figure~\ref{fig:MZR} is extracted using the metallicity and star formation histories and has been traced back to a lookback time of $1\,$Gyr rather than using the $Z_\text{gas}$ and stellar mass at $t=0$.


\bsp	
\label{lastpage}
\end{document}